%% file: bulk-14-star.tex
\documentclass[aps,prc,twocolumn,superscriptaddress,showpacs,floatfix]{revtex4-1}
\usepackage{color}
\usepackage{overpic,color}
\usepackage{subfigure}
\usepackage{amsmath,epsfig}
\usepackage{graphicx}
\usepackage{epstopdf}
\usepackage{dcolumn}
\usepackage{bm}
\usepackage{multirow}
\usepackage{hhline}
\usepackage{afterpage}
\usepackage{hyperref} 
\usepackage{float}
\usepackage{lineno}

\newcommand{\pt}{$p_{\mathrm{T}}$}
\newcommand{\snn}{$\sqrt{s_{\mathrm{NN}}}$}

\begin{document}

\title{Bulk Properties of the System Formed in Au+Au Collisions at \snn~= 14.5 GeV at STAR}

\input{author}









\begin{abstract}
We report systematic measurements of bulk properties of the system
created in Au+Au collisions at \snn~= 14.5~GeV recorded by
the STAR detector at the Relativistic Heavy Ion Collider (RHIC). The
transverse momentum spectra of $\pi^{\pm}$, $K^{\pm}$ and $p(\bar{p})$
are studied at mid-rapidity ($|y| < 0.1$) for nine centrality
intervals. The centrality, transverse momentum ($p_T$), and
pseudorapidity ($\eta$) dependence of inclusive charged particle
elliptic flow ($v_2$), and rapidity-odd charged particles directed
flow ($v_{1}$) results near mid-rapidity are also presented.  These
measurements are compared with the published results from Au+Au
collisions at other energies, and from Pb+Pb collisions at
\snn~= 2.76~TeV. The results at \snn~= 14.5 GeV
show similar behavior as established at other energies and fit well in
the energy dependence trend. These results are important as the 14.5
GeV energy fills the gap in $\mu_B$, which is of the order of 100 MeV,
between \snn~=11.5 and 19.6 GeV. Comparisons of the data
with UrQMD and AMPT models show poor agreement in general. 
\end{abstract}

\maketitle



\section{INTRODUCTION}
According to quantum chromodynamics (QCD), at very high temperature
and/or at high density, a de-confined phase of quarks and gluons is
expected to be present. At low temperature and low density, quarks and
gluons are confined inside hadrons. The exploration of the QCD phase
diagram, in the plane of temperature ($T$) and the baryon chemical
potential ($\mu_{B}$), is one of the primary objectives of high-energy
heavy-ion collision
experiments~\cite{TmuBsNN1,TmuBsNN2,TmuBsNN3,HICs1,HICs2,HICs3,HICs4}.
During the initial stages of Au+Au collisions at top RHIC energies,
there is evidence of a phase with partonic degrees of
freedom~\cite{HICs1,HICs2,HICs3,HICs4,expdof1,expdof2,expdof3,expdof4,expdof5,expdof6},
which later transits into one with hadronic degrees of
freedom~\cite{dof0,dof1,dof2,dof3}. Relevant evidence includes strong
suppression of high transverse momentum (\pt) hadron production in
Au+Au collisions relative to $p+p$
collisions~\cite{HICs1,HICs2,HICs3,HICs4,expdof1,expdof2,expdof3,expdof4},
large elliptic flow ($v_{2}$) for hadrons containing light as well as
strange and charm valence quarks, and the difference between baryon
and meson $v_{2}$ at intermediate \pt~\cite{expv2}. 

At $\mu_{B} =$ 0, the phase transition is a cross-over. This region is
well described by lattice QCD calculations~\cite{lqcd1,lqcd2}. At
larger $\mu_{B}$, a first-order phase transition is suggested by
lattice QCD~\cite{lqcd3} and various QCD-based
models~\cite{qcdb1,qcdb2,qcdb3,qcdb4}. The end point of the
first-order phase transition in the $T$,$\mu_{B}$ plane is the QCD
critical point~\cite{qcpt1,qcpt2}. To discover this critical point and
to search for the phase boundary, the Beam Energy Scan (BES-I)
program~\cite{besp1,besp2,besp3,besp5} was carried out by RHIC in the
years 2010 and 2011. Au+Au collisions were recorded at \snn~= 7.7, 11.5, 19.6, 27, and 39 GeV. In the year 2014, another Au+Au
collision energy at \snn~= 14.5 GeV was added to this BES-I
program to bridge the 100 MeV gap in $\mu_{B}$~\cite{bes_paper}
between the beam energies of 11.5 and 19.6 GeV. 

In this paper, we present bulk properties of the system, namely
\pt~spectra ($\pi$,$K$,$p$), $dN/dy$, $\left\langle p_{\mathrm T}
\right\rangle$, particle ratios, kinetic freeze-out properties,
rapidity-odd directed flow $v_{1}$ (charged hadrons), and $v_{2}$
(charged hadrons) in Au+Au collisions at \snn~= 14.5~GeV. A
systematic study of these observables as a function of \pt,
pseudo-rapidity ($\eta$), and collision centrality is discussed in
detail. Comparisons of the results with those in Au+Au collisions at
other RHIC energies and with Pb+Pb collisions at \snn~= 2.76~TeV are presented. The results are also compared to model
calculations, namely UrQMD (a hadronic transport model)~\cite{urqmd}
and AMPT (a transport model with both hadronic and partonic
interactions)~\cite{ampt}. Earlier measurements suggest that systems
at lower energies, such as 7.7 and 11.5~GeV, behave like hadron gases,
while at energies of 19.6~GeV and above, they show  partonic
behavior~\cite{Adamczyk:2013dal, Adamczyk:2014ipa, Adamczyk:2014mzf,star_v2_bes_prl, star_v2_bes_prc, star_phispectra_bes_prc, phi_v2_nasim, star_v2_200gev_prl,star_v1_bes_2,Adamczyk:2017nof}. 
The Au+Au collisions at \snn~= 14.5 GeV, lying between
these two energies, allow studies of the interplay between hadronic
and partonic phases.  

\vspace{0.2in}

\section{EXPERIMENT AND DATA ANALYSIS}
\subsection{STAR Detector}
This paper reports the results for Au+Au collisions at \snn~= 14.5 GeV taken by the STAR detector~\cite{star} at RHIC under the
BES-I program. The selected minimum-bias data triggered by the Beam
Beam Counters (BBCs)~\cite{bbc1,bbc2} are used for this analysis. The
BBCs are two scintillator-based detectors situated on both sides of
the center of STAR at pseudorapidity 3.8 $< |\eta| <$ 5.2 with full
azimuthal coverage. The detector primarily used for tracking is the
Time Projection Chamber (TPC)~\cite{tpc}. The TPC is a gas detector
filled with P10 gas ($90\%$ Argon and $10\%$ Methane). It operates at
a pressure of 2 mbar above atmospheric pressure in a constant magnetic
field of 0.5 Tesla in the longitudinal ($z$) direction. The TPC has an
acceptance of $|\eta|<1$ in pseudorapidity and $2\pi$ in
azimuth. Through ionization energy loss ($\langle dE/dx \rangle$)
measurements of the particles traversing the TPC gas, different particles can be identified.  
The Time of Flight (TOF) detector is also used for particle
identification~\cite{tof}. The TOF uses Multi Resistive Plate Chamber
(MRPC) technology. It provides full azimuthal coverage and has a
pseudorapidity acceptance of $|\eta| < 0.9$.

\vspace{0.2in}
\subsection{Event Selection}
The primary vertex of each event is determined by finding the best
common point from where most of the tracks originate. Along the beam
direction, a vertex position cut of $|V_{z}| < 30$~cm is applied to
select events for the spectra analysis.  For $v_{1}$ and $v_{2}$
analyses, a broader cut of $|V_{z}| < 70$~cm is applied to obtain reasonable statistics.  
In Au+Au collisions at 14.5 GeV, the mean vertex position for all
events is centred at ($0,-0.89$) cm in $x$-$y$ plane. 
A radial vertex position cut (defined by $V_{r} =
\sqrt{V_{x}^{2}+V_{y}^{2}}$) of $V_{r} < 1$ cm from the centre is
applied to reject interactions involving the beam pipe. After these
event cuts, the number of events analyzed for $\pi$, $K$, $p$ spectra
is nearly 10 million, while the number for inclusive charged particle
$v_{1}$ and $v_{2}$ analyses is about 17 million.  

\vspace{0.2in}
\subsection{Centrality Selection}

\begin{figure}[H]
\centering
\begin{overpic}[scale=0.4]{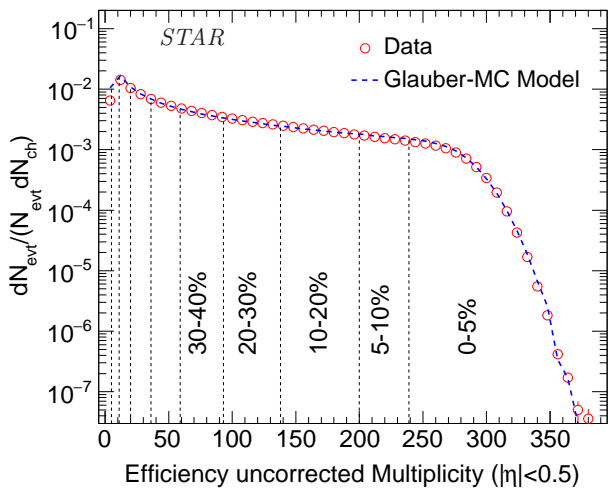}
\put (25,75) {\small \it STAR}
\end{overpic}
\small
\caption{Uncorrected charged-particle multiplicity distribution (open
  circles) measured in the TPC within $|\eta| < 0.5$ in Au+Au
  collisions at \snn~= 14.5 GeV. The blue dashed line
  represents the charged particle multiplicity distribution from a MC
  Glauber model~\cite{v2_bes_chg}. The vertical dashed lines represent
 the centrality selection criteria used.}
\label{fig:refmult}
\end{figure}

The uncorrected charged-particle multiplicity distribution is compared
and fitted with Glauber Monte-Carlo (MC) simulations as shown in
Fig.~\ref{fig:refmult}. The detailed procedure to obtain the simulated
multiplicity using Glauber Monte-Carlo is similar to that described
in~\cite{v2_bes_chg}. The minimum-bias trigger events are divided into
nine centrality classes: 0--5\%, 5--10\%, 10--20\%, 20--30\%,
30--40\%, 40--50\%, 50--60\%, 60--70\% and 70--80\%. The quoted
fractions are in terms of the total cross section obtained from the simulated events with the Glauber model. 
In addition, quantities such as average number of participating
nucleons $\left< N_{\text{part}} \right>$, number of binary collisions
$\left< N_{\text{coll}} \right>$ etc. are estimated and are listed in
Table~\ref{table:eccent}. 

\vspace{0.2in}
\subsection{Track Selection}

\begin{table}[H]
  \centering
        \caption{Track selection criteria in Au+Au collisions at \snn~= 14.5~GeV.}
        \label{tab:track-cuts}\vspace{0.1in}
        \begin{tabular}{lcc}
        \hline  
         \rule{0pt}{12pt}
		& Spectra & $v_{1}, v_{2}$  \\
	   	[3pt]
	    \hline
	    \rule{0pt}{12pt}
	    $y/\eta$ & $|y| < 0.1$ & $|\eta| < 1.0$ \\
	    [3pt]
	    \hline
	    \rule{0pt}{12pt}
	    DCA (cm) & $< 3$ &  $< 3$ \\
	    [3pt]
	    \hline
	    \rule{0pt}{12pt}
	     Number of fit points & $\geq  25$ & $\geq  15$ \\
	     [3pt]
	    \hline
	    \rule{0pt}{12pt}
	    Fraction of fit points & $\geq 0.52 $ & $\geq 0.52 $ \\
	    [3pt]
	    \hline
	    \rule{0pt}{12pt}
	    Number of $dE/dx$ points & $\geq 15$ & $\geq 15$ \\
	    [3pt]
	    \hline
        \end{tabular}
\end{table}

Details of the track cuts are tabulated in
Table~\ref{tab:track-cuts}. The distance of closest approach (DCA) of
tracks to the primary vertex is required to be less than 3 cm to
suppress tracks from secondary decays. In the spectra analysis, the
number of fit points associated with a track has to be 25 or more out
of a maximum possible 45 hits in the TPC, while for $v_{1}$ and
$v_{2}$ analyses, 15 or more hits are required.  
The cuts in these analyses are the same as the standard cuts
established in previous STAR published papers.  
The fraction of fit points on a track is required to be greater than
$52\%$ of the total possible hits to avoid split tracks. To have a
good ionization energy loss $\left< dE/dx \right>$ resolution for
tracks, the number of TPC hits used to determine $\left< dE/dx
\right>$ is required to be 15 or more. The spectra results are
obtained for tracks within the rapidity window $|y| < 0.1$. Inclusive
charged particle $v_{1}$ and $v_{2}$ analyses are carried out using
tracks within pseudorapidity $|\eta| < 1$.  

\vspace{0.2in}

\subsection{Particle Identification}

\begin{figure}[H]
\centering
\begin{overpic}[scale=0.4]{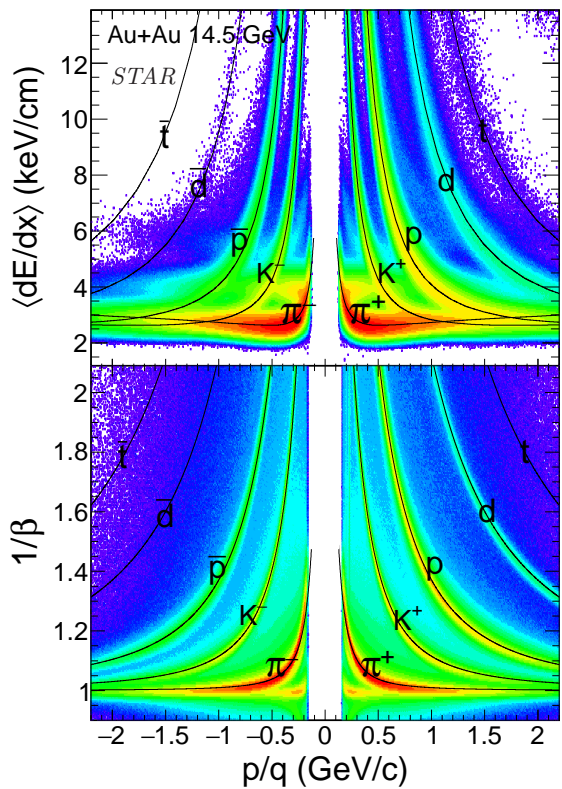}
\put (13,90) {\small \it STAR}
\end{overpic}
\caption{(Upper panel) The $\left< dE/dx \right>$ distribution of
  charged particles from the TPC as a function of momentum/charge 
for Au+Au collisions at \snn~= 14.5 GeV. The curves
represent the expected mean values of $\left< dE/dx \right>$ for the
corresponding particle species. (Lower panel) $1/\beta$ as a function
of momentum/charge from TOF in Au+Au collisions at \snn~=
14.5 GeV. The curves represent the expected values of $1/\beta$ for
the indicated particle species.}
\label{fig:dedx-beta}
\end{figure}

Particle identification in the TPC is carried out by measuring the
truncated mean of the ionization energy loss ($\left< dE/dx \right>$)
for each of the selected tracks. The measured $\left< dE/dx \right>$
of the charged particles as a function of rigidity, $p/q$ (momentum
per charge in units of the electron charge) is presented in the top
panel of Fig.~\ref{fig:dedx-beta}. The solid curves represent
theoretical values predicted by the Bichsel~\cite{bichsel}
formula. The TPC can identify pions, kaons and protons with relatively
low momentum, but the separate bands start merging at higher
momentum. The pions and kaons can be identified up to \pt~of 0.8
GeV/$c$ and protons up to 1.0 GeV/$c$.

For particle identification at relatively higher momentum, 
the TOF detector is utilized. In this analysis, TOF information is used for particle identification 
in the $p_T$ range 0.4--2.0 GeV/$c$  (0.5--2.0 GeV/$c$) for pions and kaons (protons).
The lower panel of Fig.~\ref{fig:dedx-beta} shows the characteristic
plot for TOF in which the inverse of particle velocity in units of the
speed of light, $1/\beta$, is plotted as a function of $p/q$. The
solid lines are the expected values for pions, kaons and protons. 

\begin{figure*}[!htbp]
\begin{overpic}[scale=0.4]{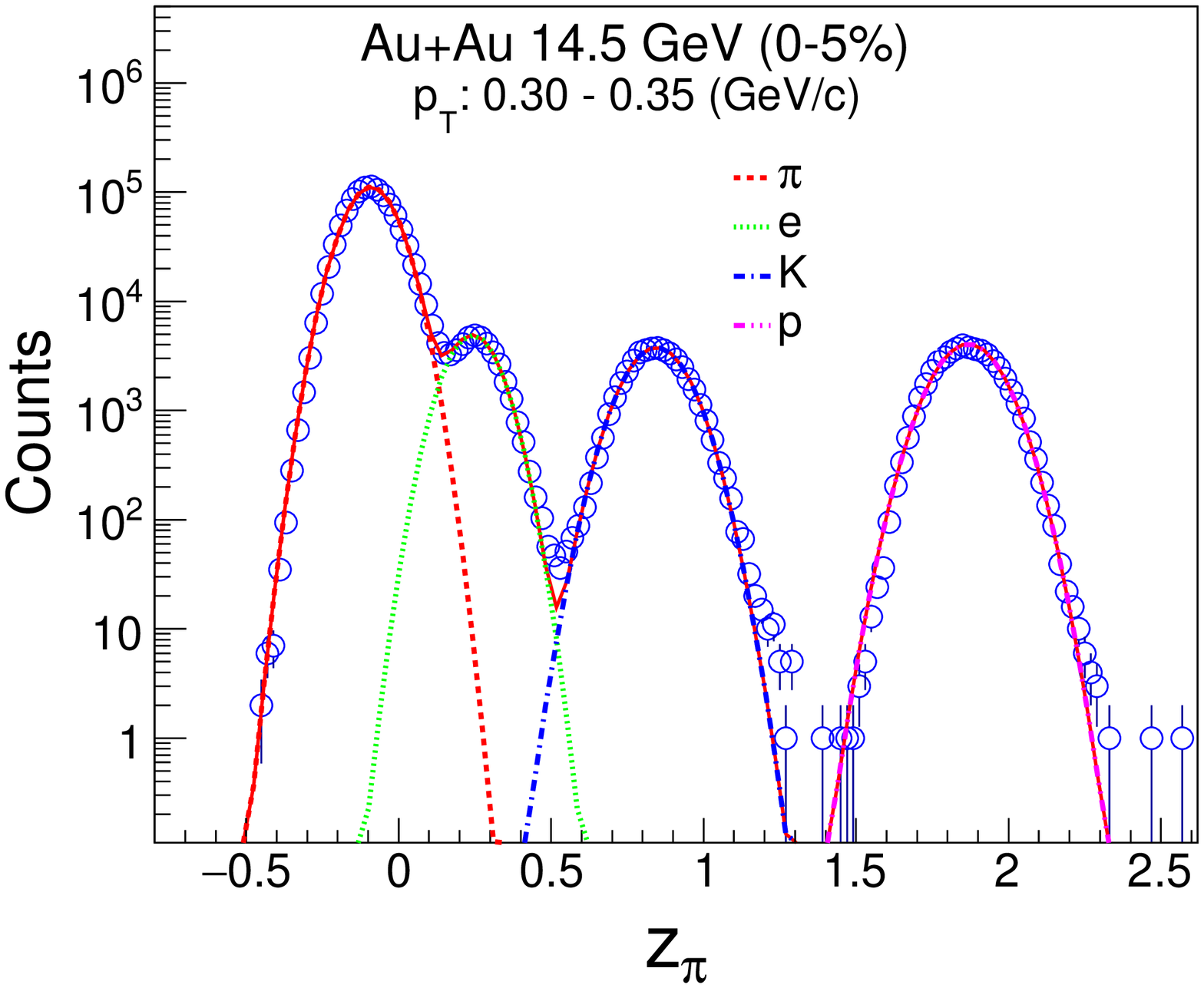}
\put (82,73) {\small \it STAR}
\end{overpic}

\begin{overpic}[scale=0.4]{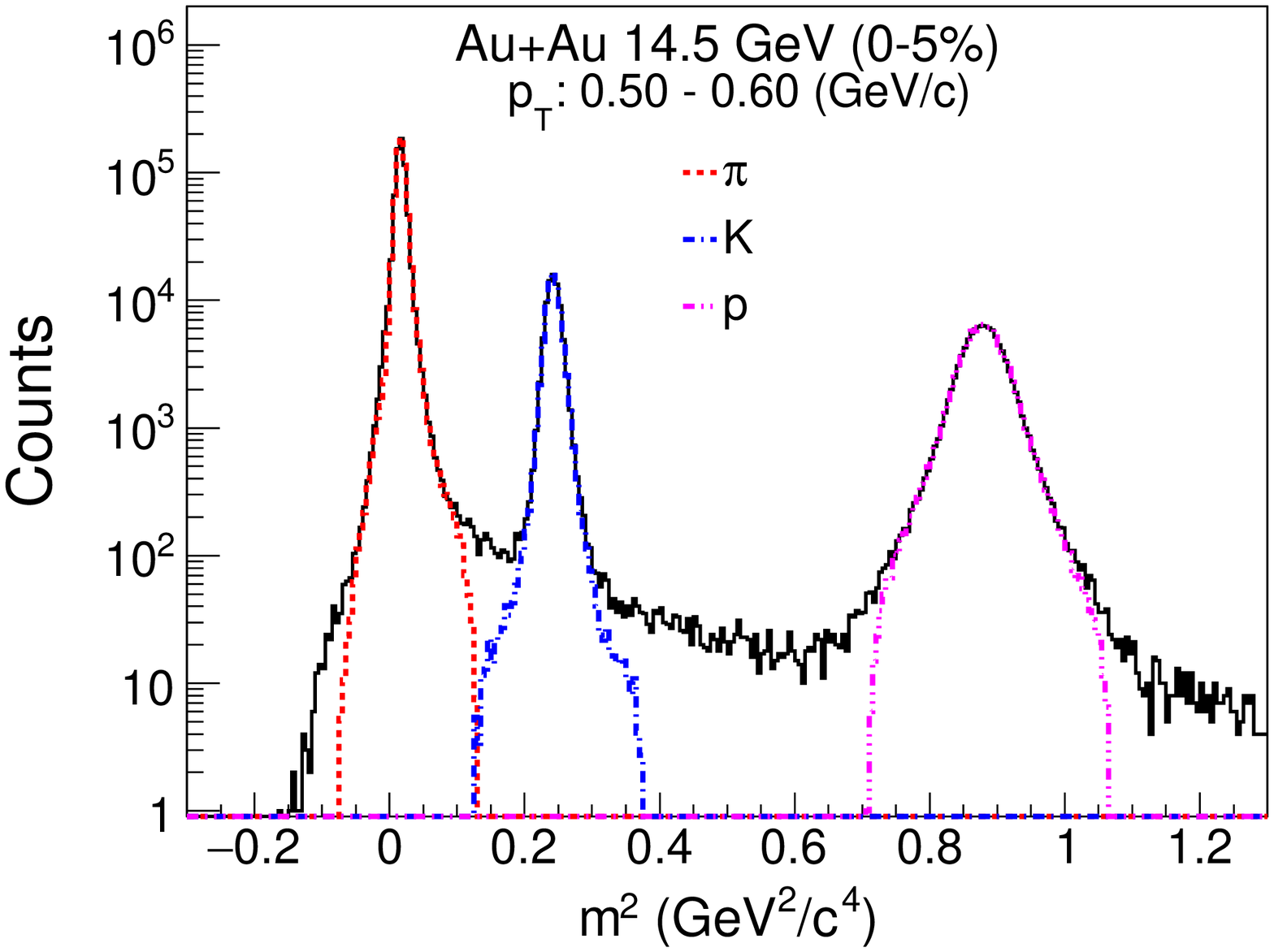}
\put (82,68) {\small \it STAR}
\end{overpic}
\caption{(Top panel) The $z_{\pi}$ distribution at midrapidity ($|y| <
  0.1$) for the \pt~range 0.30--0.35 GeV/$c$ in 0--5\% central
  Au+Au collisions at \snn~= 14.5 GeV. The curves are
  Gaussian fits representing contributions from pions (dashed red),
  electrons (dotted green), kaons (dash-dotted blue), and protons
  (dash-dot-dotted magenta). The uncertainties are statistical only.  
(Bottom panel) The $m^{2}$ distributions used to obtain the raw yields
from TOF for $\pi^{+}$ within $|y| < 0.1$ in the \pt~range
0.50--0.60 GeV/$c$. These distributions are for 0--5$\%$ centrality in
Au+Au collisions at \snn~= 14.5 GeV. The curves are fits to
the $m^{2}$ distribution, representing contributions from pions (red),
kaons (blue) and protons (magenta).}
\label{fig:zpion-m2}
\end{figure*}

The $\left< dE/dx \right>$ distribution for a specific particle type
in the TPC does not have a Gaussian shape~\cite{z-var}. It has been
demonstrated that a more appropriate Gaussian variable for a given
choice of particle type is the $z$-variable~\cite{z-var}, defined as
\\ 
\begin{equation}
z_{X} = \ln \left( \frac{\left< dE/dx \right>}{\left< dE/dx \right>_{X}^{B}} \right),
\end{equation}
where $X$ is the particle type ($e^{\pm}$, $\pi^{\pm}$, $K^{\pm}$, $p$ or $\bar{p}$ in the present analysis) and $\left< dE/dx \right>_{X}^{B}$ is the corresponding prediction of $\left< dE/dx \right>$ from the Bichsel function~\cite{bichsel}. The $z_{X}$ distribution for each particle type is expected to peak at 0.

The $z_{X}$ distributions are constructed for a particular choice of
particle, for a given \pt~range within rapidity $|y| < 0.1$. The
top panel of Fig.~\ref{fig:zpion-m2} shows the $z_{\pi}$ distribution
for positively charged tracks of $0.30 < p_{\mathrm T} < 0.35$ GeV/$c$. The
distributions are then fitted by a multi-Gaussian function to extract
the raw yield. The area under the Gaussian curve for the particle
under consideration gives the yield of that particle for that
\pt~range.  As can be noticed from Fig.~\ref{fig:zpion-m2}, the pion peak
is slightly shifted towards the negative side of zero on the $z_\pi$
axis. This could be due to issues related to calibration.  
However, any shift of the pion peak from zero does not have an impact
on the raw yield value. This method is applicable for low \pt~values,
up to the point where the distributions for pions, kaons and protons
are well separated. For higher \pt, where the distributions start to overlap, the
widths of the Gaussian distributions are constrained according to the
values at lower \pt. Following a similar procedure for each
particle type, raw yields are extracted for different \pt ranges
in nine centrality classes. 

The raw yields from TOF are obtained using the $m^{2}$ variable defined as\\
\begin{equation}
m^{2} = p^{2}\left( \frac{c^{2}T^{2}}{L^{2}} - 1 \right),
\end{equation}
where $p$, $T$, $L$, and $c$ are particle momentum, Time-of-flight,
path length of the particle, and the speed of light,
respectively. Within $|y| < 0.1$, the $m^2$ distributions are obtained
for the particle of interest in a given \pt~range, and one example
is shown in the bottom panel of Fig.~\ref{fig:zpion-m2} for the case
of $\pi^{+}$.  
To extract the raw yields using $m^2$ distributions, we follow the
same procedure as done in Refs.~\cite{bes_paper, tof-m2}. 
In this method, the $m^2$ distributions from data are fitted using the
predicted $m^2$ distributions. The predicted $m^2$  distribution is
generated by the measured time-of-flight from experimental data, thus
including the TOF detector response behavior, for a given
$dE/dx$-identified particle.  
It is observed that the predicted $m^2$  distributions do not change
much with $p_T$ and can be extended to higher $p_T$ where  
$dE/dx$ identification is limited. These predicted $m^{2}$
distributions of pions, kaons and protons are used simultaneously to
fit the measured $m^{2}$ distributions to obtain the raw yield as
shown for $\pi^{+}$  in the bottom panel of Fig.~\ref{fig:zpion-m2}
for $p_T$ range 0.5--0.6 GeV/$c$. In this way, the raw yields are
obtained for all $p_T$ bins, centralities and different particles.

\vspace{0.2in}

\subsection{Flow Analysis}

The azimuthal distribution of emitted particles with respect to the reaction plane can be decomposed in a
Fourier series~\cite{art1}. The harmonic coefficients ($v_{n}$) in this expansion are defined as,
\begin{equation}
 v_{n} = \langle \cos n(\phi-\Psi_{R}) \rangle.
 \label{eq:3a}	
\end{equation}	
Here, angle brackets denote an average over all particles in all
events for a given \pt~or $\eta$ bin, and $\Psi_{R}$ is the
azimuth of the reaction plane angle. The first harmonic coefficient is
called the directed flow ($v_{1}$), while the second is called the
elliptic flow ($v_{2}$). Since the $\Psi_{R}$ is unknown
experimentally, it is estimated from
\begin{equation}
\Psi_{n} =\frac{1}{n}\tan^{-1} \left(\frac{\sum\limits_{i=1}^{N} w_{i}\sin(n\phi_{i})}{\sum\limits_{i=1}^{N} w_{i}\cos(n\phi_{i})}\right),
\label{eq:6}
\end{equation}
where $\Psi_{n}$ is $n^{th}$-order event plane azimuth, $w_{i}$ are weights, and $N$ is the total number of
particles in an event used for the event plane
calculation~\cite{art1}. The observed $v_{n}^{\rm obs}$ is calculated
with respect to the reconstructed  event plane using
\begin{equation}
v_{n}^{\rm obs}\ =\ \langle \cos n(\phi-\Psi_{n}) \rangle.
\label{eq:7}
\end{equation}
The observed $v_{n}$ are then corrected for event plane resolution~\cite{res_cent,res_nsm}.

Two types of event plane angles are used in this analysis: the TPC
event plane~\cite{v2_bes_chg} and the BBC event
plane~\cite{bbc_ep}. In the TPC event plane method, a second-order
event plane angle ($\Psi_{2}$) is reconstructed from TPC tracks at
mid-rapidity ($|\eta| < 1$). To reduce nonflow contributions, we
utilize the $\eta$-sub method, with an additional $\eta$-gap of 0.1
between the sub-events, and then average the results from the two
sub-events~\cite{v2_bes_chg}. In the BBC event plane method, a
first-order event plane ($\Psi_{1}$), reconstructed using the hits in
both BBC detectors in opposite hemispheres (3.8$< |\eta|< $5.2), is
used to calculate $v_{1}$ and $v_{2}$.  Re-centering and shift
techniques were applied for each $\eta$ hemisphere independently to
flatten the TPC event plane and BBC event plane~\cite{art1}.  More
details about these methods can be found in a previous
publication~\cite{v2_bes_chg}. 

In addition to the event plane method, the multi-particle correlation
method~\cite{v2qc,v2gen}  is used to calculate $v_{2}$ of charged
particles. In this method, the reference flow (e.g., integrated over \pt) can be estimated both from two- and four-particle cumulants: 
\begin{eqnarray}
v_{n} \lbrace 2 \rbrace  = \sqrt{c_{n} \lbrace 2 \rbrace} ,
\label{eq:25}
\end{eqnarray}
\begin{eqnarray}
v_{n} \lbrace 4 \rbrace  = \sqrt[4]{-c_{n} \lbrace 4 \rbrace}.
\label{eq:26}
\end{eqnarray}
Here  $c_{n} \lbrace 2 \rbrace$ and $c_{n} \lbrace 4 \rbrace$ are two-
and four-particle cumulants. The two- and four-particle cumulants
without detector bias then can be formulated as 
\begin{eqnarray}
c_{n} \lbrace 2 \rbrace = \langle \langle e^{in(\phi_{1} - \phi_{2})} \rangle \rangle ,
\label{eq:23}
\end{eqnarray}
\begin{eqnarray}
c_{n} \lbrace 4 \rbrace = \langle \langle
e^{in(\phi_{1} + \phi_{2} - \phi_{3} - \phi_{4})} \rangle \rangle - 2 \times \langle \langle e^{in(\phi_{1} - \phi_{2})} \rangle \rangle .
\label{eq:24}
\end{eqnarray}
Here, double angle brackets denote an average over all events. More
details about these methods are presented in~\cite{v2_bes_chg}. 


\vspace{0.2in}

\section{CORRECTION FACTORS}
\subsection{Monte-Carlo Embedding Technique}

Several correction factors for the \pt~spectra are calculated from
MC simulations known as the embedding technique.  
The method is outlined below and more details can be found in Refs.~\cite{besp1,bes_paper,prc}.
For a given particle, MC tracks having flat rapidity and
\pt~distributions are simulated in the STAR detector using GEANT3. Those
simulated tracks are then embedded into real events at the raw data
level. The multiplicity of embedded tracks in any real event is no
more than 5\% of the measured real charged particle multiplicity of
that event. These embedded tracks are reconstructed in the same manner
as the real data reconstruction. The embedding sample is used to
calculate various correction factors such as tracking efficiency and
acceptance, and energy loss correction as discussed below.

\begin{figure*}[!tp]
\centering
\begin{overpic}[scale=0.9]{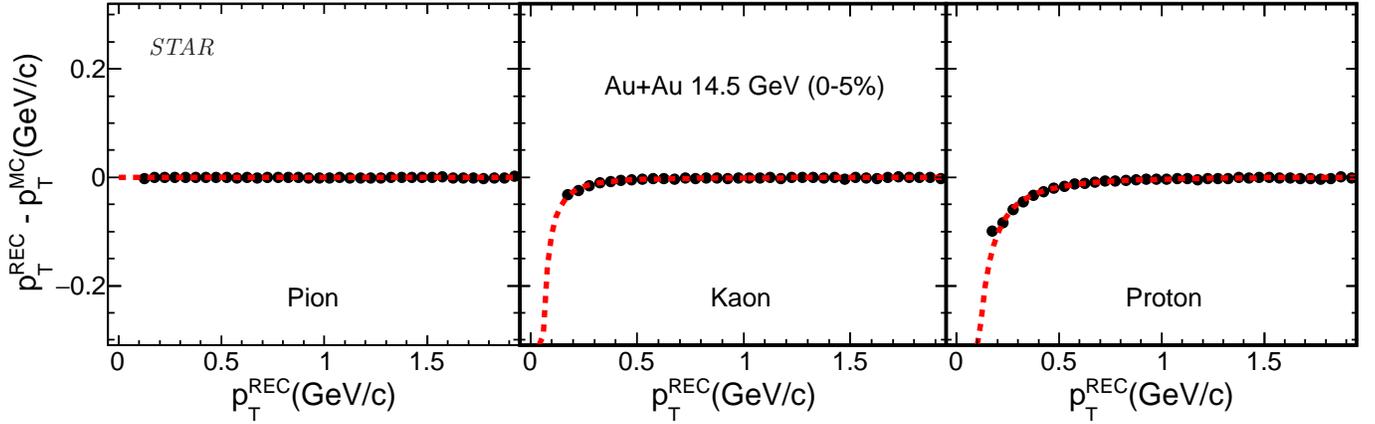}
\put (10,27) { \it STAR}
\end{overpic}
\caption{The \pt~difference of reconstructed momentum $p_{\mathrm
    T}^{\rm REC}$ and MC momentum $p_{\mathrm T}^{\rm MC}$ as a
  function of $p_{\mathrm T}^{\rm REC}$ for pions (left), kaons (middle) and protons (right) in 0-5\% central Au+Au collisions at \snn~= 14.5 GeV. 
The red curves represent the functional fit of the form $f(p_{\mathrm
  T}) = A + B \left( 1+ \frac{C}{p_{\mathrm T}^{2}} \right)^{D}$.}
\label{fig:Eloss}
\end{figure*}

\begin{figure*}[!tp]
\centering
\begin{overpic}[scale=0.9]{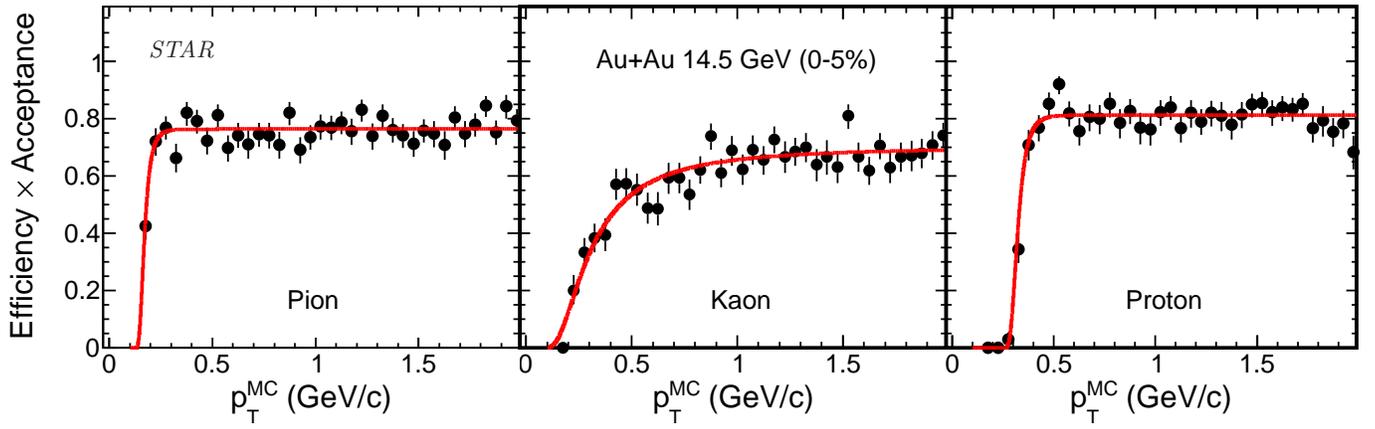}
\put (10,27) { \it STAR}
\end{overpic}
\caption{The combined tracking efficiency and acceptance as a function
  of \pt~calculated from the Monte-Carlo embedding technique for
  reconstructed pions (left), kaons (middle) and protons (right) at
  midrapidity ($|y| < 0.1$) for 0-5\% Au+Au collisions at
  \snn~= 14.5 GeV. The curves represent the functional fit
  of the form $p_0 \exp[-(p_1/p_T)^{p_2}]$. } 
\label{fig:efficiency}
\end{figure*}

\vspace{0.2in}

\subsection{Energy Loss Correction}

The TPC track reconstruction algorithm assumes the pion mass for each
particle when correcting for multiple Coulomb scattering and energy
loss in the TPC gas, which mostly affect particles of low
momenta. Therefore, a correction for momentum loss by heavier
particles like $K^{\pm}$ and $p$($\bar{p}$) is needed. This correction
is obtained from MC simulation or embedding techniques. The
distribution of momentum difference between reconstructed momentum
$p_{\mathrm T}^{\rm REC}$ and initial momentum $p_{\mathrm T}^{\rm
  MC}$ as a function of $p_{\mathrm T}^{\rm REC}$ gives the amount of energy loss correction for each track. The relevant plot for Au+Au collisions at \snn~= 14.5 GeV, showing the \pt~dependence of energy loss, is presented in Fig.~\ref{fig:Eloss} for pions (left), kaons (middle) and protons (right). The red curve represents a functional fit to the data points in the case of kaons and protons,
\begin{equation}
f(p_{\mathrm T}) = A + B \left( 1+ \frac{C}{p_{\mathrm T}^{2}} \right)^{D}, \label{Eq:Eloss}
\end{equation}
where $A, B, C$ and $D$ are fit parameters. This energy loss fraction is the same for all centrality classes for a particular particle type. All the results shown in this paper have been corrected for this energy loss effect.

\begin{figure*}[!tp]
\centering
\begin{overpic}[scale=0.9]{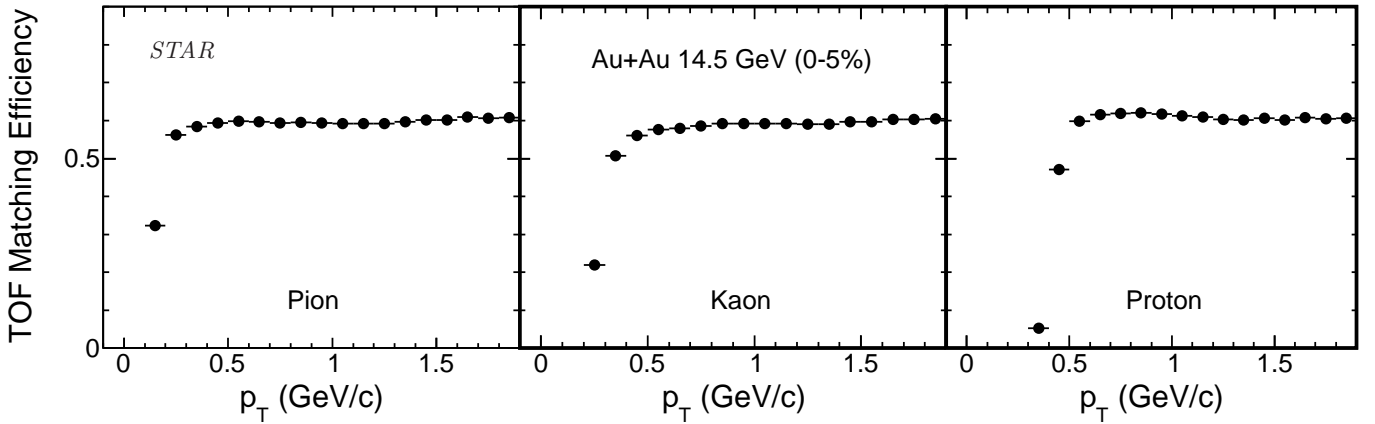}
\put (10,27) { \it STAR}
\end{overpic}
\caption{TOF matching efficiency as a function of \pt~for pions (left), kaons (middle) and protons (right) at midrapidity ($|y| < 0.1$) in 0-5\% Au+Au collisions at \snn~= 14.5 GeV.}
\label{fig:tof-eff}
\end{figure*}

\vspace{0.2in}

\subsection{Tracking Efficiency and Acceptance}

A correction for detector efficiency and acceptance needs to be
applied to the \pt~spectra of analyzed particles. This correction
factor is obtained from the embedding technique described above. The
combined tracking efficiency and acceptance is the ratio of the
distribution of reconstructed to original Monte-Carlo tracks as a
function of \pt~in the rapidity range of interest. This functional
dependence of combined tracking efficiency and acceptance on \pt~is presented in Fig.~\ref{fig:efficiency} for reconstructed pions (left), kaons (middle), and protons (right) in Au+Au collisions at \snn~= 14.5 GeV. 
The curves represent the functional fit of the form $p_0 \exp[-(p_1/p_T)^{p_2}]$, used to parameterize the efficiency.
This correction factor is thus calculated for each particle type in nine centrality classes and each \pt~spectrum is divided by this fraction. 

\vspace{0.2in}

\subsection{TOF Matching Efficiency}

The TOF detectors form a curved cylindrical surface surrounding the TPC and have a reduced geometric acceptance compared to the TPC. 
Circumstances arise where TPC tracks are not detected in TOF, especially
at low momentum, either because the track is out of the TOF acceptance or because of the TOF inefficiency. 
As a result, the yield of particles identified by TOF needs to be corrected, in addition to
the TPC track reconstruction efficiency. 
This is referred to as the TOF matching efficiency or TOF particle identification efficiency for TPC tracks. 
This efficiency is 
calculated from the STAR data
as the ratio of the number of tracks detected in TOF to the total number of tracks in the TPC within the acceptance under study. 
It is shown 
as a function of \pt~in Fig.~\ref{fig:tof-eff} for pions (left), kaons (middle) and protons (right) in Au+Au collisions at \snn~= 14.5 GeV for 0--5\% centrality.
The raw yields obtained using TOF are divided by the matching efficiency for each centrality, \pt~range, and for each particle type.  

\vspace{0.2in}

\subsection{Pion Background Correction}

The measured pions get a contribution from the feed-down of weak decays, muon
contamination, and background pions produced in the detector material.
Therefore, it is important to remove these background contributions from the total pion yield. 
To obtain this correction,
we used the same approach as done in Refs.~\cite{bes_paper,prc}.
The Monte-Carlo simulated events are generated by HIJING~\cite{hijing} and are processed through the STAR detector simulated by GEANT3~\cite{geant}. 
These events are reconstructed in the same manner as real data. 
In the MC sample, the pion background fraction is estimated since different contributions to the total pion yield are known.
The pion background fraction decreases exponentially with \pt. Its value at low \pt (=0.225 GeV/$c$) is $\sim$ 16$\%$ and becomes negligible above 0.6 GeV/$c$. It shows negligible centrality dependence, hence the same correction is applied for all centrality classes. 

\vspace{0.2in}

\subsection{Proton Background Correction}

The yield of protons has a significant contribution from background protons coming from interactions of highly energetic particles with the detector material. 
In order to estimate the proton background fraction, we follow the same procedure as used in Refs.~\cite{prc,pbkgrnd,bes_paper}. 
This fraction is estimated by comparing the proton and antiproton DCA distributions obtained from real data. 
The difference between the proton and antiproton DCA distribution gives the estimate of proton background contribution. 
The proton background fraction decreases as a function of \pt~and decreases from peripheral to central collisions. At $p_T= $ 0.60--0.65 GeV/$c$, its typical value is about 40\% in peripheral collisions and 2\% in central collisions.  
For minimum bias collisions, proton background at low \pt~is around
$15\%$ and becomes almost negligible for $p_{\mathrm T} > 1.2$ GeV/$c$. 
The (anti)protons also have a contribution of feed-down from weak decays of hyperons,
which include particles like $\Sigma$ which has not been measured. Contrary to pions, the
analysis cut of DCA $<$ 3 cm includes almost all daughter particles from hyperon
decays~\cite{pbkgrnd_2,200gevprl}. Thus, (anti)proton yields presented here are all inclusive similar to
those at other RHIC energies~\cite{bes_paper,prc}.

\vspace{0.2in}

\section{Systematic Uncertainties}

To estimate the size of systematic uncertainties associated with the \pt~spectra of the particles under study, we
vary the event and track cuts, and the quality of fits to $\left< dE/dx \right>$  measurements. 
The following parameters are varied: the event $V_{z}$ range (from $|V_{z}| < 30$ cm to $|V_{z}| < 50$ cm), track cuts like DCA (from 3 cm to 2 cm), number of fit points (from 25 to 20), and number of $\left< dE/dx \right>$ points (from 15 to 10).
We have also varied the fit range for the $z$ distribution 
and the PID selection (using $\langle dE/dx \rangle$) cut of a given particle 
used for the predicted $m^{2}$ distribution.  

\begin{table}[H]
  \centering
	\caption{Systematic uncertainties related to $\pi$, $K$ and $p(\bar{p})$ integrated particle yields in Au+Au collisions at \snn~= 14.5~GeV.}
	\label{tab:tot-sys}\vspace{0.1in}
	\begin{tabular}{lccc}
	\hline	
	\rule{0pt}{12pt}
	~~
	& $\pi$ & $K$ & $p\,\bar{p}$ \\
	[3pt]
	\hline
	$V_{z}$ & 1$\%$ & $1\%$ & $1\%$ \\
	\hline
	Track Cuts & $4\%$ & $4\%$ & $6\%$ \\
	\hline
	PID & $6\%$ & $8\%$ & $7\%$ \\
	\hline
	Extrapolation & $5\%$ & $4\%$ & $6\%$ \\
	\hline
	Corrections & $5\%$ & $5\%$ & $5\%$ \\
	\hline
	Proton Background & -  & - & 5--6\% \\
	\hline 
	Total & $10\%$ & $10\%$ & $12\%$ \\
	\hline
	\end{tabular}
\end{table}

Apart from these systematic uncertainties for the case of \pt~spectra, an additional error of 5$\%$ is added in quadrature due to detector tracking efficiency and acceptance~\cite{bes_paper, prc, 200gevprl}. The pion feed-down correction and the proton background fraction also contribute to the systematic uncertainty; however, the former is negligible and the latter contributes about 5--6\% only at low \pt. All the sources of systematic uncertainty are added in quadrature and are tabulated in Table~\ref{tab:tot-sys}. The total systematic uncertainties on pion, kaon and proton yields are $10\%$, $10\%$ and $12\%$, respectively. 

The calculation of \pt~integrated particle yields ($dN/dy$) and
$\left< p_{\mathrm T} \right>$ requires a fitting function to
extrapolate the \pt~spectra to the unmeasured \pt~ region. Thus,
another important source of systematic uncertainty in $dN/dy$ and
$\left< p_{\mathrm T} \right>$ is extrapolation.
For pions, kaons and protons, the default fit functions used to extract yields are Bose-Einstein, $m_{T}$-exponential, and double exponential, respectively. To estimate the systematic uncertainty, these fit functions for pions, kaons and protons are changed to \pt-exponential, Boltzmann and $m_{T}$-exponential functions, respectively. The relevant functional forms are 
\begin{itemize}
\item Bose-Einstein:		~$\propto 1/(\exp(m_{T}/T_{\rm BE})-1)$
\item $p_T$-exponential:	~$\propto \exp(-p_{\mathrm
    T}/T_{p_{\mathrm T}})$
\item $m_T$-exponential:	~$\propto \exp(-m_{T}/T_{m_{T}})$
\item Boltzmann:		~$\propto m_{T}\exp(-m_{T}/T_{B})$
\item Double-exponential:	~$A\, e^{-p_T^2 / T_1^2} + B\, e^{-p_T^2 / T_2^2}$  
\end{itemize}

The systematic uncertainty on mean \pt~mainly comes from the errors
associated with extrapolation of \pt~spectra. The fitting range of the
fit function also affects the value of $\left< p_{\mathrm T}\right>$,
which is included as a source of systematic uncertainty. The
systematic uncertainty on $\left< p_{\mathrm T}\right>$ for pions, kaons and protons is $5\%$, $2\%$ and $6\%$, respectively.

The systematic uncertainty on integrated particle ratios is calculated from the systematic uncertainty on $dN/dy$. The systematic uncertainty due to tracking efficiency cancels in particle ratios. The error associated with extrapolation mostly cancels in the case of particle to antiparticle ratios, but does not cancel for the ratios of different particle species. 

\begin{figure*}[!tp]
\centering
\begin{overpic}[scale=0.9]{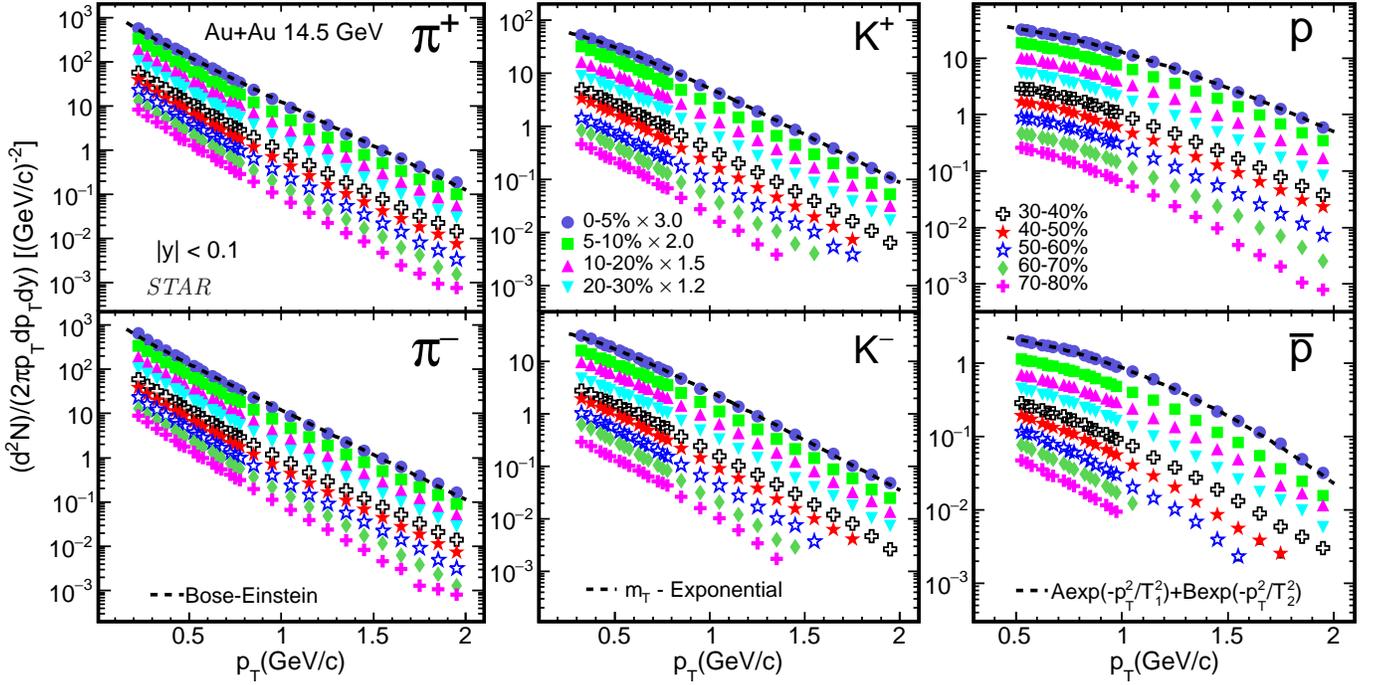}
\put (10,28.5) { \it STAR}
\end{overpic}
\caption{The \pt~spectra of $\pi^{\pm}$, $K^{\pm}$, $p$ ($\bar{p}$)
  measured at midrapidity ($|y| < 0.1$) in Au+Au collisions at
  \snn~= 14.5 GeV. Spectra are plotted for nine centrality
  classes, with some spectra multiplied by a scale factor to improve
  clarity, as indicated in the legend. The data points shown for
  $p_{\mathrm T} = $ 0.4--2.0 GeV/$c$ for pions and kaons, and for 0.5--2.0 GeV/$c$ for protons, are obtained using both TPC and TOF. Data points measured using only the TPC are shown for \pt~in the range 0.2--0.8, 0.3--0.8 and 0.5--1.0 GeV/$c$ for pions, kaons and protons, respectively. The \pt~range 0.4--0.8 GeV/$c$, 0.4--0.8 GeV/$c$ and 0.5--1.0 GeV/$c$ for pions, kaons and protons, respectively, is the overlap region containing data measurements in both categories, namely, TPC only, and TPC+TOF. The \pt-spectra are fitted with a Bose-Einstein function for pions, an $m_T$-exponential for kaons, and a double exponential for (anti)protons. Statistical and systematic uncertainties are added in quadrature.}
\label{fig:pT-spectra}
\end{figure*}

The Blast-Wave fit~\cite{bw} to particle \pt~spectra provides the kinetic freeze-out parameters. The point-to-point systematic uncertainty associated with the \pt~spectra propagates to the systematic uncertainties on the kinetic freeze-out parameters. The \pt~ranges used for fitting also affect the results. These variations are included in the systematic uncertainty on kinetic freeze-out parameters.

The systematic uncertainties for $v_{1}$ and $v_{2}$ measurements are estimated by varying event and track cut parameters from their default values. The $z$-position of the primary vertex is varied between 60 and 80 cm. The DCA of the primary tracks is varied between 2.0 cm and 3.0 cm. The number of fit points is varied from 18 to 25. In the case of $v_{2}$ measurements, the $\eta$-gap is varied between 0.05 and 0.075. In total, about 100 combinations of such cut variations are considered and the RMS of the variation is taken as the systematic uncertainty. A maximum of 2\% relative systematic uncertainty due to event cuts, and 1\% due to track cuts, is found for the various centrality classes and for the various \pt~and $\eta$ bins.

\section{RESULTS AND DISCUSSIONS}
\subsection{Transverse Momentum Spectra}

The transverse momentum spectra for $\pi^{+}$, $\pi^{-}$, $K^{+}$,
$K^{-}$, $p$, and $\bar{p}$ in Au+Au collisions at \snn~=
14.5 GeV are presented in Fig.~\ref{fig:pT-spectra}. The spectra are
plotted for nine collision centralities 0--5\%, 5--10\%, 10--20\%,
20--30\%, 30--40\%, 40--50\%, 50--60\%, 60--70\%, 70--80\%. Further
information can be extracted from the particle \pt~spectra through
functional fitting in terms of $dN/dy$ and $\left< p_{\mathrm T} \right>$.  As mentioned earlier the functions used for this purpose are Bose-Einstein, $m_{T}$-exponential and double exponential for pions, kaons and protons, respectively. 
It can be inferred that the invariant yields exhibit a \pt~dependence
(decrease with increasing \pt) as well as a centrality dependence (decrease towards the peripheral collisions). 
The shapes of the kaon and (anti)proton spectra show a gradual flattening from peripheral to central collisions. The trend is less pronounced for pions.
This flattening reflects a stronger effect of radial flow on particles with higher mass and for events with increasing centrality. 
\begin{figure*}[!tp]
\centering
\begin{overpic}[scale=0.9]{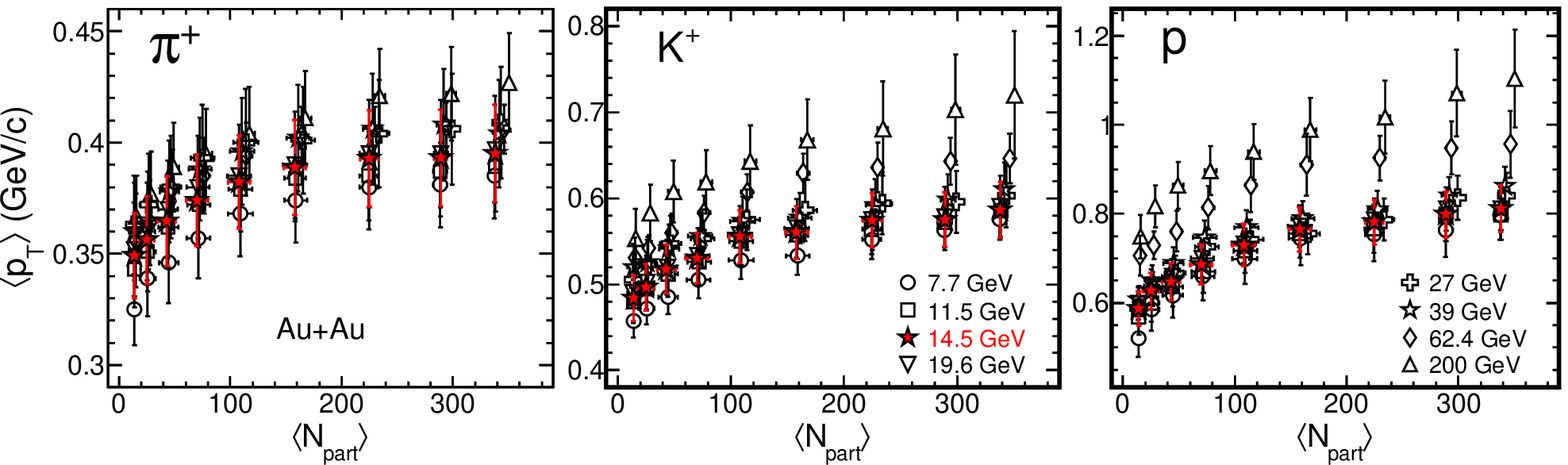}
\put (27,8.0) { \it STAR}
\end{overpic}
\caption{$\left< p_{\mathrm T} \right>$ of $\pi^{+}$, $K^{+}$ and $p$ as a function of $\left< N_{\text{part}} \right>$ for Au+Au collisions at \snn~= 14.5 GeV. These averages are compared with the corresponding results from Au+Au collisions at \snn~= 7.7, 11.5, 19.6, 27, 39, 62.4, and 200 GeV measured by STAR in earlier runs~\cite{bes_paper, prc, 200gevprl}. Statistical and systematic uncertainties have been added in quadrature.}
\label{fig:mean-pT}
\end{figure*}

\begin{figure*}[!tp]
\centering
\begin{overpic}[scale=0.9]{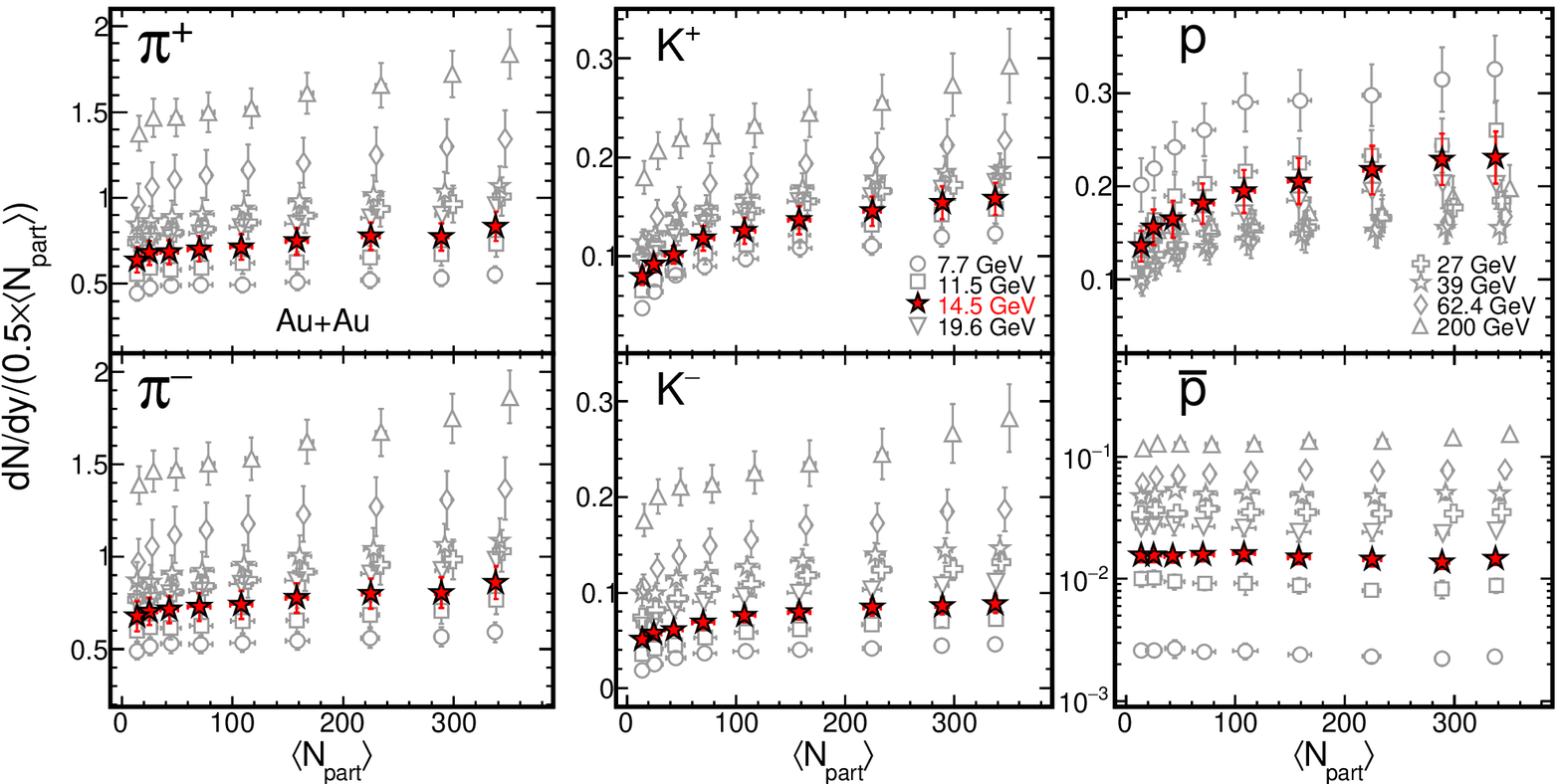}
\put (28,6.5) { \it STAR}
\end{overpic}
\caption{$dN/dy$ of $\pi^{+}$, $\pi^{-}$, $K^{+}$, $K^{-}$, $p$ and
  $\bar{p}$ scaled by ($0.5 \times \left< N_{\text{part}} \right>$) as
  a function of $\left< N_{\text{part}} \right>$ for Au+Au collisions
  at \snn= 14.5 GeV. These yields are compared with the corresponding
  results from Au+Au collisions at \snn~= 7.7, 11.5, 19.6, 27, 39, 62.4, and 200 GeV measured by STAR in earlier runs~\cite{bes_paper, prc, 200gevprl}. Statistical and systematic uncertainties have been added in quadrature.}
\label{fig:dNdy}
\end{figure*}

\vspace{0.2in}

\subsection{Average Transverse Momenta}

Average transverse momenta quantitatively reflect the slopes of the
measured \pt~spectra of the particles. i.e., the transverse dynamics
influences $\left< p_{\mathrm T} \right>$. The dependence of $\left<
  p_{\mathrm T} \right>$ on the number of nucleon participants $\left<
  N_{\text{part}}\right>$ is shown in Fig.~\ref{fig:mean-pT} for Au+Au
collisions at \snn~= 14.5 GeV. These averages are compared
with the corresponding results from Au+Au collisions at \snn~= 7.7, 11.5, 19.6, 27, 39, 62.4, and 200 GeV measured by STAR in
earlier runs~\cite{expdof1, expdof2, expdof3, expdof4,
  bes_paper,prc,200gevprl}. It is seen from the figure that $\left<
  p_{\mathrm T}\right>$ of $\pi^{\pm}$, $K^{\pm}$ and $p(\bar{p})$ increases with increasing $\left< N_{\text{part}}\right>$. 
This indicates an increase of radial flow from peripheral to central collisions~\cite{Heinz:2013th}.
Mean $p_T$ and inferred radial flow also increase from pions to kaons,
and from kaons to protons. The behavior of $\left< p_{\mathrm
    T}\right>$ as a function of $\left< N_{\text{part}}\right>$ in
Au+Au collisions at \snn~= 14.5 GeV is similar within error
bars to what is observed at other BES-I energies, although it slowly
increases with collision energy. The values of $\left< p_{\mathrm T}\right>$ for $\pi^{+}$, $\pi^{-}$, $K^{+}$, $K^{-}$, $p$, and $\bar{p}$ are listed in Table~\ref{tab:mean-pT} for Au+Au collisions at \snn~= 14.5 GeV.

\vspace{0.2in}

\subsection{Particle Yields}
The particle production in a collision centrality interval is defined
as $dN/dy$ or particle yield, which we measure at midrapidity ($|y| <
0.1$) and is obtained by integrating over \pt. The measured $dN/dy$ is shown in Fig.~\ref{fig:dNdy} for $\pi^{+}$, $\pi^{-}$, $K^{+}$, $K^{-}$, $p$ and $\bar{p}$, normalized with 0.5$\times \left< N_{\text{part}} \right>$, as a function of $\left< N_{\text{part}} \right>$ in Au+Au collisions at \snn~= 14.5 GeV. These yields are compared with the corresponding results from Au+Au collisions at \snn~= 7.7, 11.5, 19.6, 27, 39, 62.4, and 200 GeV measured by STAR in earlier runs~\cite{expdof1, expdof2, expdof3, expdof4, bes_paper, prc, 200gevprl}. The values of $dN/dy$ for $\pi^{+}$, $\pi^{-}$, $K^{+}$, $K^{-}$, $p$, and $\bar{p}$ are also tabulated in Table~\ref{tab:dndy} for Au+Au collisions at \snn~= 14.5 GeV. 

\begin{figure*}[!tbp]
\centering
\begin{overpic}[scale=0.9]{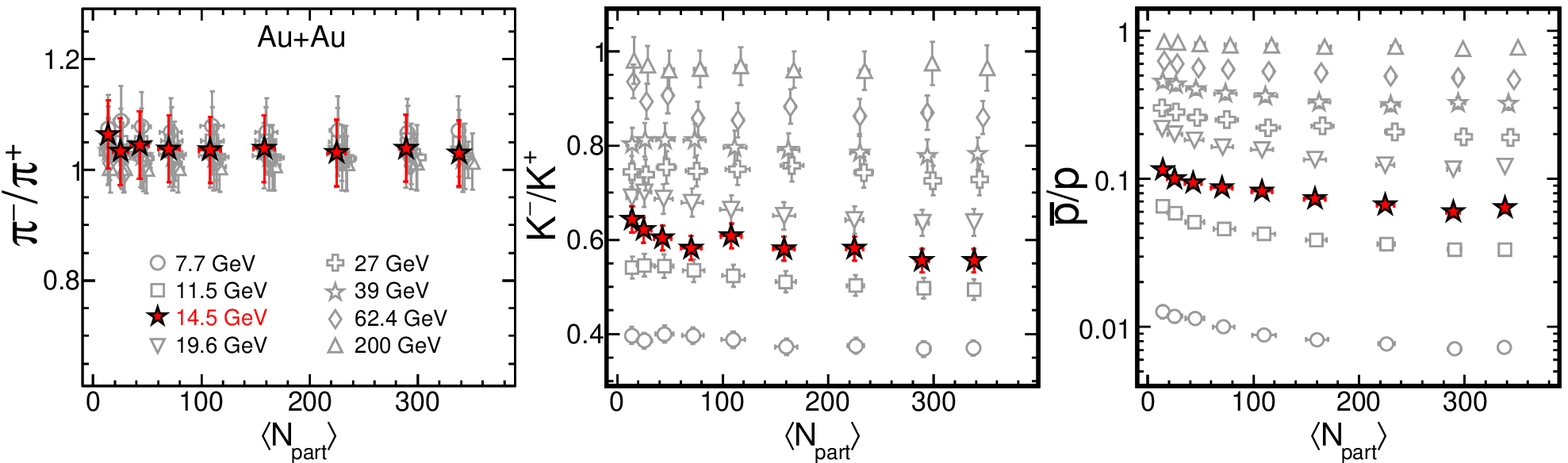}
\put (26.5,26.5) { \it STAR}
\end{overpic}

\caption{$\pi^{-}$/$\pi^{+}$, $K^{-}$/$K^{+}$ and $\bar{p}/p$ ratios as a function of $\left< N_{\text{part}} \right>$ in Au+Au collisions at \snn~= 14.5 GeV. These ratios are compared with the corresponding results from Au+Au collisions at \snn~= 7.7, 11.5, 19.6, 27, 39, 62.4, and 200 GeV measured by STAR in earlier runs~\cite{bes_paper, prc, 200gevprl}.  Statistical and systematic uncertainties have been added in quadrature.}
\label{fig:like-ratios}
\end{figure*}

\begin{figure*}[!tbp]
\centering
\begin{overpic}[scale=0.85]{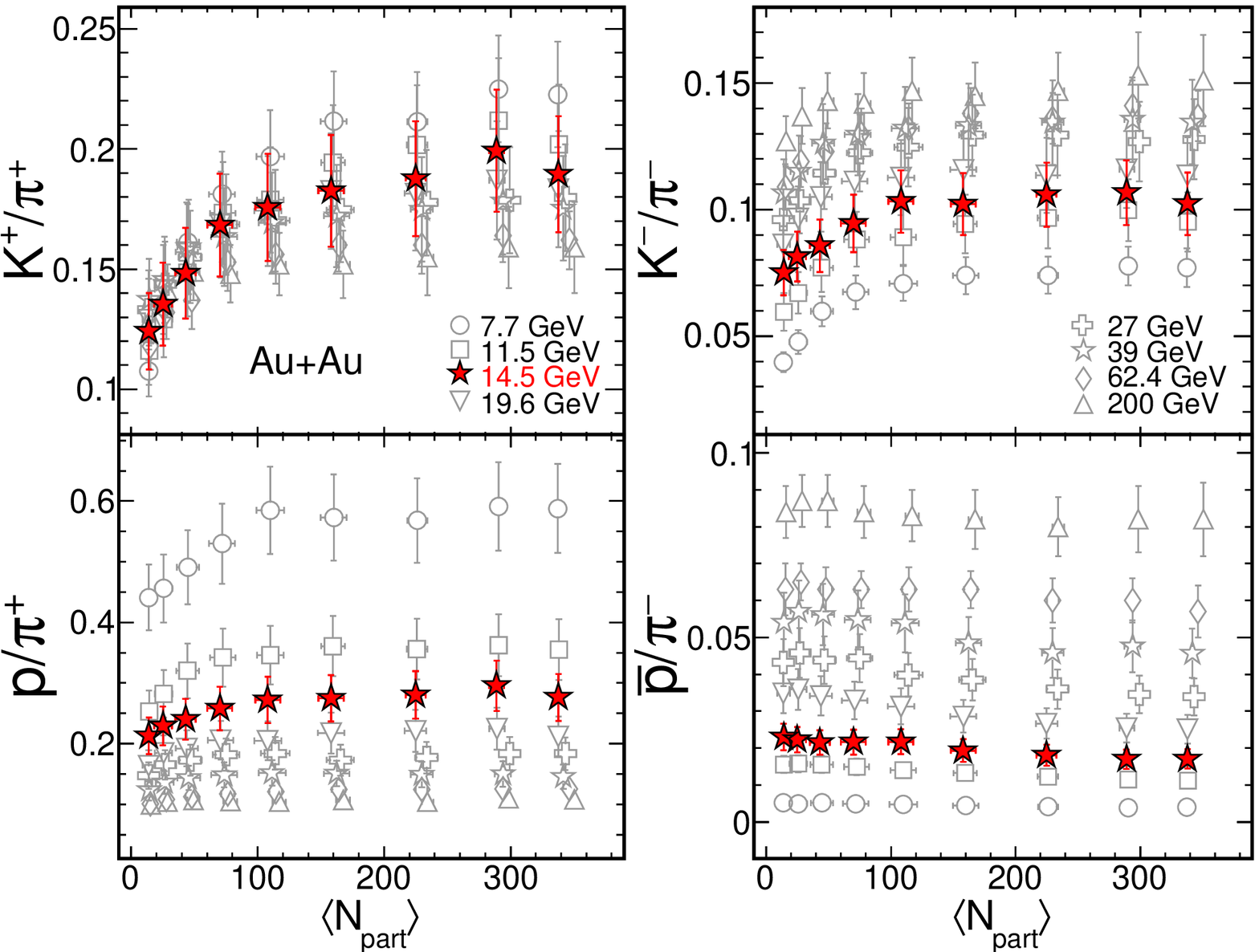}
\put (12,72.5) { \it STAR}
\end{overpic}
\caption{$K^{+}$/$\pi^{+}$, $K^{-}$/$\pi^{-}$, $p/\pi^{+}$ and $\bar{p}$/$\pi^{-}$ ratios as a function of $\left< N_{\text{part}} \right>$ in Au+Au collisions at \snn~= 14.5 GeV. These ratios are compared with the corresponding results from Au+Au collisions at \snn~= 7.7, 11.5, 19.6, 27, 39, 62.4, and 200 GeV measured by STAR in earlier runs~\cite{bes_paper, prc, 200gevprl}. Statistical and systematic uncertainties have been added in quadrature.}
\label{fig:unlike-ratios}
\end{figure*}
The pion, kaon, and proton yields slowly increase from peripheral to central collisions. This may indicate contributions from hard processes which depend on the number of nucleon-nucleon binary collisions, increasing with $N_{\rm{part}}$ more than linearly~\cite{Wang:2000bf}.
The antiproton yields remain almost flat with centrality. This may be due to an increasing baryon-antibaryon annihilation effect with increasing centrality.
The yields of pions, kaons, and antiprotons all increase with increasing collision energy. 
However, the yield of protons shows the opposite trend and decreases up to 39 GeV after which it starts to increase. 
This reflects an increase in baryon density due to baryon stopping at lower energies~\cite{Randrup:2006nr,bes_paper,besp1}. The results in Au+Au collisions at \snn~= 14.5 GeV show a similar behaviour as observed by STAR at other energies~\cite{expdof1, expdof2, expdof3, expdof4, bes_paper, prc, 200gevprl}. 

\vspace{0.2in}

\subsection{Particle Ratios}

Particle ratios provide additional information about particle production and the system evolution in high-energy heavy-ion collisions. In this context, we have analyzed particle ratios in Au+Au collisions at \snn~= 14.5 GeV, and compared to published results for Au+Au collisions at other collision energies~\cite{expdof1,expdof2,expdof3,expdof4,bes_paper,prc,200gevprl}. 

Figure~\ref{fig:like-ratios} shows the dependence of the antiparticle to particle ratios on $\left< N_{\text{part}} \right>$ in
Au+Au collisions at \snn~= 14.5 GeV. These ratios are compared with the corresponding results from Au+Au collisions at \snn~= 7.7, 11.5, 19.6, 27, 39, 62.4, and 200 GeV measured by STAR in earlier runs~\cite{expdof1,expdof2,expdof3,expdof4,bes_paper,prc,200gevprl}.  The $\pi^{-}/\pi^{+}$ ratio has no significant centrality dependence and hovers around unity for all energies. At lower energies, including Au+Au collisions at \snn~= 14.5 GeV, this ratio is slightly greater than one, which is due to isospin conservation and 
the contribution from decays of resonances like $\Delta$ baryons~\cite{bes_paper,besp1}. This effect is more visible at lower energies due to the comparatively smaller yield of pions. The $K^{-}/K^{+}$ ratio is almost flat within uncertainties across all centralities.
However, this $K^{-}/K^{+}$ ratio shows an increase with increasing beam energy. This is because at lower energies, associated production is the dominant mechanism, producing only $K^{+}$, whereas with increasing energy, pair production dominates, producing both $K^{+}$ and $K^{-}$~\cite{bes_paper,besp1}. 
The $\bar{p}/p$ ratio shows a modest increase from central to peripheral collisions, which could be attributed to an increase in proton yields
as a result of baryon stopping and/or a decrease in antiproton yields due to annihilation in central collisions~\cite{bes_paper,besp1}. 
This ratio also increases with increasing collision energy. All these antiparticle-to-particle ratios in Au+Au collisions at \snn~= 14.5 GeV follow the same general patterns as observed at other energies~\cite{expdof1, expdof2, expdof3, expdof4, bes_paper, prc,200gevprl}.

Various ratios of different particle species such as $K^{+}$/$\pi^{+}$, $K^{-}$/$\pi^{-}$, $p/\pi^{+}$ and $\bar{p}$/$\pi^{-}$ are shown in Fig.~\ref{fig:unlike-ratios} for Au+Au collisions at \snn~= 14.5 GeV. Previously published results from the STAR experiment at other beam energies~\cite{expdof1,expdof2,expdof3,expdof4,bes_paper,prc,200gevprl} are also shown for comparison. Both $K^{+}$/$\pi^{+}$ and $K^{-}$/$\pi^{-}$ ratios increase from peripheral to mid-central collisions and then remain almost independent of $\left< N_{\text{part}} \right>$. This pattern is due to strangeness equilibrium described in various thermodynamical models~\cite{model1,model2} and is also impacted by baryon stopping at midrapidity. 
The results from Au+Au collisions at \snn~= 14.5 GeV fit well in the energy dependence trend. 
The $p/\pi^{+}$ ratio increases slowly from peripheral to central collisions, whereas the $\bar{p}/\pi^{-}$ ratio stays flat across all values of $\left< N_{\text{part}} \right>$. Also, there is a decrease in the $p/\pi^{+}$ ratio and an increase of the $\bar{p}/\pi^{-}$ ratio with increasing collision energy, which together can be attributed to baryon stopping at lower energies being prominent for central collisions.

\vspace{0.2in}

\subsection{Kinetic Freeze-out Properties} 

The invariant yields and \pt~spectra of particles give us tools to
study the freeze-out properties of the system. There are two
freeze-out stages observed in high-energy heavy-ion collision
experiments: chemical freeze-out and kinetic freeze-out.   
First, inelastic collisions among the particles cease, defining the
chemical freeze-out stage. After that point, there is no further
production of new particles, and the yields of particle types becomes
fixed. Various thermodynamic models are widely applied to extract the
information of this stage in terms of chemical freeze-out temperature
and baryon chemical potential~\cite{HICs4, bes_paper, thermus,
  th_model}. Thereafter, the particles collide only elastically. After
further expansion of the system, as the inter-particle separation
becomes large, such elastic collisions between particles also cease,
leading to the kinetic freeze-out stage. The momenta of the particles
are fixed after this point, and the particles freely propagate to the
detector. The particle \pt~spectra thus contain information about the kinetic freeze-out stage. Hydrodynamics inspired models such as the Blast-Wave model are used to extract the kinetic freeze-out properties~\cite{HICs4,bes_paper,prc,bw}. This stage is characterized by the kinetic freeze-out temperature $T_{k}$ and radial flow velocity $\beta$, which carry signatures of the transverse expansion of the system. Here, we follow the previously adopted procedures to study the kinetic freeze-out properties in Au+Au collisions at \snn~= 14.5 GeV. The chemical freeze-out properties are not discussed in this paper as the final measurements for strange hadrons yields for $\Lambda$ and $\Xi$ are not available. These will be reported in a future STAR paper. \\

The calculation of kinetic freeze-out parameters is carried out
through a Blast-Wave model ~\cite{bw} fit to the measured particle
\pt~spectra in Au+Au collisions at \snn~= 14.5 GeV. It is a hydrodynamics inspired model in which the particles are assumed to be locally thermalized at the kinetic freeze-out temperature $T_{k}$ and move with a common radial flow velocity $\beta$. For such a radially boosted uniform hard sphere, the transverse momentum distribution of the produced particles can be written as~\cite{bw,bes_paper,prc}

\begin{eqnarray}
\frac{dN}{p_{\mathrm T}dp_{\mathrm T}} \propto \int_{0}^{R} r dr
  \,m_{T} I_{0} \left( \frac{p_{\mathrm T}\sinh\rho(r)}{T_{k}} \right) \nonumber \\
 \times K_{1} \left( \frac{m_{T}\cosh\rho(r)}{T_{k}} \right), \label{eq:bw}
\end{eqnarray}
where $m_{T} = \sqrt{p_{\mathrm T}^{2}+m^{2}}$ is the transverse mass of the particle, $\rho (r) = \tanh^{-1} \beta$, and $I_{0}$ and $K_1$ are modified Bessel functions.  A flow velocity profile of the following form is used~\cite{bw,bes_paper,prc}.
\begin{equation}
\beta = \beta_{s} (r/R)^n, \label{beta}
\end{equation}
where $\beta_{S}$ is the surface velocity, $r/R$ is the radial position in the thermal source with radius R, and the exponent $n$ in the flow velocity profile is a parameter. The average transverse radial flow velocity $\left< \beta \right>$ is given by $\left< \beta \right> = \frac{2}{2+n} \beta_{S}$.

\begin{figure}[H]
\centering
\begin{overpic}[scale=0.4]{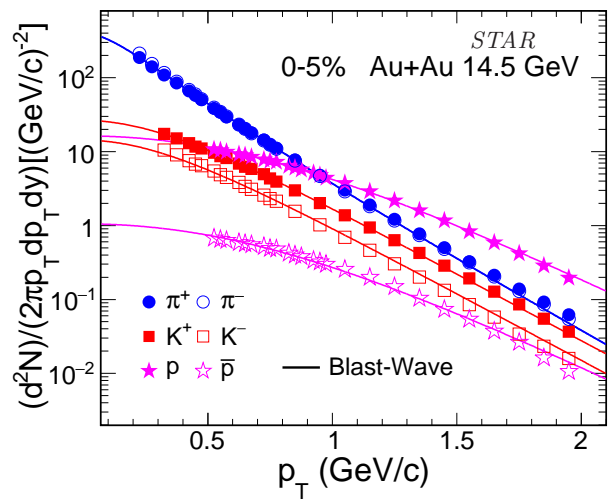}
\put (75,75) { \it STAR}
\end{overpic}
\caption{Simultaneous Blast-Wave model fits to the \pt-spectra of
  $\pi^{\pm}$, $K^{\pm}$, $p$($\bar{p}$) from 0--5\% central Au+Au
  collisions at \snn~= 14.5 GeV. Uncertainties on experimental data represent statistical and systematic uncertainties added in quadrature, mostly smaller than the symbol size.}
\label{fig:bw-fit}
\end{figure}

\begin{figure*}[!tp]
\centering
\begin{overpic}[scale=0.9]{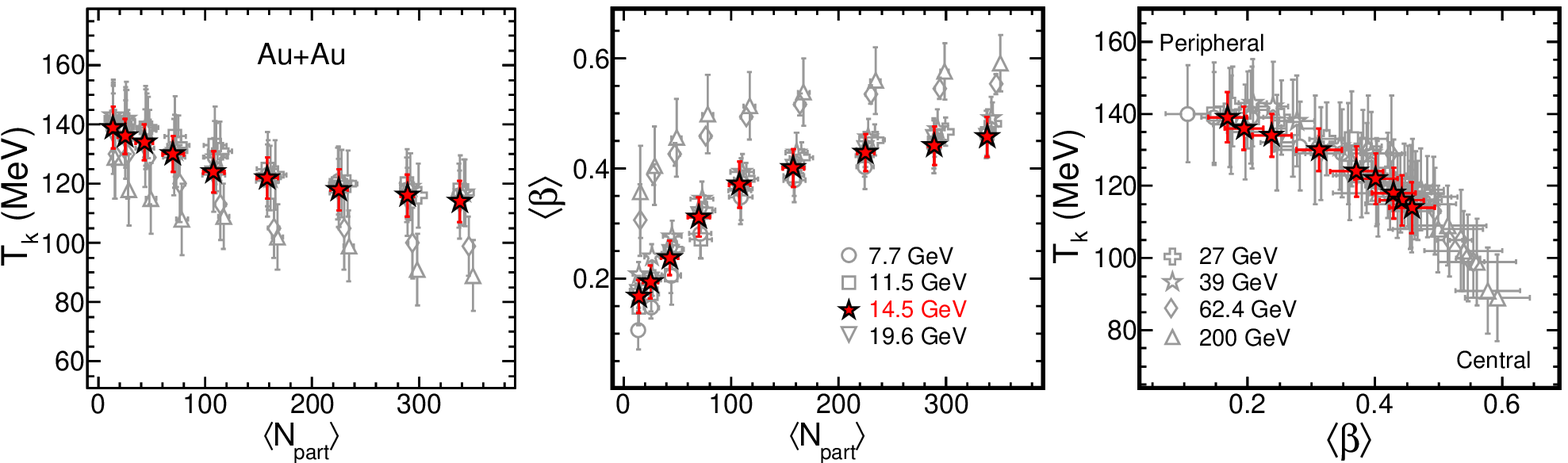}
\put (7,7) { \it STAR}
\end{overpic}
\caption{Left panel: $T_{k}$ as a function of $\left< N_{\rm part} \right>$. Middle panel: $\beta$ as a function of $\left< N_{\rm part} \right>$. Right panel: variation of $T_{k}$ with $\beta$.  In all three panels, present results for Au+Au collisions at \snn~= 14.5 GeV are shown in comparison with the same quantities for Au+Au collisions at \snn~= 7.7, 11.5, 19.6, 27, 39, 62.4, and 200 GeV measured by STAR in earlier runs \cite{bes_paper}. Systematic uncertainties are shown. Statistical uncertainties are much smaller than systematic ones.}
\label{fig:kfo-para}
\end{figure*}

To extract the kinetic freeze-out parameters, simultaneous Blast-Wave
model fits to the $\pi^{\pm}$, $K^{\pm}$ and $p(\bar{p})$ spectra are
performed~\cite{bw,bes_paper,prc} as plotted in Fig.~\ref{fig:bw-fit}
for central Au+Au collisions at \snn~= 14.5 GeV. The
low-\pt~region of the pion spectra is affected by resonance decays,
and therefore the pion spectra above $p_{\mathrm T} > 0.5$ GeV/$c$ are
used for fitting. The Blast-Wave model is not very suitable for
fitting the high \pt~region of the \pt~spectra~\cite{fit-bw1}. 
Hence, the Blast-Wave model fits are very sensitive to the \pt~range used~\cite{fit-bw2}. The previously optimized \pt~ranges~\cite{bes_paper,prc,fit-bw2} are used for Au+Au collisions at \snn~= 14.5 GeV to extract the kinetic freeze-out parameters. The fit ranges used for pions, kaons, and protons are 0.5--1.35 GeV/$c$, 0.3--1.35 GeV/$c$, and 0.5--1.25 GeV/$c$, respectively.

Figure~\ref{fig:kfo-para} presents the kinetic freeze-out parameters $T_{k}$ (left) and $\left< \beta \right>$ (middle) as a function of $N_{\text{part}}$, and presents the correlation between $T_{k}$ and $\beta$ (right) for Au+Au collisions at \snn~= 14.5 GeV. These results are compared with published data for Au+Au collisions at \snn~= 7.7, 11.5, 19.6, 27, 39, 62.4, and 200 GeV measured by STAR in earlier runs~\cite{bes_paper,prc}. $T_{k}$ shows a dependence on $N_{\text{part}}$, decreasing from peripheral to central collisions. This observation supports the prediction of a short-lived fireball in the case of peripheral collisions~\cite{Tk}. The average flow velocity $\left< \beta \right>$, on the other hand, increases from peripheral to central collisions. This indicates a higher rate of expansion of the system in central collisions. It is also seen that higher RHIC energies such as 62.4 and 200 GeV, have comparatively higher $\beta$ than other BES-I energies. Lastly, the correlation plot between $T_{k}$ and $\beta$ confirms 
an anti-correlation between these two quantities, i.e. as $T_{k}$ decreases, $\beta$ increases. The behavior of the kinetic freeze-out parameters in Au+Au collisions at \snn~= 14.5 GeV is consistent with previous observations~\cite{bes_paper,prc}. The extracted  fit parameters $T_{k}$, $\left< \beta \right>$, and $n$  along with the $\chi^{2}$/ndf values from the Blast-Wave model fits in Au+Au collisions at \snn~= 14.5 GeV are reported in Table~\ref{tab:bw-para}.

\vspace{0.2in}

\subsection{Azimuthal Anisotropy}

\vspace{0.2in}

\subsubsection{The event plane resolution}

\vspace{0.2cm}
\begin{figure*}[!htbp]
\centering
\begin{overpic}[scale=0.8]{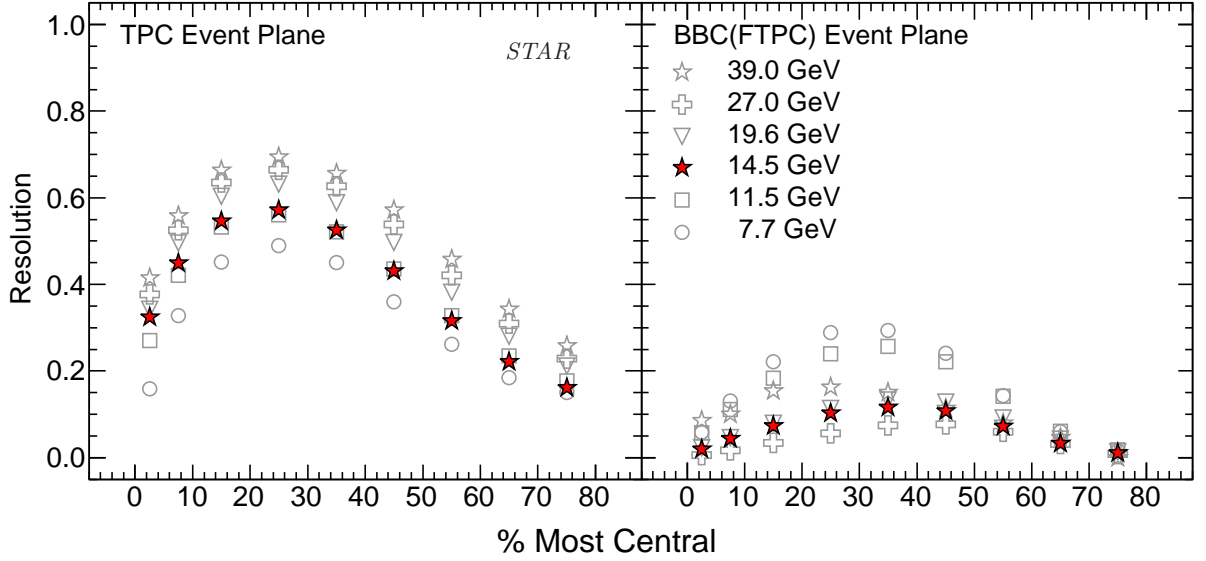}
\put (42,42) { \it STAR}
\end{overpic}
\caption{The event plane resolution calculated for Au+Au collisions
  at \snn~= 14.5~GeV (solid star) as a function of
  centrality. The current results are compared with those for 7.7, 11.5,
  14.5, 19.6, 27 and 39~GeV.  The left panel shows the second-order event plane resolution 
  reconstructed by using the TPC tracks ($|\eta|<$1). The right panel shows the second-order
  event plane resolution for 39 GeV from the FTPC (2.5$<|\eta|<$4.0)
  and the first-order event plane resolution from the inner tiles of the BBC (3.8$<|\eta|<$5.2).}
\label{fig:resol-tpc-bbc}
\end{figure*}

\begin{figure*}[!htbp]
\centering
\begin{overpic}[scale=0.8]{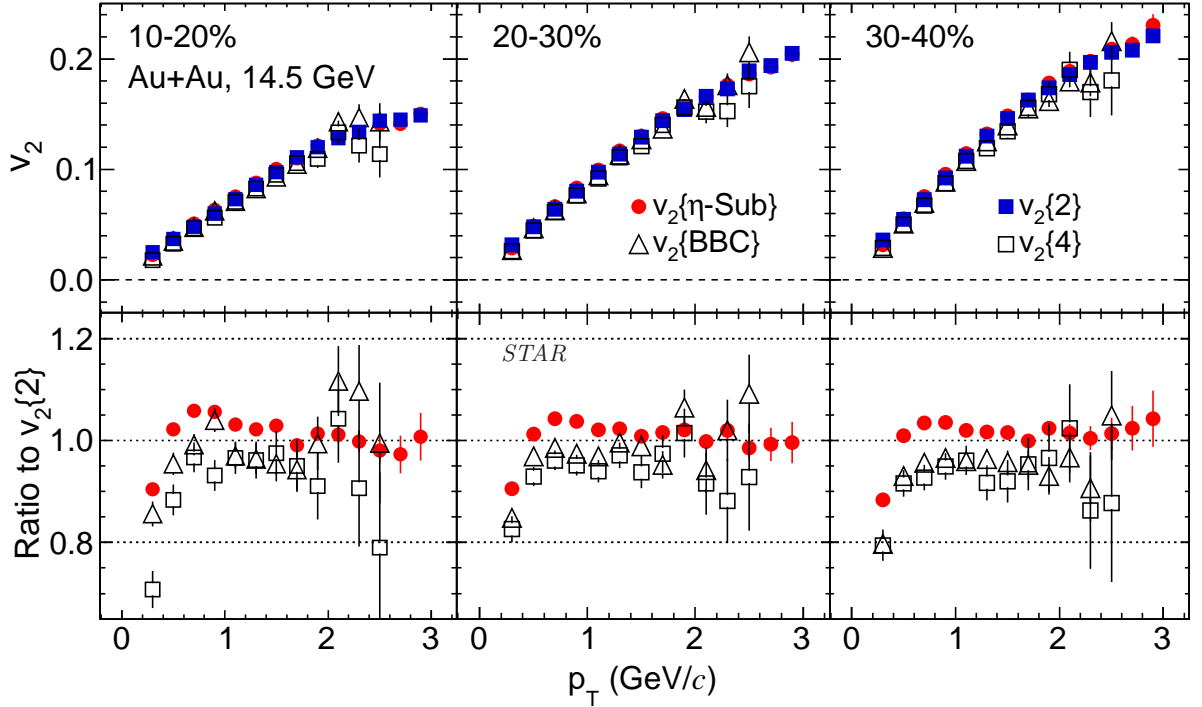}
\put (42,29) { \it STAR}
\end{overpic}
 \caption{Inclusive charged particle $v_{2}$ at mid-pseudorapidity ($|\eta|<$1.0) as a function of
     \pt~for 10--20\% (left), 20--30\% (middle) and 30--40\% (right)
     centralities in Au + Au collisions at \snn~= 14.5 GeV.  Results are shown for the $\eta$-sub event plane
     method (circles), BBC event plane (open triangles), 2-particle (solid squares) and 4-particle 
     (open squares) cumulants. The bottom panels show the ratio of $v_{2}$ measured using 
     the various methods with respect to the 2-particle cumulant result, $v_{2}\lbrace 2 \rbrace$. Errors are statistical. Systematic uncertainties are small ($\sim 2\%$).}
\label{fig:v2mthd}
\end{figure*}

Due to the finite multiplicity in each event, the event plane angle ($\Psi_{n}$)
deviates from the reaction plane azimuthal angle ($\Psi_{R}$). Hence a
resolution correction needs to be performed to obtain the correct
measurement of the flow coefficients ($v_{n}$) \cite{art1}. For this analysis, 
the event planes are determined from the TPC in the midrapidity region, and 
from the BBC at forward rapidity. 

Figure~\ref{fig:resol-tpc-bbc} shows the second-order event-plane resolution from 
the TPC (left panel) and the first-order event-plane resolution from the BBC (right 
panel) as a function of centrality in Au+Au collisions at \snn~= 14.5 
GeV. The event plane resolution has been calculated for nine collision centralities: 
0--5\%, 5--10\%, 10--20\%, 20--30\%, 30--40\%, 40--50\%, 50--60\%, 60--70\% and 
70--80\%. As the event plane resolution depends on the number of particles used 
for event plane reconstruction, it shows a tendency to increase from
peripheral to central collisions. On the other hand, the event
plane is calculated using the anisotropic flow of the event itself, and therefore it
tends to decrease towards central collisions where flow values are
small. Because of these two competing effects, the overall resolution
first increases from peripheral to mid-central collisions and then
decreases. Figure~\ref{fig:resol-tpc-bbc} includes event plane resolutions for the 
same methods at other BES-I energies studied previously by STAR. 
Due to limited statistics and poor BBC resolution, the FTPC~\cite{ftpc_star} event plane is used instead of BBC at 39 GeV. As expected the resolution of the TPC and BBC event planes decreases as the collision energy increases, since the resolution depends on the multiplicity and on the
$v_{2}$ signal~\cite{art1}. The 14.5 GeV resolution does not lie between that observed 
at the adjacent beam energies above and below, and instead is slightly lower than a 
smooth trend would predict. This is a consequence of the additional material of the 
Heavy Flavor Tracker close to the beam pipe, which was present only during the 2014 
run at 14.5 GeV.  
The event plane resolution corrections to the observed $v_{n}$ are applied on an event-by-event 
basis~\cite{res_cent,res_nsm}. In this method, the resolution correction has been applied by 
dividing the flow coefficient of each track, $\cos n(\phi - \Psi_{n})$, by the event-plane resolution 
$\langle R \rangle$ for the appropriate centrality class. 

\vspace{0.2in}

\subsubsection{Comparison of $v_{2}$ from different methods}

\begin{figure*}[!tp]
 \centering
\begin{overpic}[scale=0.5]{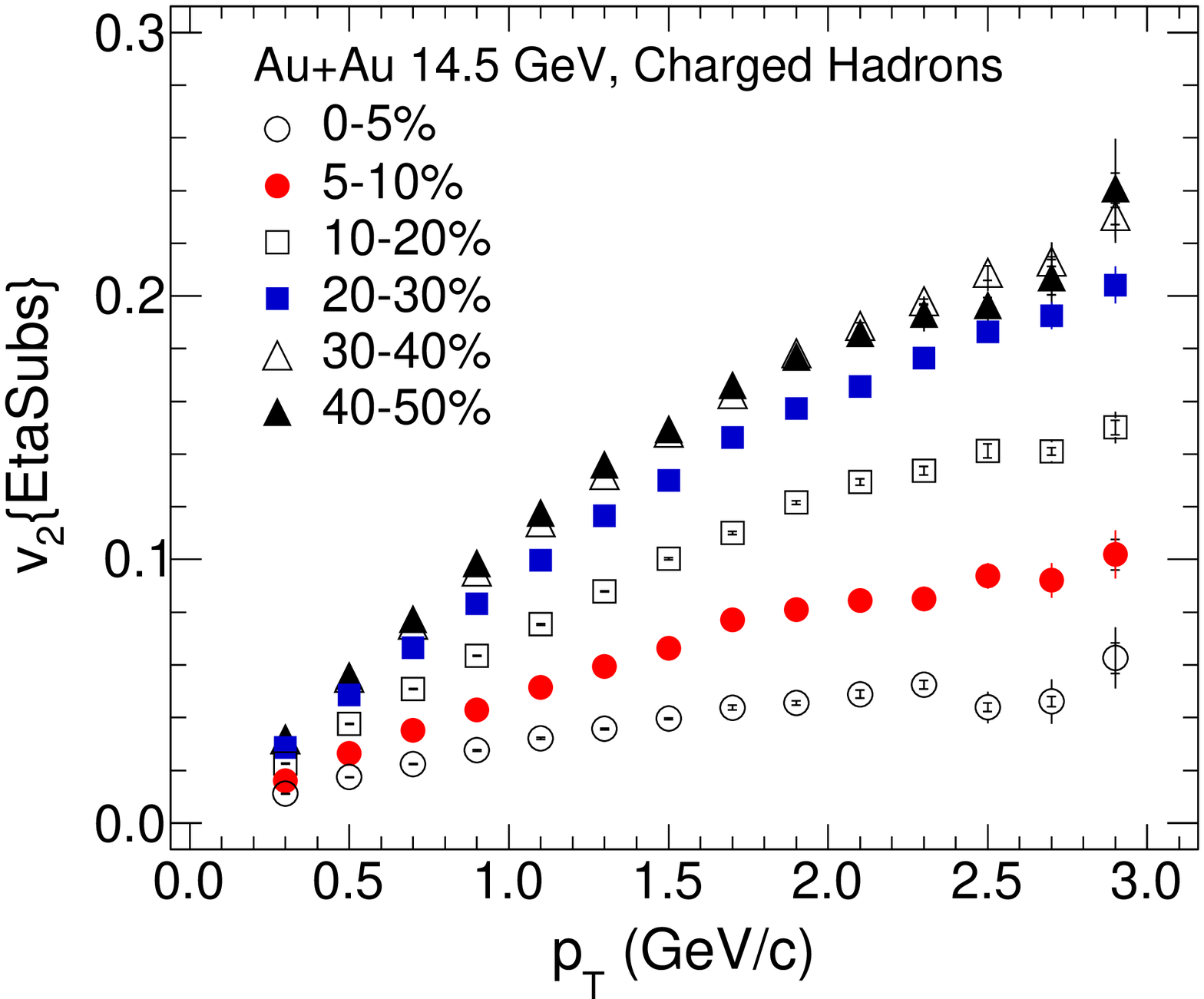}
\put (22,39) { \it STAR}
\end{overpic}
\begin{overpic}[scale=0.5]{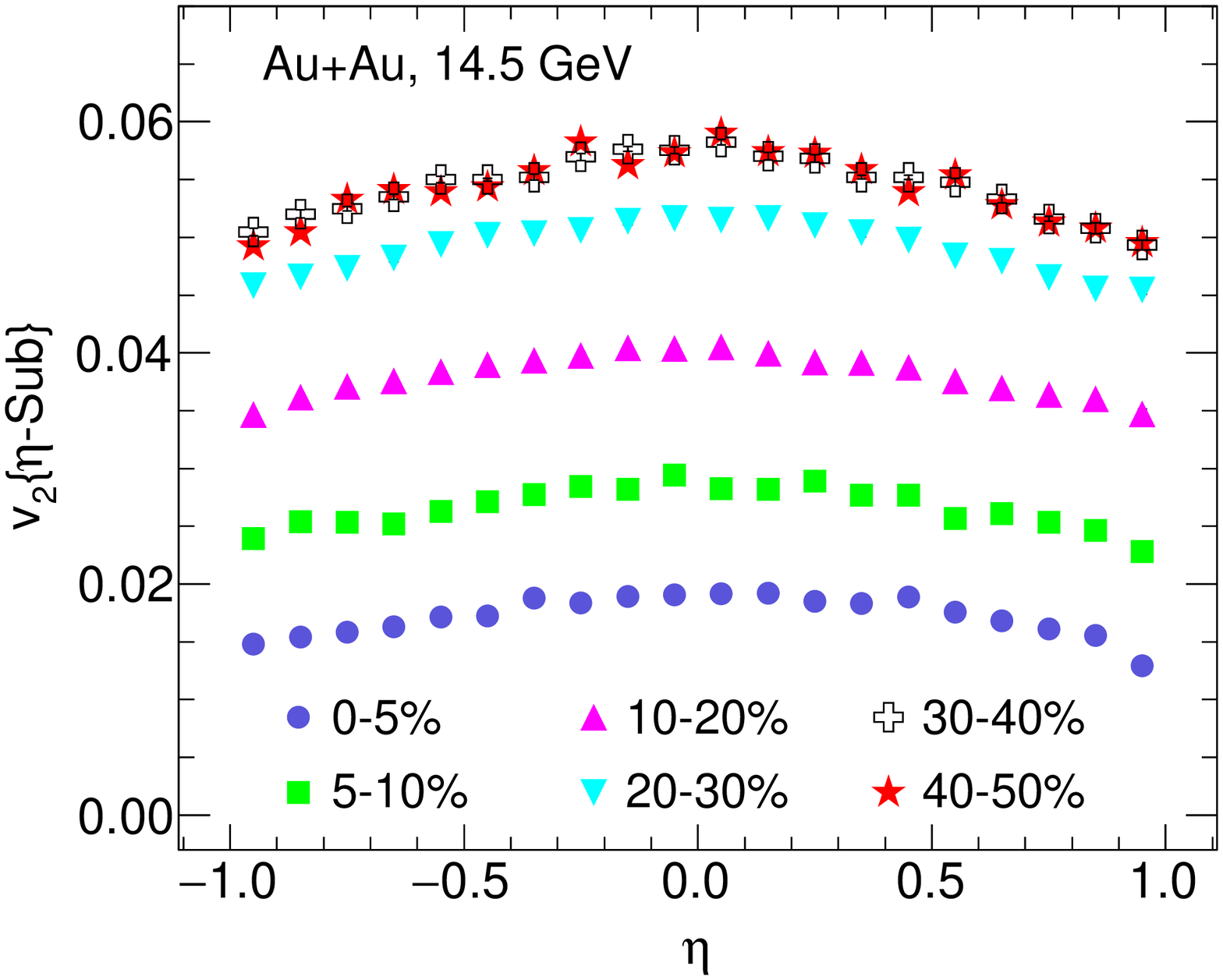}
\end{overpic}
\small
 \caption{\small{Inclusive charged particle elliptic flow $v_{2}$ at mid-pseudorapidity 
 ($|\eta|<$1.0) as a function of transverse momentum \pt~(top) and 
the \pt-integrated $v_2(\eta)$ for six
centrality classes (bottom), obtained using the
 $\eta$-sub event plane method in Au+Au collisions at
 \snn~= 14.5 GeV. Statistical uncertainties are shown
     by error bars, while systematic uncertainties are smaller and are
     plotted as caps.}}
\label{fig:v2pt-eta}
\end{figure*} 

Figure~\ref{fig:v2mthd} presents inclusive charged particle $v_{2}
(p_{\mathrm T})$, using various 
methods, for Au+Au collisions at \snn~= 14.5 GeV. The methods have different 
sensitivities to nonflow effects and $v_{2}$ fluctuations. For the purpose of exact comparisons, 
$v_{2}$ for each method is divided by the elliptic flow based on the two-particle cumulant 
method (denoted $v_2\{2\}$) and the ratios are shown in the lower panels of Fig.~\ref{fig:v2mthd}. 
The difference of $v_{2}\lbrace 2 \rbrace$ compared to
$v_{2}\lbrace {\rm BBC} \rbrace$, $v_{2}\lbrace 4 \rbrace$, and
$v_{2}\lbrace \eta{\rm-sub} \rbrace$ depends on the \pt~range. A larger 
difference is observed in the low-\pt~region ($p_{\mathrm T} < $
1 GeV/$c$). From $p_{\mathrm T} \sim 1$ GeV/$c$ and above, the difference stays roughly constant. The difference between $v_{2}\lbrace{\rm BBC} \rbrace$ and $v_{2}\lbrace 4 \rbrace$ is relatively small, and is less
dependent on \pt. The results suggest that nonflow contributions to the
event plane and two-particle correlation methods depend on \pt. They
also indicate that the use of the first-order reaction plane 
(BBC event plane) to study the second harmonic flow reduces flow
fluctuations which are not correlated between different harmonics. 

\vspace{0.2in}

\subsubsection{Dependence of $v_2$ on transverse momentum, pseudorapidity and centrality}

\vspace{0.2cm}
\begin{figure}[!h]
\begin{overpic}[scale=0.4]{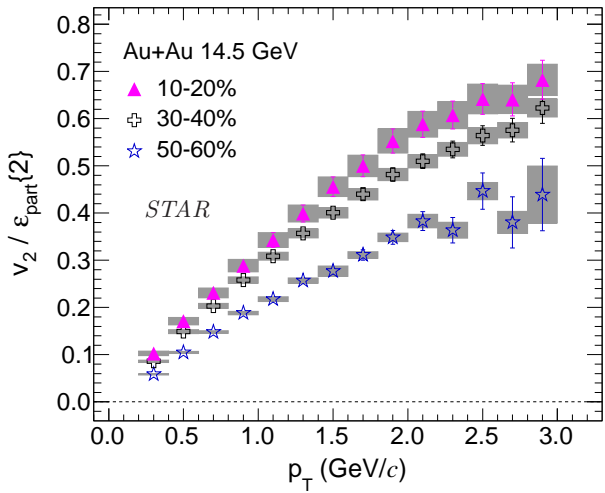}
\put (22,46) { \it STAR}
\end{overpic}
  \caption{The ratio $v_{2}/\epsilon_{\rm part}\{2\}$ for inclusive
    charged particle elliptic flow $v_{2}$ at mid-pseudorapidity as a
    function of \pt~for 10--20\%, 30--40\%, and 50--60\% collision centralities in Au+Au collisions at \snn~= 14.5~GeV. The $v_{2}$ data are from the $\eta$-sub event plane method, and the spatial eccentricity $\epsilon_{\rm part}\{2\}$ is based on a Glauber calculation. The error bars and shaded boxes present the statistical and systematic uncertainties, respectively.}
\label{fig:v2e2}
 \end{figure}

Results for charged particle $v_{2}$ as a
function of \pt~for six collision centrality intervals
are presented in the top panel of Fig.~\ref{fig:v2pt-eta}. The $v_{2}$
shows a monotonically increasing trend with increasing \pt~for Au+Au
collisions at \snn~= 14.5 GeV. The differential $v_{2}$ also exhibits 
centrality dependence. The trends of $v_{2}$($p_{\mathrm T}$) are similar to
those observed at other BES-I energies. The bottom panel of Fig.~\ref{fig:v2pt-eta} presents the 
\pt-integrated $v_{2} (\eta)$ for six centrality classes. The $v_{2}$ has a
weak dependence on $\eta$. Also, there is a clear centrality
dependence observed in $v_{2}$. The trend of $v_{2}(\eta)$ is similar to that for
other BES-I energies~\cite{expdof1,expdof2,expdof3,expdof4,bes_paper,prc,200gevprl}. 

The larger magnitude of $v_{2}$ in peripheral collisions can be attributed to the larger initial
eccentricity in coordinate space for peripheral
collisions. The participant eccentricity is the initial configuration
space eccentricity of the participating nucleons. The
root-mean-square participant eccentricity ($\epsilon_{\rm part}\{2\}$) is
calculated from a MC Glauber model~\cite{glauber1,glauber2} and reported in Table~\ref{table:eccent}. 

In Fig.~\ref{fig:v2e2}, the centrality dependence of 
$v_{2}(p_T)$ over eccentricity $\epsilon_{\rm part}\{2\}$ is shown for Au + Au collisions at
\snn~= 14.5~GeV for 10--20\%, 30--40\%, and 50--60\% collision centralities. 
Central collisions have higher $v_{2}/\epsilon_{\rm part}\{2\}$
than peripheral collisions. This finding is consistent with a picture where
collective interactions are stronger in collisions with a larger number
of participants. The centrality dependence of $v_{2}/\epsilon_{\rm part}\{2\}$ is
observed to be similar to that reported previously by 
STAR~\cite{expdof1,expdof2,expdof3,expdof4,bes_paper,prc,200gevprl}.  

\vspace{0.2in}

\subsubsection{Beam energy dependence of $v_{2}$}

\begin{figure*}[!htbp]
 \centering
\begin{overpic}[scale=0.30]{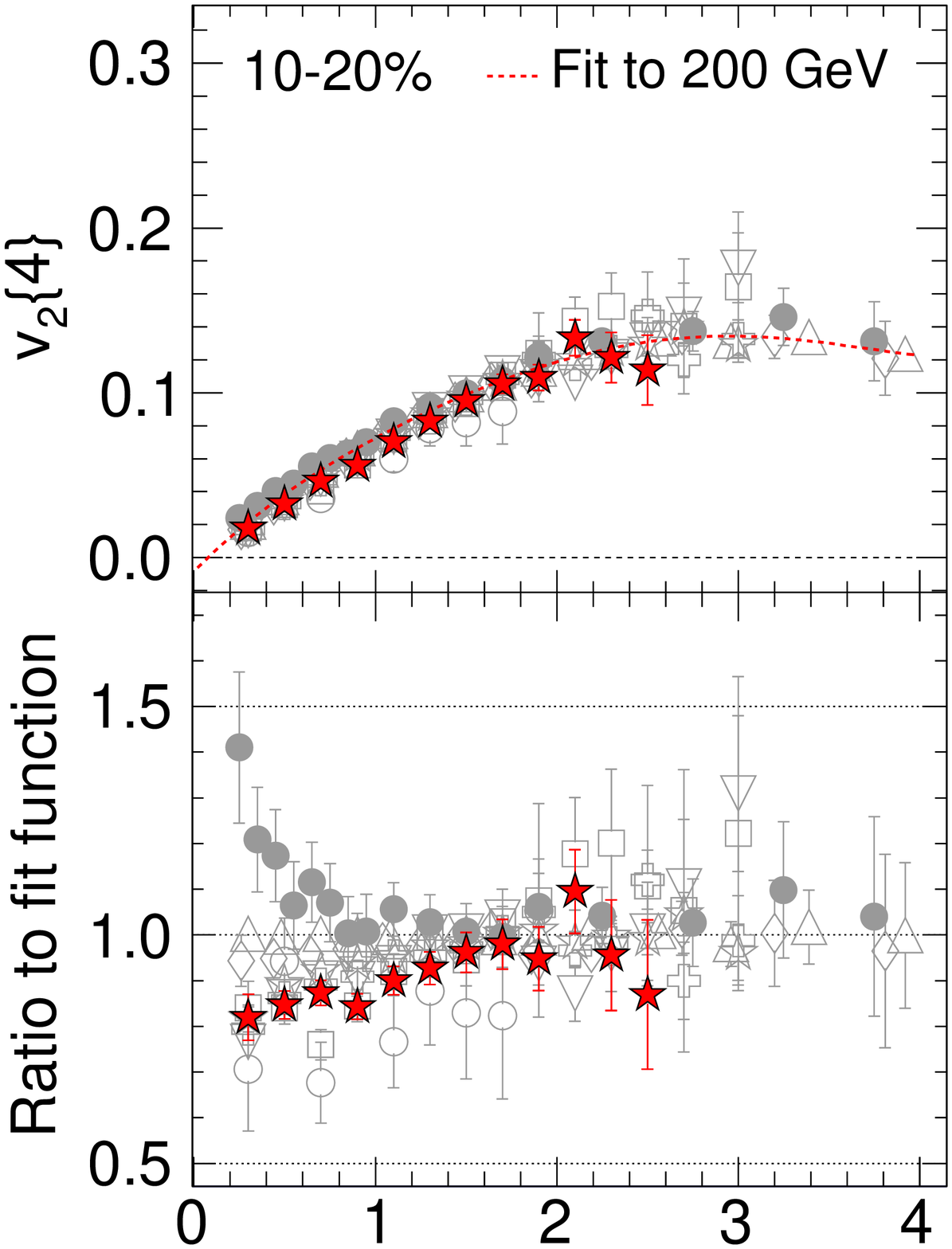}
\put (20,50) { \it STAR}
\end{overpic}
\begin{overpic}[scale=0.30]{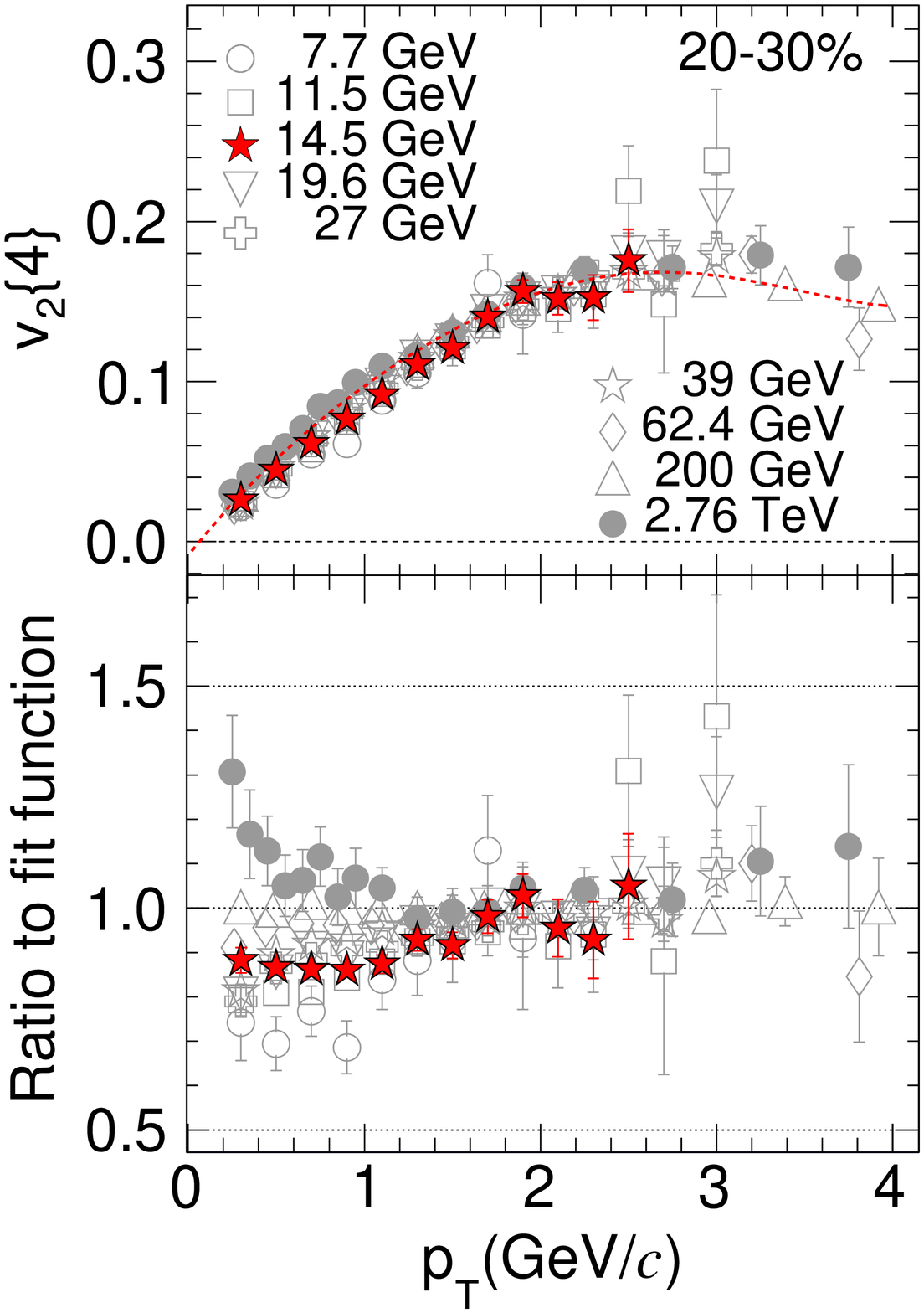}
\put (20,50) { \it STAR}
\end{overpic}
\begin{overpic}[scale=0.30]{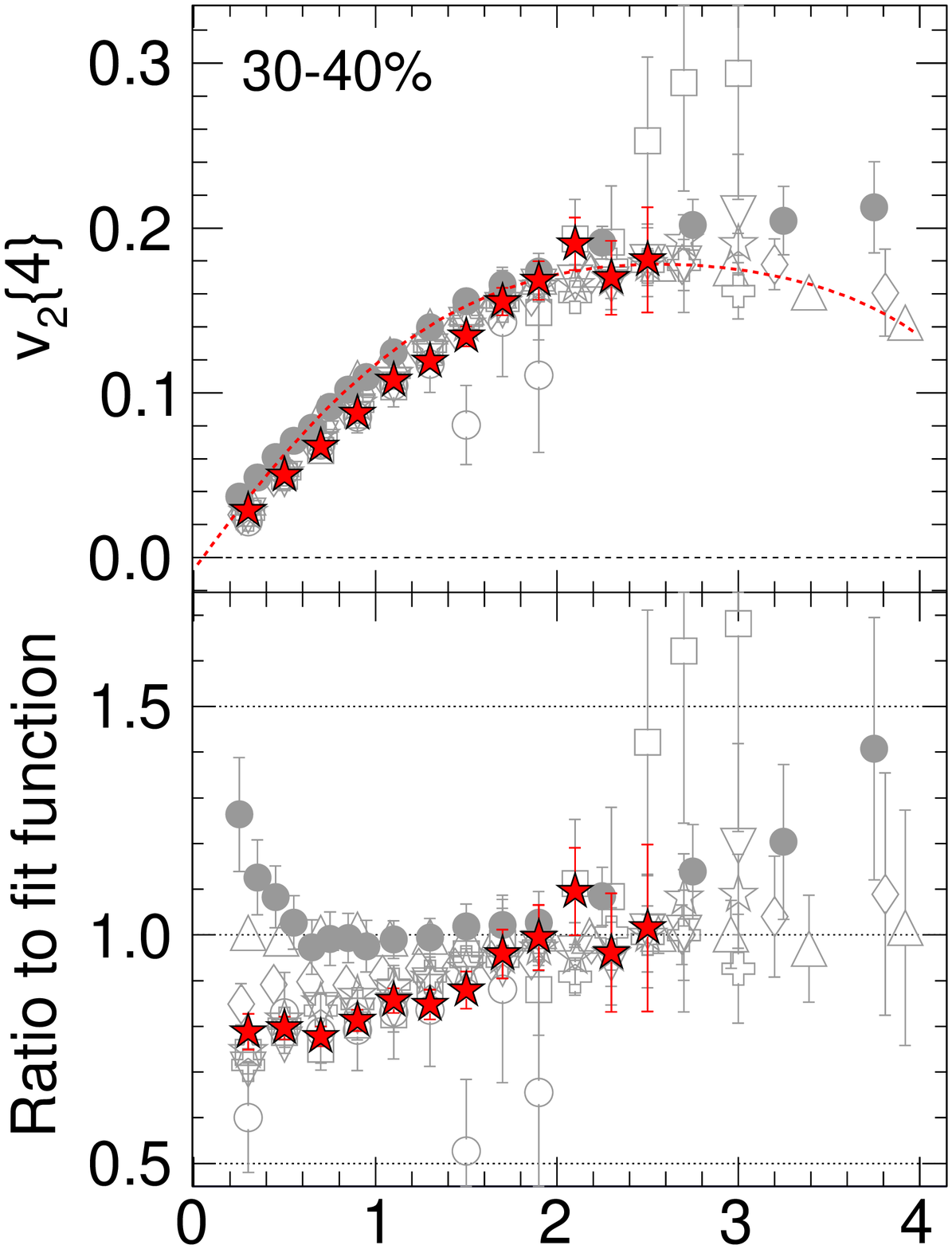}
\put (20,50) { \it STAR}
\end{overpic}
 \small 
 \caption{\small{The upper panels show inclusive charged particle
     elliptic flow $v_{2}\lbrace4\rbrace$ versus \pt~for various
     collision energies (\snn~= 7.7 GeV to 2.76 TeV) at three
     centralities: 10--20\%, 20--30\%, and 30--40\%. The present
     results at 14.5 GeV (and also for other energies except 2.76 TeV)
     are for mid-pseudorapidity ($|\eta| < 1.0$). The measurement of
     $v_{2}$ at 2.76 TeV was done at mid-pseudorapidity ($|\eta| <
     0.8$).  
 Furthermore, all results for \snn~= 7.7 to 200 GeV are for Au+Au collisions and those for 2.76 TeV are for Pb+Pb collisions. The dashed red curves show 5th-order polynomial function fits to the results from Au+Au collisions at \snn~= 200 GeV. The lower panels show the ratio of $v_{2}\lbrace4\rbrace$ versus \pt~for all \snn~with respect to the fit curve. Error bars are shown only for statistical uncertainties. Systematic uncertainties are small ($\sim 2\%$)}}
\label{fig:v24Edep_pt}
 \end{figure*}
 
\begin{figure*}[!htb]
 \centering
\begin{overpic}[scale=0.4]{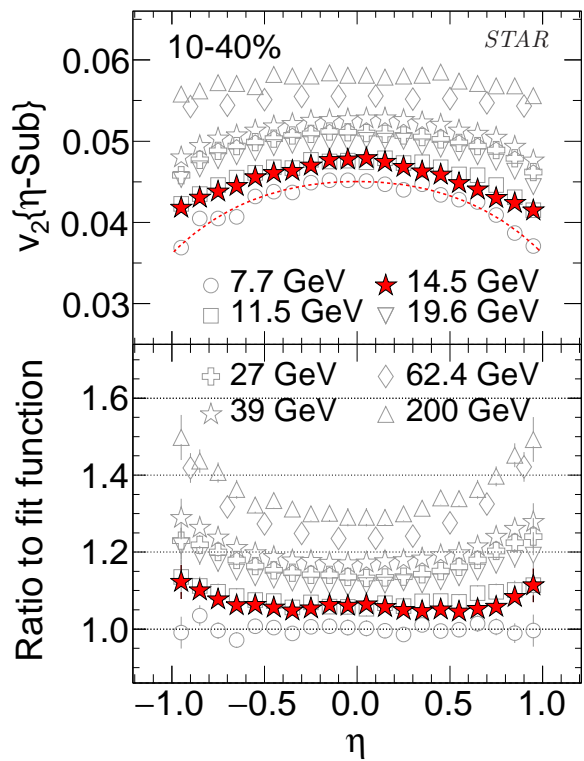}
\put (60,95) { \it STAR}
\end{overpic}
 \small
 \caption{\small{The upper panel shows inclusive charged particle elliptic flow $v_{2}$($\eta$-sub) versus $\eta$ at mid-pseudorapidity for various collision energies (\snn~= 7.7 GeV to 200 GeV). The dashed red curve shows an empirical fit to the result from Au+Au collisions at \snn~= 7.7 GeV. The lower panels shows the ratio of $v_{2}\lbrace4\rbrace$ versus $\eta$ for all \snn~with respect to the fit curve. The results are shown for 10--40\% collision centrality. Error bars are shown only for statistical uncertainties.}}
\label{fig:v24Edep_eta}
 \end{figure*}
 
\begin{figure*}[!tbp]
\centering
\begin{overpic}[scale=0.7]{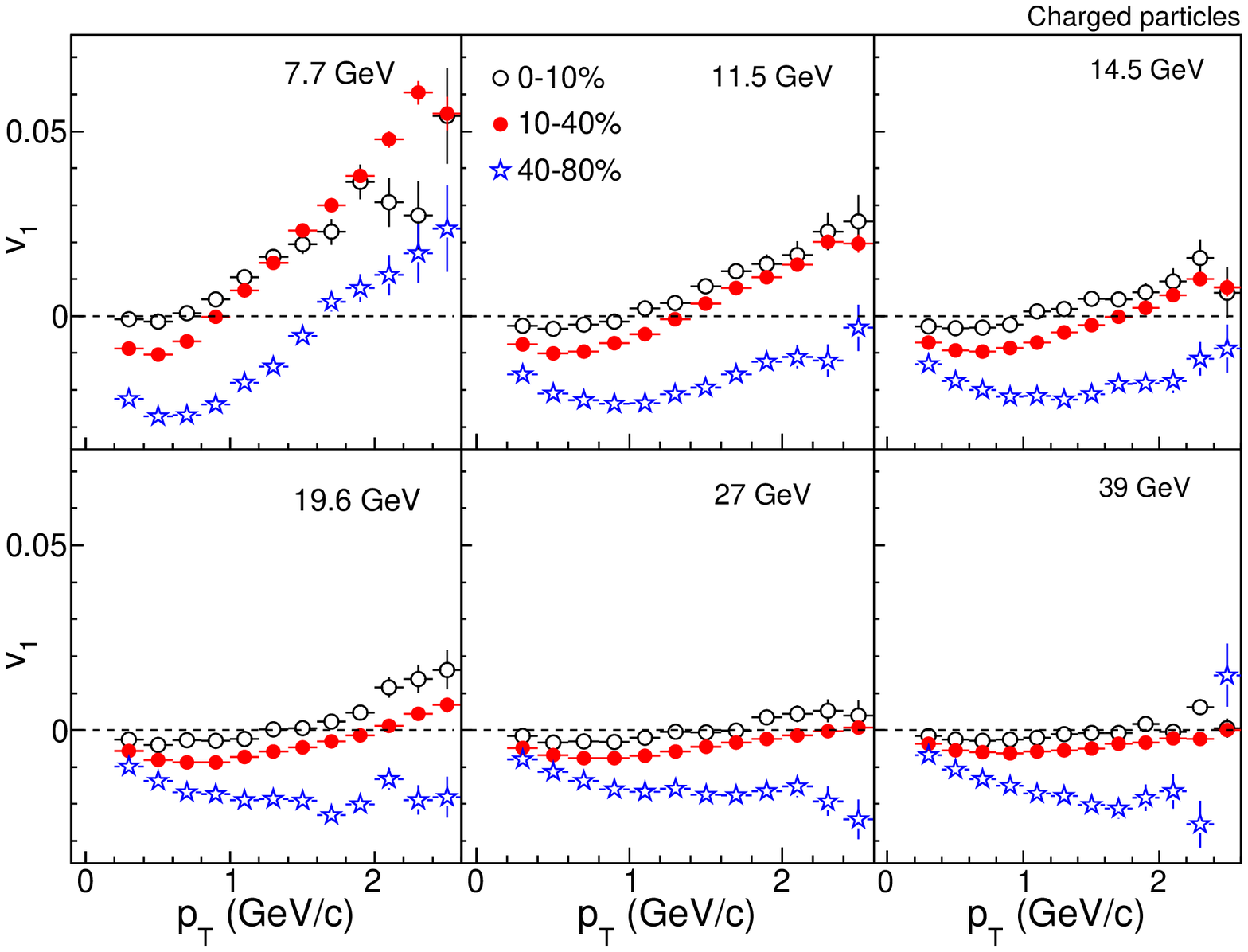}
\put (70,35) { \it STAR}
\end{overpic}
\caption{Charged particle $v_{1}$ as a function of \pt~in Au+Au
  collisions at \snn~= 7.7 -- 39~GeV for 0--10\%, 10--40\% and 40--80\% centrality intervals.  Error bars are shown only for statistical uncertainties. Systematic uncertainties are small ($\sim 2\%$).}
\label{fig:v1-eta-pt-snn-1}
\end{figure*}

\begin{figure*}[!tbp]
\centering
\begin{overpic}[scale=0.7]{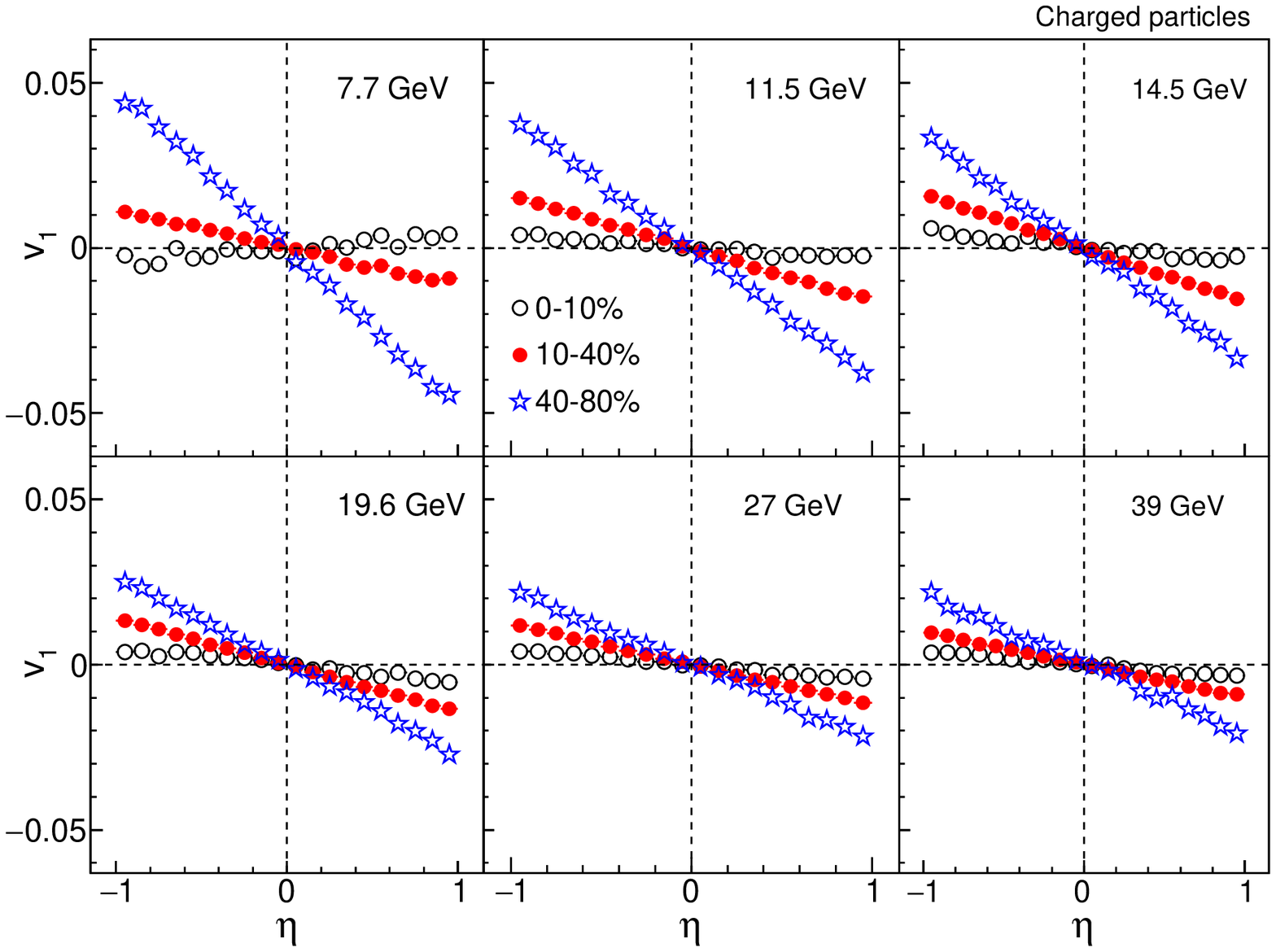}
\put (70,35) { \it STAR}
\end{overpic}
\caption{Charged particle $v_{1}$ as a function of $\eta$ in Au+Au
  collisions at \snn~= 7.7 -- 39~GeV for 0--10\%, 10--40\% and 40--80\% centrality intervals.  Error bars are shown only for statistical uncertainties. Systematic uncertainties are small ($\sim 2\%$). }
\label{fig:v1-eta-pt-snn-2}
\end{figure*} 

\begin{figure*}[!tbp]
\centering
\begin{overpic}[scale=0.42]{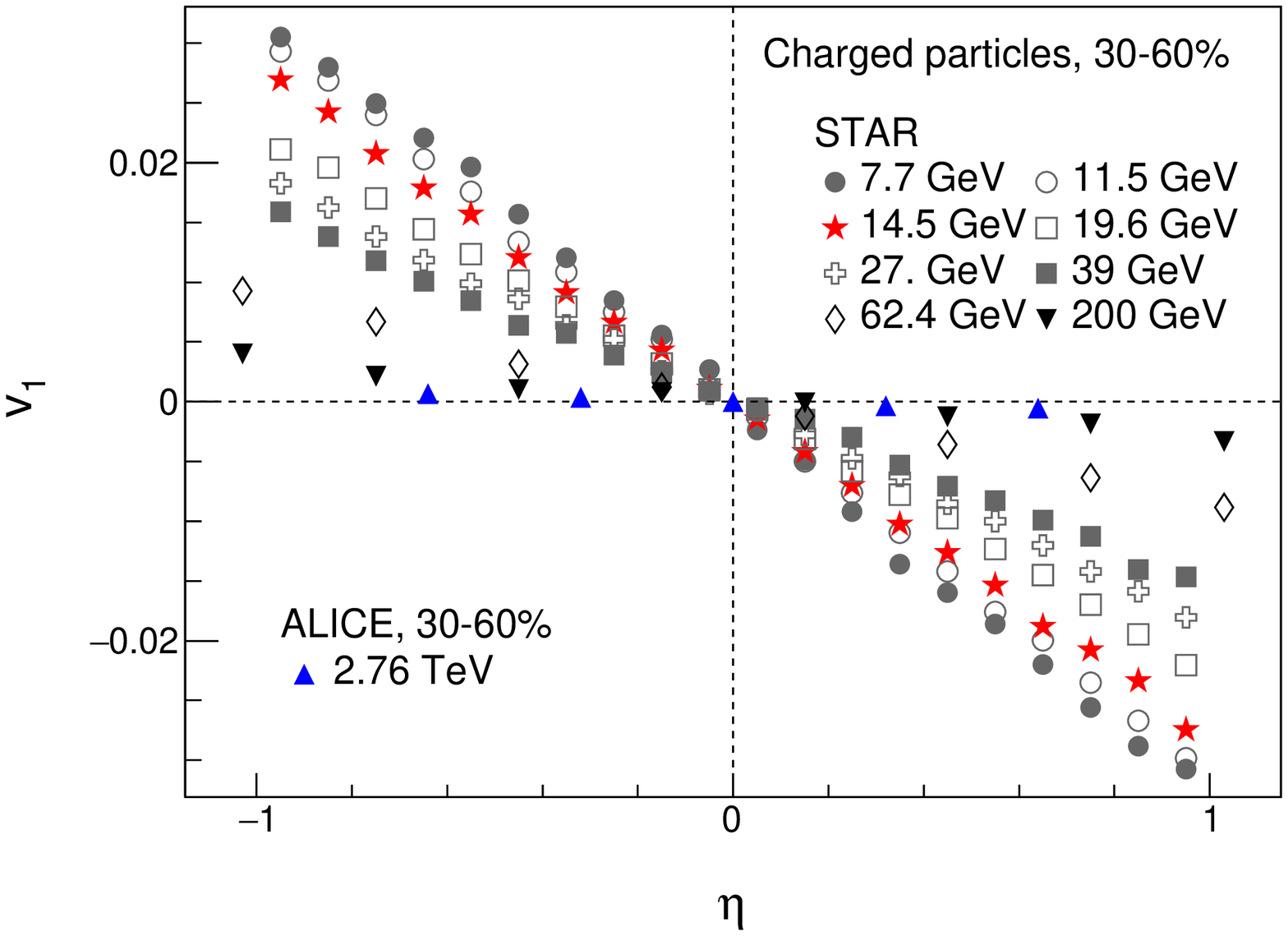}
\end{overpic}
\begin{overpic}[scale=0.42]{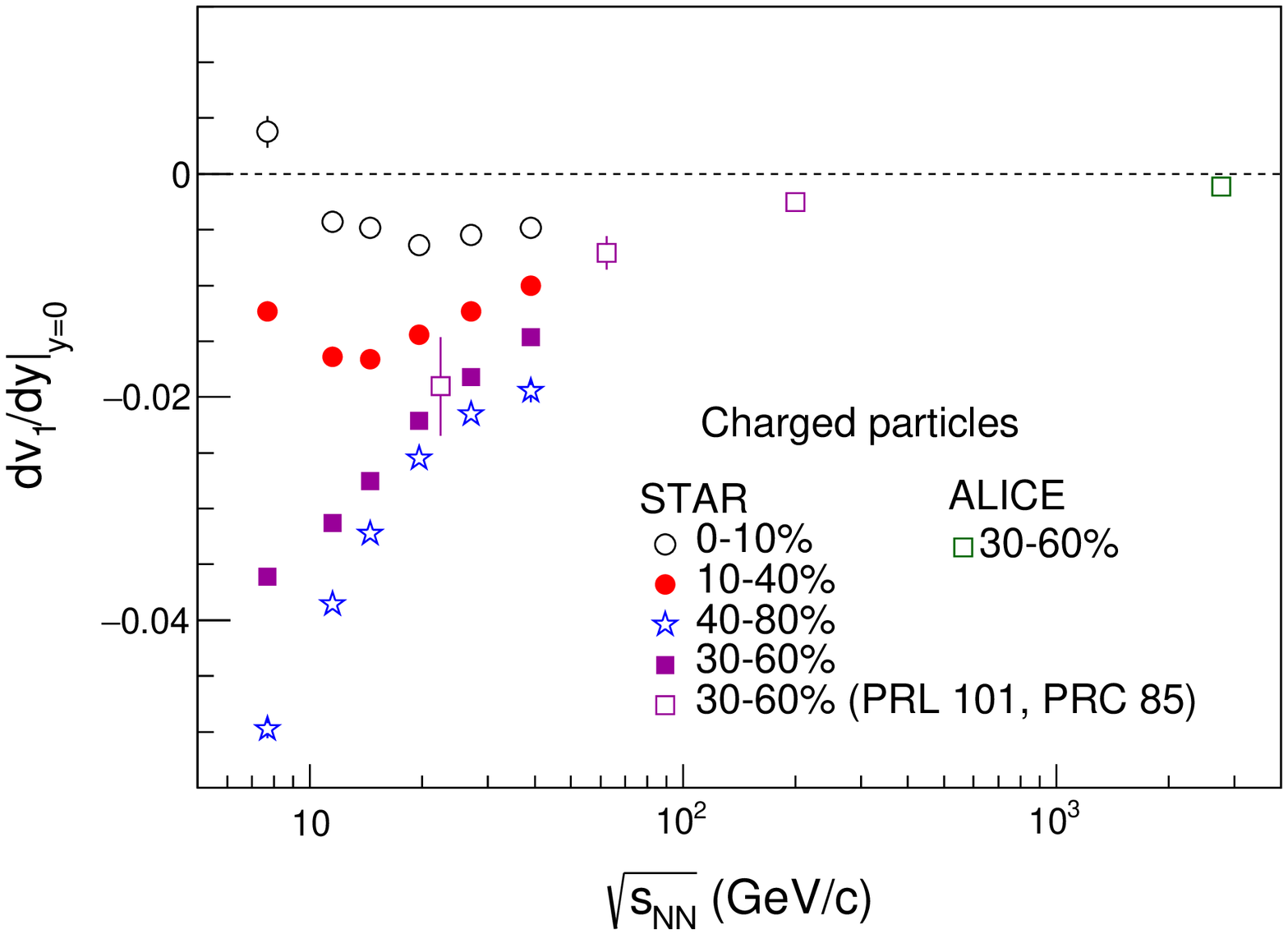}
\end{overpic}

\caption{Left panel: Charged particle $v_{1}$ as a function of $\eta$ in Au+Au collisions at \snn~= 7.7--39~GeV for 30--60\% centrality interval. Results are compared to 62.4 and 200 GeV Au+Au collisions at RHIC and to 2.76 TeV Pb+Pb collisions at the LHC. Right panel: Charged particle $v_{1}$ slope, $dv_{1}/dy$, at mid-pseudorapidity as a function of \snn~for different centralities. Error bars are shown only for statistical uncertainties. Systematic uncertainties are small ($\sim 2\%$).}
\label{fig:v1-eta-pt-snn-lhc}
\end{figure*} 

The BES-I data from the STAR experiment offer an opportunity to study the collision energy dependence of $v_{2}$ using a 
wide-acceptance detector at midrapidity. Figure~\ref{fig:v24Edep_pt} shows the 
comparison of the \pt~dependence of
$v_{2}\lbrace4\rbrace$ for \snn~= 14.5~GeV at 10--20\%,
20--30\%, and 30--40\% centralities with other published results
from STAR~\cite{expdof1,expdof2,expdof3,expdof4,bes_paper,prc,200gevprl,v2_62gev,v2_62_200gev}
and ALICE~\cite{v2_2.76TeV}. One reason to select the 
$v_{2}\lbrace4\rbrace$ results for this comparison is 
to keep the method for $v_{2}$ extraction consistent
with the published results from ALICE. Another reason is that
$v_{2}\lbrace4\rbrace$ is found to have low sensitivity to nonflow correlations. The
200 GeV data are empirically fit by a fifth-order polynomial function.  
For comparison, the $v_{2}$ from other energies are divided by the fit
function and shown in the lower panels of Fig.~\ref{fig:v24Edep_pt}.  We choose 
the 200 GeV data as a reference because its statistical uncertainties are
smallest. For \pt~below 2 GeV/$c$, the $v_{2}$ values rise with
increasing collision energy. Above $p_{\mathrm T} \sim$ 2 GeV/$c$, the $v_{2}$
values are comparable within statistical uncertainties. The increase
of $v_{2}$($p_{\mathrm T}$) as a function of energy can be attributed to
the change of chemical composition from low to high energies
\cite{bes_paper} and/or larger collectivity at the higher collision
energies.

Figure~\ref{fig:v24Edep_eta} presents the $\eta$ dependence of $v_2\{\eta
{\rm-sub}\}$ for \snn~= 7.7 to 200 GeV. The 14.5 GeV
data points are plotted as solid red stars. The dashed red line in the
upper panel in Fig.~\ref{fig:v24Edep_eta} shows an empirical fit to
the results from Au+Au collisions at \snn~= 7.7 GeV. The
bottom panel presents the ratio of $v_{2}(\eta)$ from all beam
energies with respect to this fitted curve. The $v_{2}(\eta)$
changes shape as the beam energy decreases. The $v_{2}(\eta)$ shape 
at 14.5 GeV follows the trend of other beam energies.

\vspace{0.2in}

\subsubsection{Transverse momentum, pseudorapidity, and centrality dependence of $v_{1}$}
 Measurements of the charged particle $v_{1}(p_T)$ in three centralities (0-10\%, 10--40\% and 40-80\%) in Au+Au 
collisions at \snn~= 7.7 -- 39~GeV are presented in Fig.~\ref{fig:v1-eta-pt-snn-1}. This work focuses only on the (pseudo)rapidity-odd component of the first harmonic coefficient (directed flow).  Since this component by definition 
has the property $v_1(-\eta, p_T) = -v_1(\eta, p_T)$, the integral of $v_1(\eta, p_T)$ 
over any symmetric $\eta$ range is zero.  Therefore, in presenting the above-mentioned 
pseudorapidity-integrated $p_T$ dependence, the $v_1$ at negative $\eta$ is multiplied 
by $-1$.  By definition, $v_{1} (p_{\mathrm T})$ must approach zero as \pt~approaches zero.
The observed $v_{1}$ starts from a negative value, then crosses zero
around $p_{\mathrm T} \sim$
1--2 GeV depending on collision energy and centrality. We see that, in more 
peripheral collisions and/or at higher energies, the sign change might occur at higher \pt~compared to more central collisions and lower energies.
The values of the charged particle $v_{1}(p_T)$ for three centralities (0-10\%, 10--40\% and 40-80\%) in Au+Au Au+Au collisions at \snn~= 14.5 GeV are listed in Table~\ref{table:v1pt_data}. 

Figure~\ref{fig:v1-eta-pt-snn-2} presents the charged particle $v_{1}$ as
a function of $\eta$ for three centrality classes in Au+Au
collisions at \snn~= 7.7 -- 39~GeV. The $v_{1}$ slope at midrapidity for charged 
particles increases from central to peripheral collisions. The trend
in $v_{1}(p_{\mathrm T}, \eta)$
as a function of centrality for \snn~= 7.7 -- 39~GeV shows a similar behavior as observed in other STAR 
published data~\cite{expdof1,expdof2,expdof3,expdof4,bes_paper,prc,200gevprl}. 
The values of the charged particle $v_{1}$ as a function of $\eta$ for three centrality classes in Au+Au Au+Au collisions at $\sqrt{s_{NN}} =$ 14.5 GeV are listed in Table~\ref{table:v1eta_data}. 

\vspace{0.2in}

\subsubsection{Beam-energy dependence of $v_{1}$}
The left panel in Fig.~\ref{fig:v1-eta-pt-snn-lhc} shows a comparison of $v_{1}(\eta)$ at 30--60\% centrality for Au+Au collisions at
\snn~= 7.7 -- 200~GeV and for Pb+Pb collisions at \snn~=2.76 TeV. We
observe a clear energy dependence in the $v_{1}(\eta)$ for \snn~= 7.7 GeV - 2.76 TeV. To calculate the slope of $v_{1}$, we fit the data with a function $F1 \times y + F3 \times y^{3}$. The linear term in this function ($F1$) gives the $v_{1}$ slope ($dv_{1}/dy$).
 The right panel in Fig.~\ref{fig:v1-eta-pt-snn-lhc} shows the beam energy dependence
of $dv_{1}/dy$ for 0-10\%, 10--40\% and 40-80\%  centralities for Au+Au collisions at
\snn~= 7.7 -- 200~GeV. The $dv_{1}/dy$  for 30-60\%
centrality for the above energies are compared with the same from the
published data from Au+Au collisions at \snn~= 200 GeV at
RHIC and Pb+Pb collisions at \snn~= 2.76 TeV  at the LHC.
We observe a smooth increase in the magnitude of $dv_{1}/dy$ at mid-pseudorapidity with decreasing
beam energy for 30-60\% centrality.

\vspace{0.2in}

\subsection{Model Comparisons}

Measurements from STAR suggest that at 7.7 GeV and 11.5 GeV, particle
production is dominated by hadronic processes, whereas at energies
around 20 GeV and above, partonic degrees of freedom become more
important~\cite{star_v2_bes_prl, star_v2_bes_prc,
  star_phispectra_bes_prc, phi_v2_nasim, star_v2_200gev_prl}. The
\snn~= 14.5 GeV Au+Au collisions analyzed here thus lie in
a transition region of great interest. Various bulk properties of the
system like $\left< p_{\mathrm T} \right>$, $dN/dy$, particle ratios, elliptic flow $v_{2}$, and directed flow $v_{1}$ measured in Au+Au collisions at \snn~= 14.5 GeV are compared with calculations from AMPT (version 2.25t7d)~\cite{ampt} and UrQMD (version 3.3p1)~\cite{urqmd}. The initial parameter settings for the models follow the recommendations in the cited papers. The UrQMD model treats only hadronic interactions whereas AMPT has two versions --- a string melting version (denoted AMPT-SM) which allows for both partonic and hadronic interactions among the particles, while the default version of AMPT treats only hadronic interactions. 
Recently, there have been studies with the AMPT-SM model to explain the particle multiplicity and flow measurements at RHIC and LHC 
using different values of the parton cascade scattering cross-section such as 1.5 mb and 10 mb. It was found that a 1.5 mb scattering cross-section gives a better description of data at these energies~\cite{Xu:2011fi,Ye:2017ooc}. We have generated AMPT-SM events with two different partonic cross sections (1.5 mb and 10 mb), denoted as AMPT 1.5mb and AMPT 10mb. The larger the partonic cross section, the later the hadronic cascade begins. 

\vspace{0.1in}

\subsubsection{\mdseries{\textit{Mean transverse momentum}}}

\begin{figure*}[!tp]
\centering
\resizebox{0.92 \textwidth}{!}{%
\includegraphics{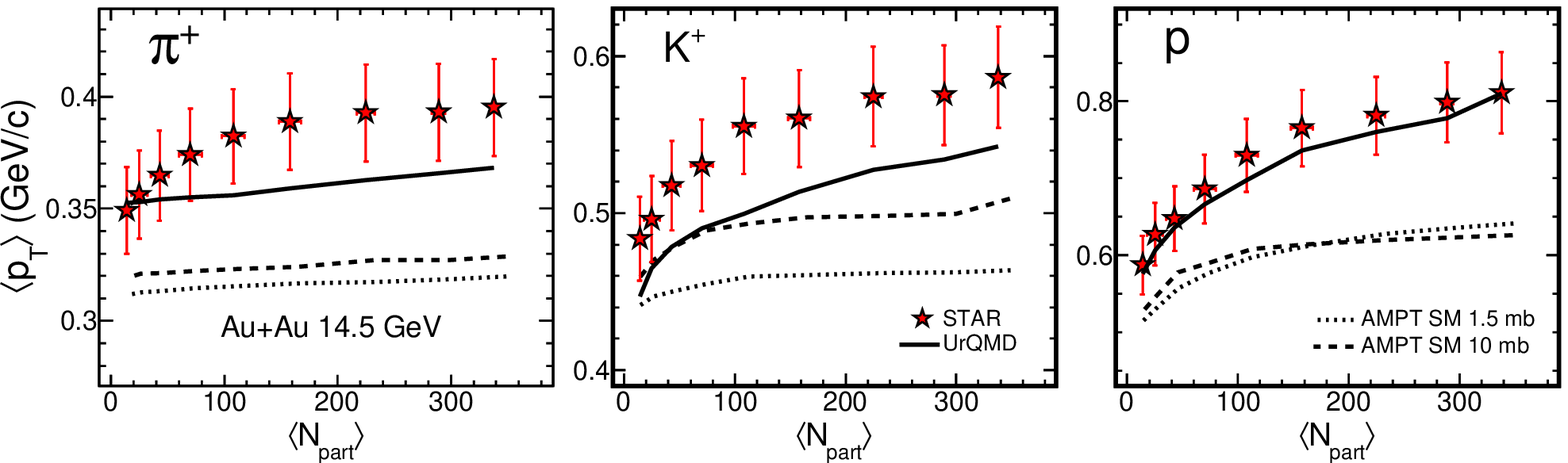} 
}
\caption{$\left< p_{\mathrm T}\right>$ of $\pi^{+}$, $K^{+}$ and $p$ as a function of $\left< N_{\text{part}} \right>$ for Au+Au collisions at \snn~= 14.5 GeV in STAR. These measurements are compared with UrQMD, AMPT 1.5mb and AMPT 10mb.}
\label{fig:pT-model}
\end{figure*}

\begin{figure*}[!tp]
\centering
\resizebox{0.92 \textwidth}{!}{%
\includegraphics{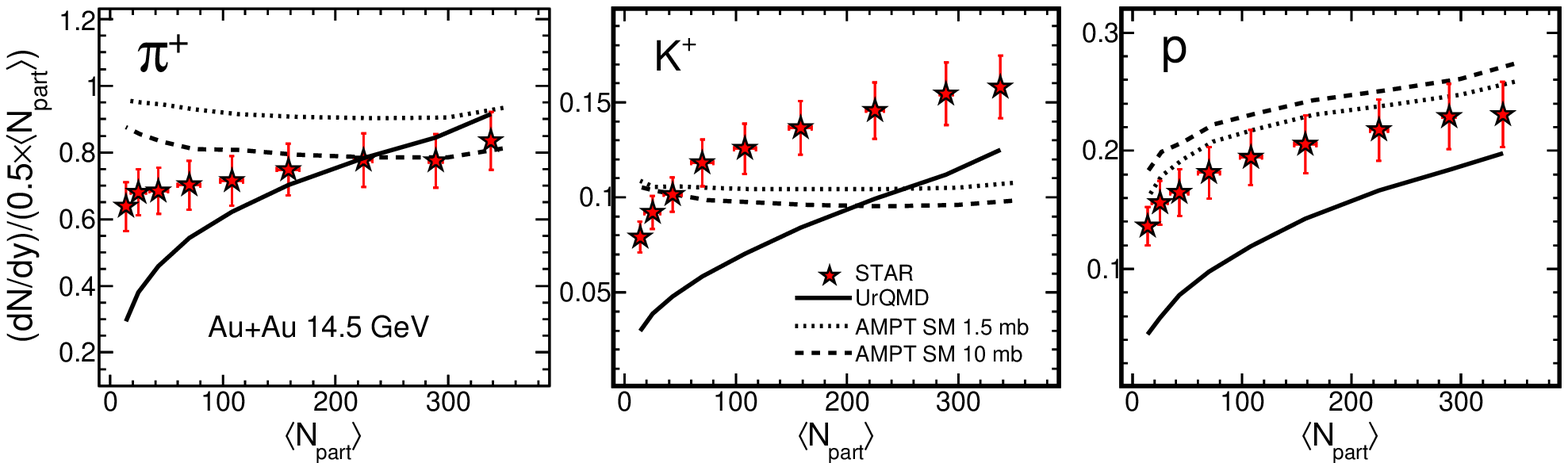}
}
\caption{$(dN/dy)/(0.5\times \left< N_{\text{part}}\right>)$ for $\pi^{+}$, $K^{+}$ and $p$ as a function of $\left< N_{\text{part}} \right>$ in Au+Au collisions at \snn~= 14.5 GeV in STAR. These measurements are compared with UrQMD, AMPT 1.5mb and AMPT 10mb.}
\label{fig:dndy-model}
\end{figure*}

The average \pt~of $\pi^{+}$, $K^{+}$ and $p$ as a function of $\left<
  N_{\text{part}} \right>$ obtained from UrQMD, AMPT 1.5mb and AMPT
10mb model calculations are compared with STAR measurements for Au+Au
collisions at \snn~= 14.5 GeV in
Fig.~\ref{fig:pT-model}. The value of $\left< p_{\mathrm T}\right>$
for all the studied particles is found to be too low in all AMPT-SM
calculations.  UrQMD is generally too low in $\left< p_{\mathrm T}\right>$ also, but is closer to the data, and shows good agreement for protons. 

\begin{figure*}[!tp]
\centering
\resizebox{0.92 \textwidth}{!}{%
\includegraphics{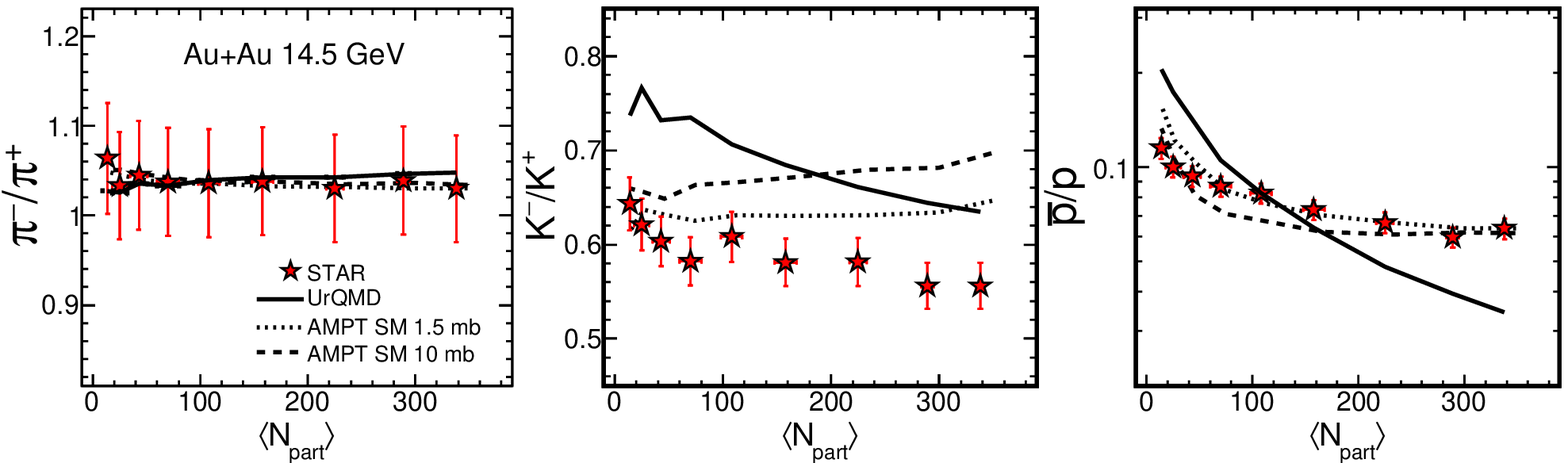}
}
\caption{$\pi^{-} / \pi^{+}$, $K^{-} / K^{+}$ and $\bar{p}/p$ ratios as a function of $\left< N_{\text{part}} \right>$ in Au+Au collisions at \snn~= 14.5 GeV in STAR. These experimental ratios are compared with UrQMD, AMPT 1.5mb and AMPT 10mb. }
\label{fig:like-ratios-model}
\end{figure*}

\begin{figure*}[!tp]
\centering
\resizebox{0.65 \textwidth}{!}{%
\includegraphics{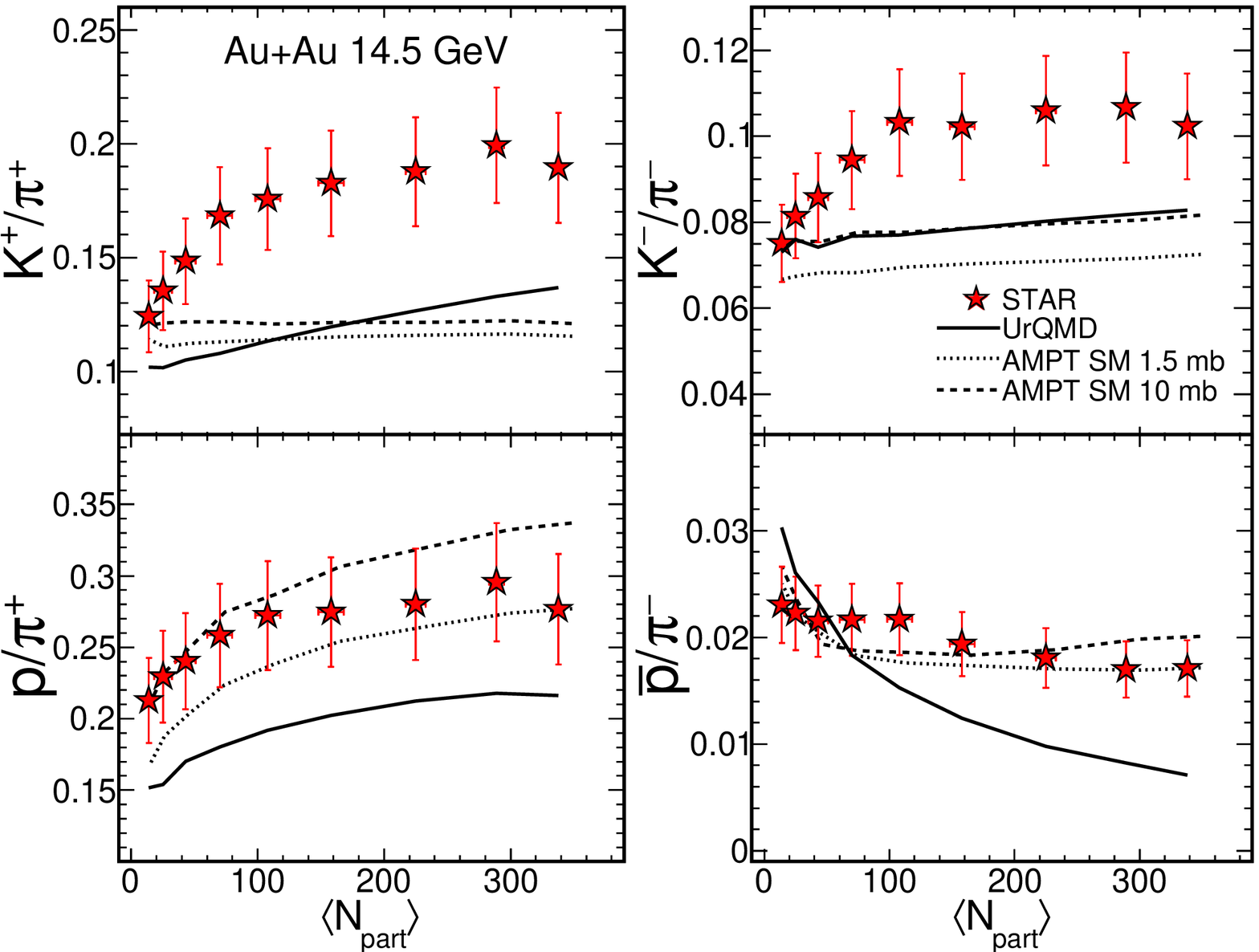}
}
\caption{$K^{+} / \pi^{+}$, $K^{-} / \pi^{-}$, $p/\pi^{+}$ and
  $\bar{p} / \pi^{-}$ ratios as a function of $\left< N_{\text{part}}
  \right>$ in Au+Au collisions at \snn~= 14.5 GeV in STAR. These experimental ratios are compared with UrQMD, AMPT 1.5mb and AMPT 10mb.}
\label{fig:mixed-ratios-model}
\end{figure*}

\vspace{0.1in}

\subsubsection{\mdseries{\textit{Particle yields}}} 

Figure~\ref{fig:dndy-model} shows $dN/dy$ divided by $0.5\times 
\left< N_{\text{part}}\right>$ versus $\left< N_{\text{part}}\right>$ 
for $\pi^{+}$, $K^{+}$ and $p$ from Au+Au collisions at \snn~= 14.5 GeV.  The STAR measurements are compared with UrQMD and 
with AMPT 1.5mb and AMPT 10mb.  UrQMD and AMPT are close to the 
$\pi^{+}$ data for central collisions, but deviate for peripheral collisions. 
All models disagree markedly with $K^{+}$ measurements.  In the case of
protons, AMPT-SM is close, with AMPT 1.5mb being slightly but consistently 
closer, while UrQMD lies well below the data at all centralities.  

\vspace{0.1in}

\subsubsection{\mdseries{\textit{Particle ratios}}}

Antiparticle to particle ratios ($\pi^{-}/\pi^{+}$,
$K^{-}/K^{+}$ and $\bar{p}/p$) as a function of $\left<
  N_{\text{part}} \right>$ in Au+Au collisions at \snn~= 14.5 GeV are shown in Fig.~\ref{fig:like-ratios-model}. These 
measured ratios are compared with UrQMD and AMPT-SM
calculations. The pion ratios from all models are in close agreement 
with experiment, while AMPT gets the wrong trends for the kaons. The proton ratios from 
AMPT-SM are in good agreement with experiment, while UrQMD 
shows poor agreement.  

Figure~\ref{fig:mixed-ratios-model} shows STAR measurements of $K^{+}/\pi^{+}$, $K^{-}/\pi^{-}$, $p/\pi^{+}$ and $\bar{p}/\pi^{-}$ ratios as a function of $\left< N_{\text{part}} \right>$ in Au+Au collisions at \snn~= 14.5 GeV, along with UrQMD and AMPT-SM model calculations.  $K^{+}/\pi^{+}$ and $K^{-}/\pi^{-}$ ratios are under-predicted by all model calculations. In the case of $p/\pi^{+}$, AMPT-SM straddles the data, and in the case of $\bar{p}/\pi^{-}$, AMPT-SM shows good agreement. On the other hand, the latter two ratios are not tracked by UrQMD. 

\vspace{0.1in}

\subsubsection{\mdseries{\textit{Elliptic flow}}}

\begin{figure*}[!htbp]
 \centering
 \resizebox{0.32 \textwidth}{!}{%
 \includegraphics{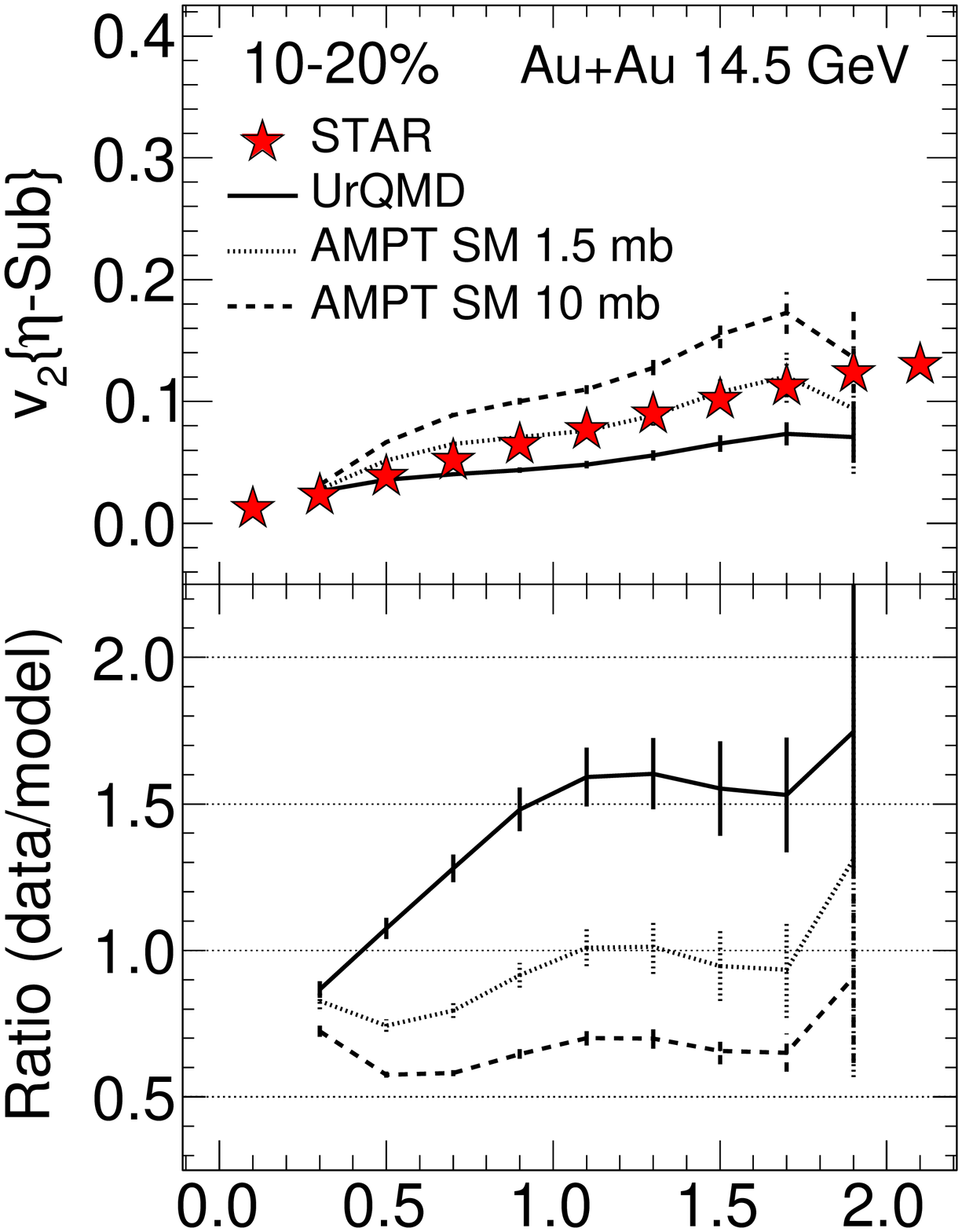}
 }
 \resizebox{0.32 \textwidth}{!}{%
 \includegraphics{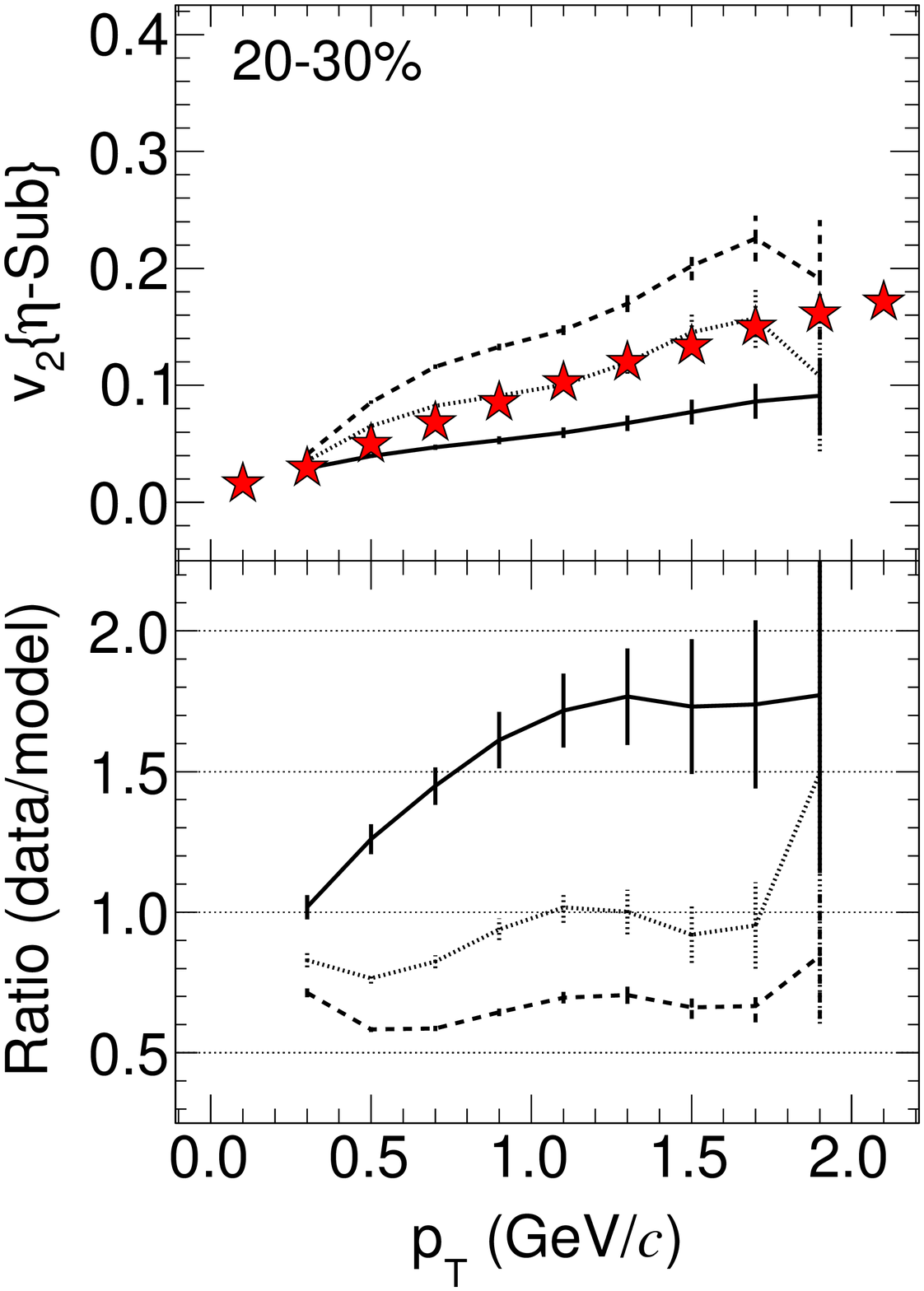}
 }
 \resizebox{0.32 \textwidth}{!}{%
 \includegraphics{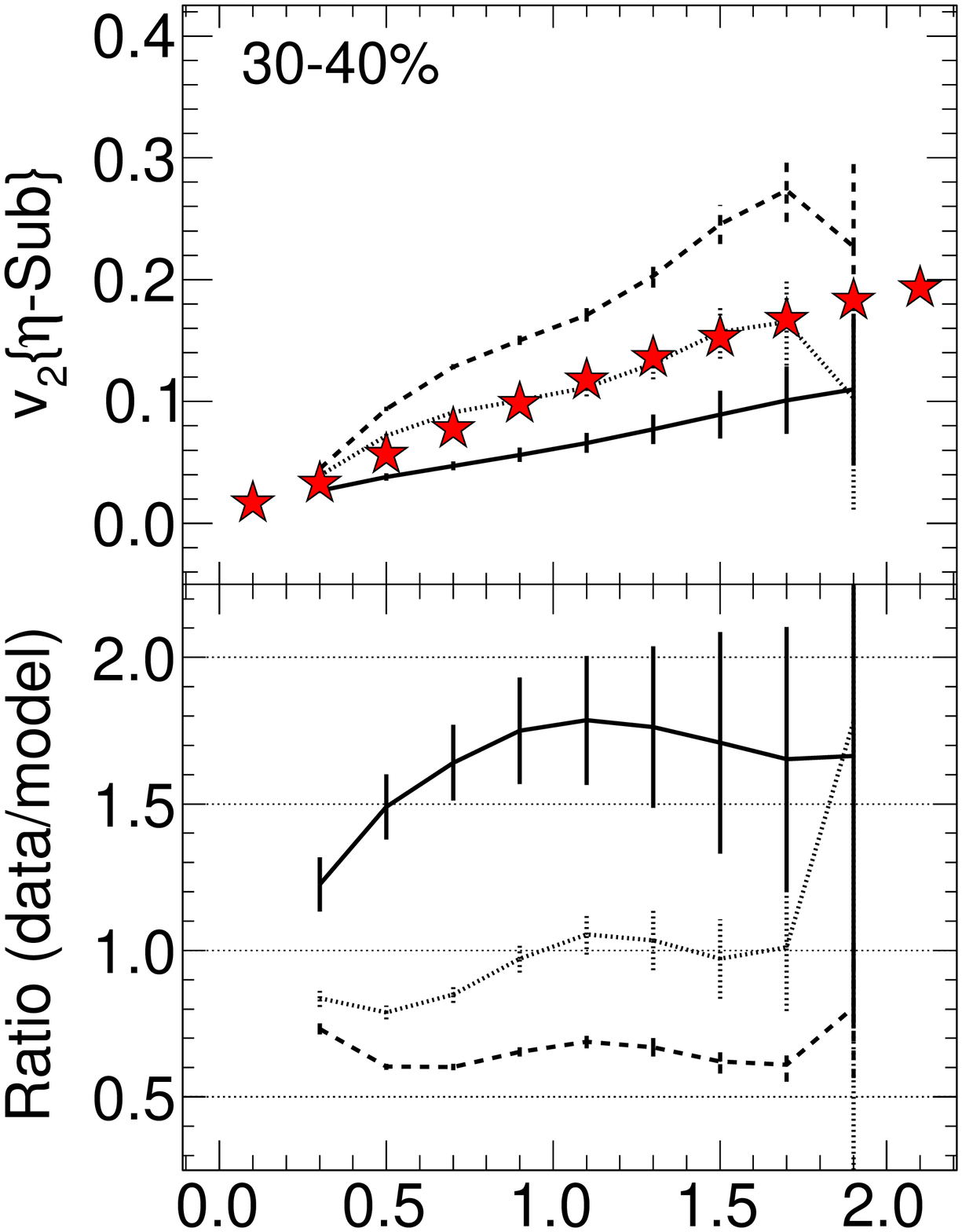}
 }
 \small
 \caption{\small{Upper panels: $p_T$ dependence of $v_2\{\eta{\rm-sub}\}$ for 14.5 GeV Au+Au collisions at 10--20\%, 20--30\% and 30--40\% centralities, as measured by STAR.  Calculations from UrQMD, AMPT 1.5mb, and AMPT 10mb are also plotted.  Lower panels: ratios of the experimental data to each model calculation.}}
 \label{fig:v2ep_pt_model}
\end{figure*}

\begin{figure*}[!htbp]
 \centering
 \resizebox{0.32 \textwidth}{!}{%
 \includegraphics{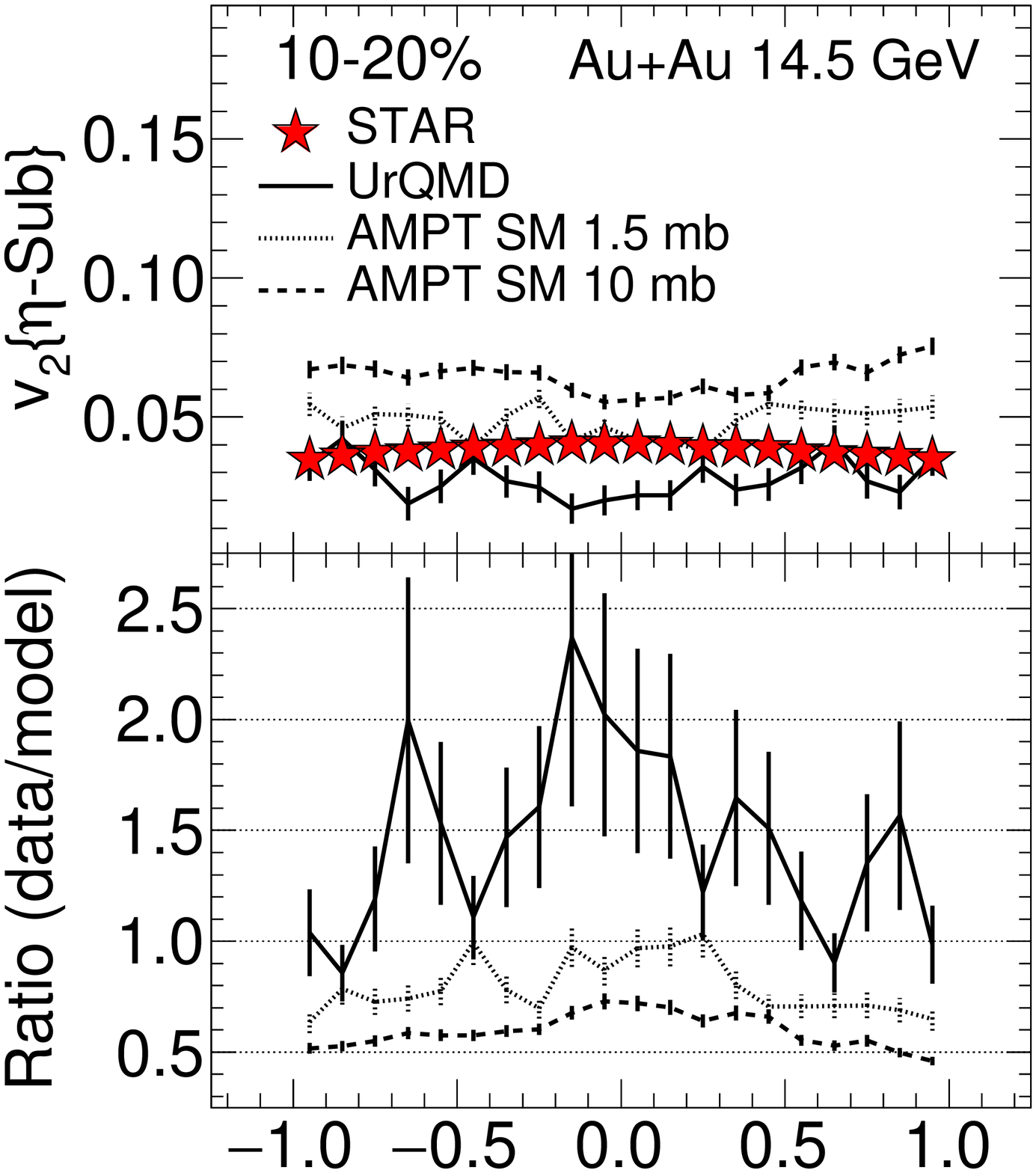}
 }
 \resizebox{0.32 \textwidth}{!}{%
 \includegraphics{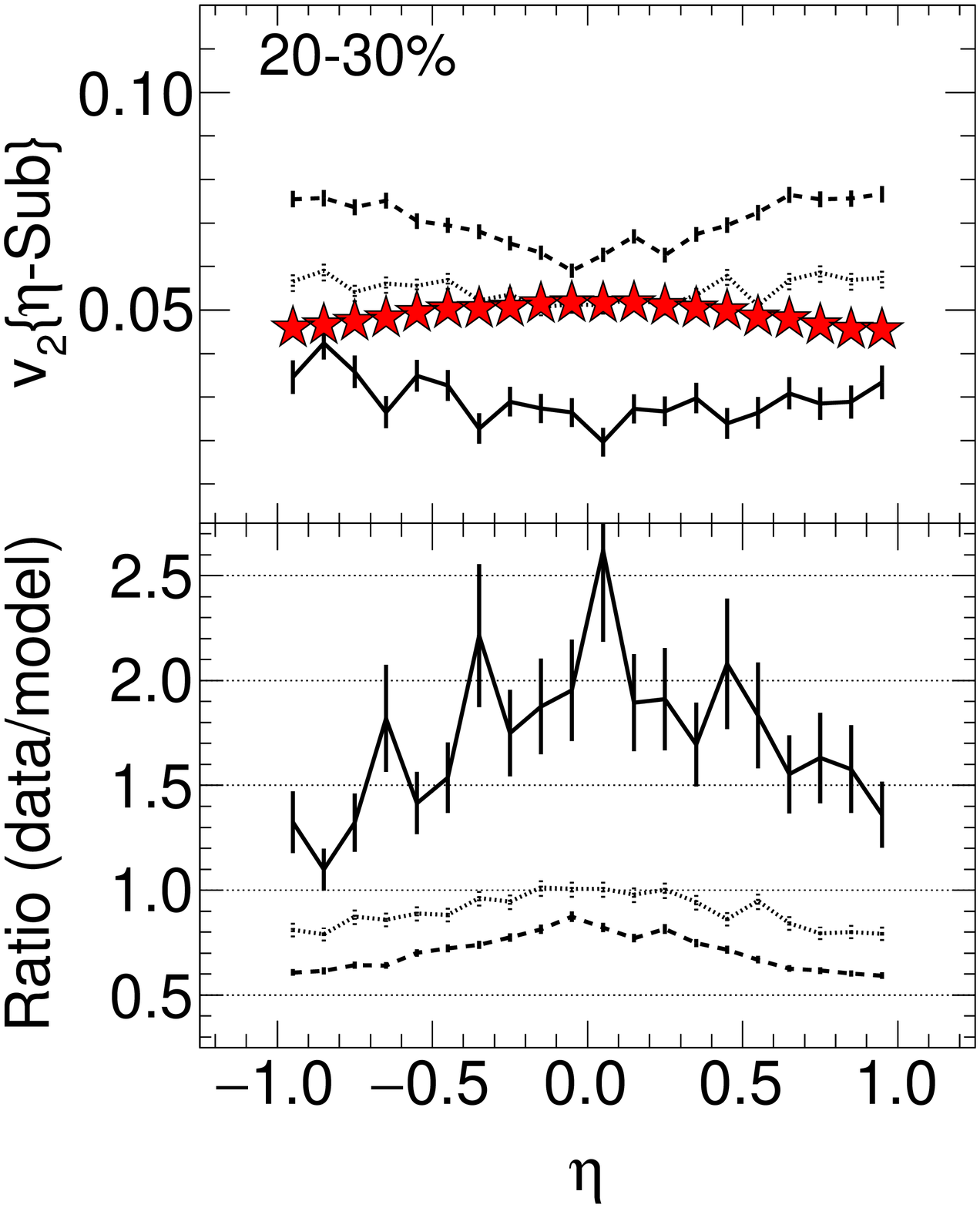}
 }
 \resizebox{0.32 \textwidth}{!}{%
 \includegraphics{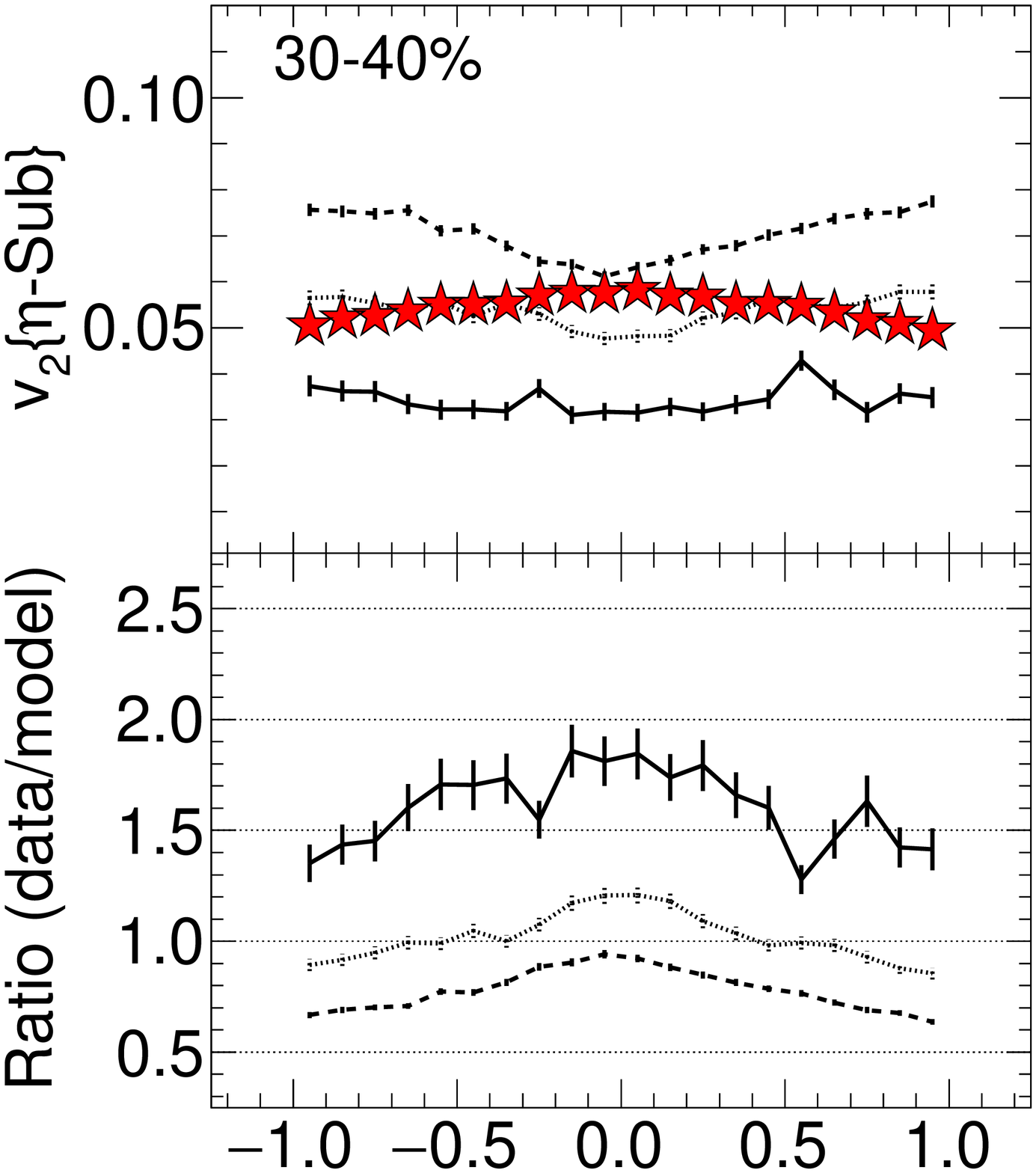}
 }
 \small
 \caption{\small{Upper panels: pseudorapidity ($\eta$) dependence of $v_2\{\eta{\rm-sub}\}$ for 14.5 GeV Au+Au collisions at 10--20\%, 20--30\% and 30--40\% centralities, as measured by STAR.  Calculations from UrQMD, AMPT 1.5mb, and AMPT 10mb are also plotted.  Lower panels: ratios of the experimental data to each model calculation.}}
 \label{fig:v2ep_eta_model}
\end{figure*}

The upper panels of Fig.~\ref{fig:v2ep_pt_model} present the $p_T$ dependence of $v_2\{\eta{\rm-sub}\}$ for 14.5 GeV Au+Au collisions at 10--20\%, 20--30\% and 30--40\% centralities. The STAR measurements are compared with UrQMD, AMPT 1.5mb, and AMPT 10mb.  The lower panels of Fig.~\ref{fig:v2ep_pt_model} present the ratio of the experimental data to each model calculation.  The AMPT 1.5mb calculation exhibits the best agreement, with AMPT 10mb being consistently too high and UrQMD consistently too low. 
Figure~\ref{fig:v2ep_eta_model} presents very similar comparisons as Fig.~\ref{fig:v2ep_pt_model}, except 
transverse momentum dependence is replaced by pseudorapidity dependence.  Here, again we observe similar behavior, i.e. the AMPT 1.5mb calculation exhibits the better agreement.

\vspace{0.4cm}

\vspace{0.1in}

  \subsubsection{\mdseries{\textit{Directed flow}}}

\begin{figure*}[!htbp]
\centering
\resizebox{0.6\textwidth}{!}{%
\includegraphics{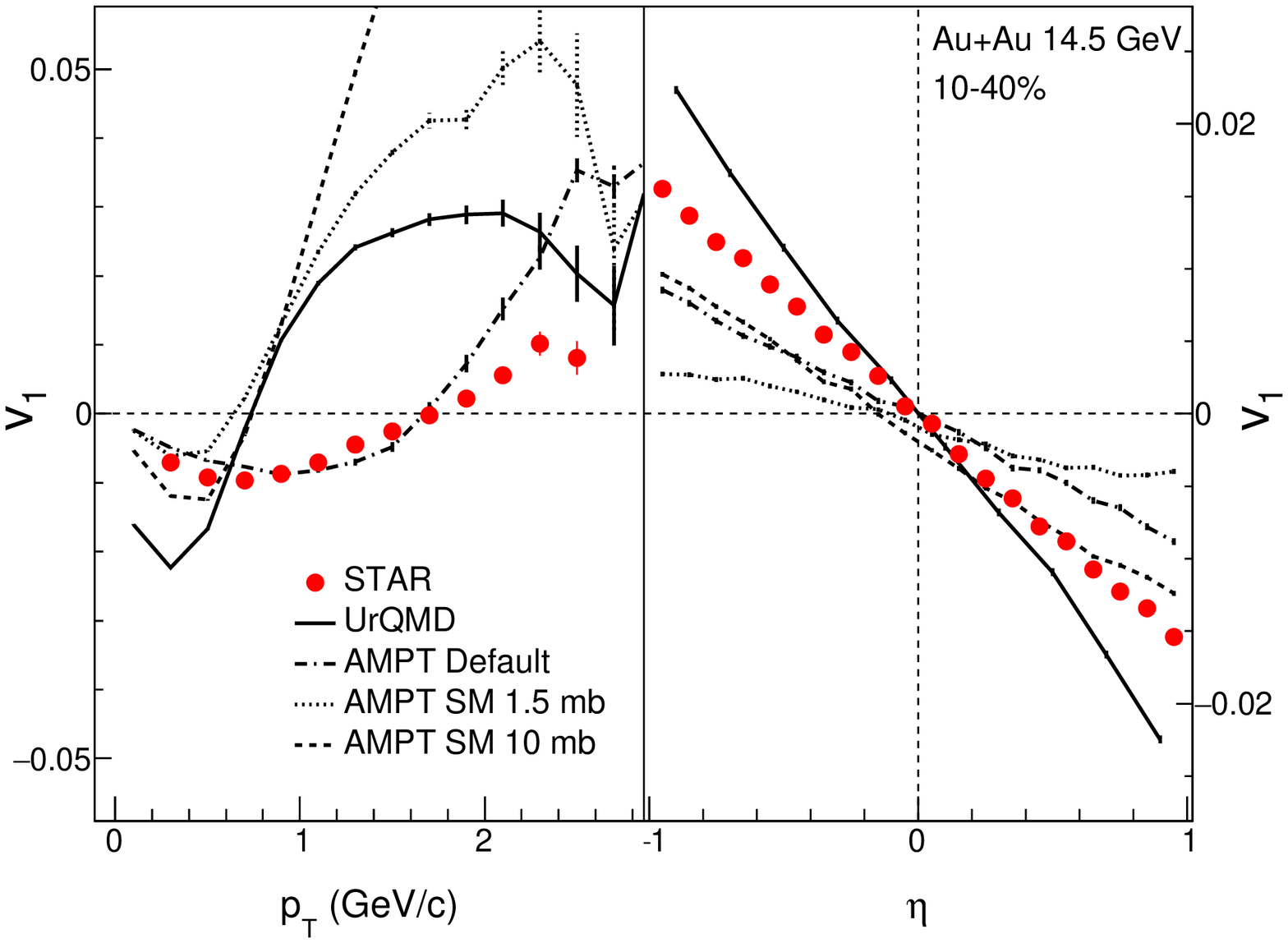}
}
\caption{Charged particle $v_1(p_T)$ (left panel) and $v_1(\eta)$ (right panel) for 
10--40\% centrality Au+Au collisions at \snn~= 14.5 GeV. The measured 
directed flow is compared to UrQMD, AMPT, AMPT 1.5mb, and AMPT 10mb model 
calculations.}
\label{fig:v1etapt_urqmd}
\end{figure*}

Figure~\ref{fig:v1etapt_urqmd} presents charged particle $v_1(p_T)$ (left 
panel) and $v_1(\eta)$ (right panel) for 10--40\% centrality Au+Au collisions 
at \snn~= 14.5 GeV.  These STAR measurements are compared 
to UrQMD, AMPT, AMPT 1.5mb, and AMPT 10mb model calculations.   

UrQMD shows poor agreement with the $v_1$ measurements, whereas the default 
AMPT roughly tracks $v_1(p_T)$ up to $p_T \sim 1.8$ GeV/$c$. AMPT 1.5mb and 
AMPT 10mb are both significantly worse than the default AMPT for $v_1(p_T)$.  No 
model calculation agrees with the measured data for $v_1(\eta)$. The latest state-of-the-art 
models do not show even qualitative agreement with $v_1$ measurements for identified
particles at BES-I energies~\cite{v1_review}.   It should be noted 
that in both AMPT-SM options, antibaryons violate $v_1 = 0$ at $y = 0$, as required 
by symmetry. This is a known artefact of the implementation of the quark coalescence 
mechanism in AMPT.  A recent paper has shown that this artefact can be fixed with a 
modified quark coalescence prescription~\cite{zun_xu}.  
\vspace{0.2in}

\section{SUMMARY}

We have presented basic observables for identified particles in Au+Au
collisions at \snn~= 14.5 GeV. The transverse momentum
spectra of $\pi$, $K$, $p$, and $\bar{p}$ at midrapidity ($|y| < 0.1$)
are measured for nine centralities: 0--5\%, 5--10\%, 10--20\%,
20--30\%, 30--40\%, 40--50\%, 50--60\%, 60--70\% and 70--80\%. Bulk
properties such as average transverse momentum $\left< p_{\mathrm T} \right>$, particle yields $dN/dy$, particle ratios, chemical and kinetic freeze-out properties, charged particle elliptic and directed flow ($v_2$ and $v_1$) are extracted for Au+Au collisions at \snn~= 14.5 GeV. All the results are compared with the published measurements at other collisions energies. 

The mean $\left< p_{\mathrm T} \right>$ values for $\pi$, $K$ and $p$ increase from peripheral to central collisions 
-- an indication of increasing radial flow in more central collisions. The centrality dependence of radial flow is more pronounced in kaons compared to pions, and in protons compared to kaons.

Midrapidity particle yields for $\pi$, $K$ and $p$ show a mild
centrality dependence, while no dependence on centrality is observed
for $\bar{p}$. The $(dN/dy)/(0.5\times \left< N_{\text{part}}\right>)$ for $\pi$, $K$ and $\bar{p}$ increase with collision energy, while for $p$ it decreases with collision energy up to 39 GeV and then increases. This effect is attributed to baryon stopping at lower RHIC energies.

No significant centrality dependence is observed in the case of $\pi^{-}/\pi^{+}$ and $K^{-}/K^{+}$ ratios. $\pi^{-}/\pi^{+}$ is slightly greater than unity, which is due to isospin conservation and the contribution from decays of resonances like the $\Delta$.  The $\bar{p}/p$ ratio slightly decreases from peripheral to central collisions as a consequence of increasing baryon stopping in central collisions. The $K^{+}/ \pi^{+}$ and $K^{-}/ \pi^{-}$ ratio increases with increasing centrality and follow the energy dependence trend established at other energies. The energy dependence is due to the dominance of pair production over associated production at higher energies. The $p/\pi^{+}$ ratio increases from peripheral to central collisions, but no significant dependence on centrality is observed in the case of $\bar{p}/\pi^{-}$ ratio.

Kinetic freeze-out parameters are obtained from simultaneous Blast-Wave model fits to the \pt~spectra for $\pi^{\pm}$, $K^{\pm}$ and $p\,(\bar{p})$. The kinetic freeze-out temperature $T_{k}$ decreases from peripheral to central collisions, which is suggestive of a short-lived fireball in peripheral collisions.
On the other hand, average flow velocity $\left< \beta \right>$ increases from peripheral to central collisions, indicating larger radial flow for central collisions, consistent with $\left< p_T \right>$ results.

The measured bulk observables are compared to UrQMD and AMPT model
calculations. Values of $\left< p_{\mathrm T}\right>$ 
are underestimated by both the UrQMD and AMPT models. 
The AMPT model agrees with the measured $dN/dy$ for pions ($\langle N_{\rm{part}} \rangle>100$) and protons, but does not reproduce kaon $dN/dy$. 
The UrQMD model mostly shows poor agreement with $dN/dy$ for all the measured particles. $\pi^{-}/\pi^{+}$  ratios are reproduced within uncertainties by both AMPT and UrQMD models. All models show poor agreement with STAR data of $K^{-}/K^{+}$ and $K/\pi$ ratios. The $\bar{p}/p$ ratio is well described by AMPT but not by UrQMD.  The measured $p/\pi^{+}$ and $\bar{p}/\pi^{-}$ ratios are poorly reproduced by UrQMD, while AMPT does better.  
The dependence  of $v_{2}$ of charged particles at midrapdity, on \pt~and $\eta$ is similar to that 
observed at other BES-I energies. The $v_{2}$ in peripheral
collisions is larger than in central collisions.  A clear centrality
dependence is observed in $v_{2}$. A weak dependence of the 
\pt-integrated charged particle $v_{2}$ on $\eta$ is observed. 
The shape of $v_{2}(\eta)$ at 14.5 GeV resembles that reported at 
other beam energies. 

The magnitude of charged particle $v_{1}$ increases from central to 
peripheral collisions at 14.5 GeV,
and a  
similar pattern is observed at other beam energies. The magnitude of
$v_1$ decreases with increasing beam energy. 

The UrQMD model underpredicts the STAR measurement of charged particle $v_{2}$ 
at 14.5 GeV.  
The AMPT string melting option with 10 mb parton-parton 
interaction cross section overpredicts the data, while the 1.5 mb option 
is closer. The UrQMD model shows poor agreement with both 
$v_1(p_T)$ and $v_1(\eta)$ at 14.5 GeV. 

The measured observables ($\left< p_{\mathrm T} \right>$, $dN/dy$,
particle ratios, chemical and kinetic freeze-out parameters, $v_{2}$
and $v_{1}$) in Au+Au collisions at \snn~= 14.5 GeV
conform to the smooth trend of beam energy dependence reported in 
prior publications. 
The results at 14.5 GeV 
are important since they fill the gap in $\mu_B$ of the order of about 100 MeV between beam energies 11.5 GeV and 19.6 GeV. The results will help in tuning the parameters of various models intended to explain the low energy data.  
Previous measurements by the STAR collaboration have revealed interesting trends related to the dominance of hadronic interaction and partonic interactions in observables such as higher moments of conserved quantities~\cite{Adamczyk:2013dal}, 
$v_1$~\cite{Adamczyk:2014ipa}, correlations~\cite{Adamczyk:2014mzf}, 
azimuthal anisotropy~\cite{star_v2_bes_prl}, and $R_{CP}$~\cite{Adamczyk:2017nof} 
between \snn~= 11.5 and 19.6 GeV. 
The data set at 14.5 GeV has provided a clearer understanding of the beam-energy dependence of bulk observables.

\vspace{0.2in}

\section{ACKNOWLEDGEMENTS}

We thank the RHIC Operations Group and RCF at BNL, the NERSC Center at LBNL, and the Open Science Grid consortium for providing resources and support.  This work was supported in part by the Office of Nuclear Physics within the U.S. DOE Office of Science, the U.S. National Science Foundation, the Ministry of Education and Science of the Russian Federation, National Natural Science Foundation of China, Chinese Academy of Science, the Ministry of Science and Technology of China and the Chinese Ministry of Education, the National Research Foundation of Korea, Czech Science Foundation and Ministry of Education, Youth and Sports of the Czech Republic, Hungarian National Research, Development and Innovation Office (FK-123824), New National Excellency Programme of the Hungarian Ministry of Human Capacities (UNKP-18-4), Department of Atomic Energy and Department of Science and Technology of the Government of India, the National Science Centre of Poland, the Ministry  of Science, Education and Sports of the Republic of Croatia, RosAtom of Russia and German Bundesministerium fur Bildung, Wissenschaft, Forschung and Technologie (BMBF) and the Helmholtz Association.



\newpage

\begin{table*}[!h]
\centering
\small
\caption{Summary of centrality bins, average number of participants $N_{\rm part}$, number of binary collisions $N_{\rm coll}$, reaction plane eccentricity $\epsilon_{\rm RP}$, participant eccentricity $\epsilon_{\rm part}$, root-mean-square of the participant eccentricity $\epsilon_{\rm part}\lbrace2\rbrace$, and transverse area $S_{\rm part}$ from MC Glauber simulations at  $\sqrt{s_{NN}} = $14.5 GeV. The errors are systematic uncertainties.}

\vspace{0.5cm}
\begin{tabular}{c|c|c|c|c|c|c} 
\hline
\rule{0pt}{15pt}
\textbf{Centrality (\%)}	& $\langle N_{\rm part}\rangle$	& $\langle N_{\rm coll}\rangle$   & $\langle\epsilon_{\rm RP}\rangle$  & $\langle\epsilon_{\rm part}\rangle$ & $\epsilon_{\rm part}\lbrace2\rbrace$ & $\langle S_{\rm part}\rangle$ 
\\
[5pt]
\hline
$0-5$	 	 &    338 $\pm$ 2	  	&	788 $\pm$ 30	&	 0.04	$\pm$ 0.01	 &		0.10 $\pm$ 0.01	&	0.12 $\pm$ 0.01	&	25.5 $\pm$ 0.6 \\
$5-10$	 &    289 $\pm$ 6	  	&	634 $\pm$ 20	&	 0.11	$\pm$ 0.01	 &		0.15 $\pm$ 0.01	&	0.16 $\pm$ 0.01	&	22.9 $\pm$ 0.7 \\
$10-20$	 &    226 $\pm$ 8	  	&	454 $\pm$ 24	&	 0.18	$\pm$ 0.01	 &		0.22 $\pm$ 0.01	&	0.24 $\pm$ 0.01	&	19.3 $\pm$ 0.8 \\
$20-30$	 &    159 $\pm$10		&	283 $\pm$ 24	&	 0.27	$\pm$ 0.01	 &		0.30 $\pm$ 0.01	&	0.32 $\pm$ 0.01	&	15.5 $\pm$ 0.9 \\
$30-40$	 &    108 $\pm$ 10	&	168 $\pm$ 22	&	 0.32	$\pm$ 0.01	 &		0.37 $\pm$ 0.01	&	0.40 $\pm$ 0.01	&	12.4 $\pm$ 1.0 \\
$40-50$	 &     70 $\pm$ 8  		&	  94 $\pm$ 18	&	 0.37	$\pm$ 0.01	 &		0.44 $\pm$ 0.01	&	0.47 $\pm$ 0.01	&	  9.8 $\pm$ 1.1 \\
$50-60$	 &     44 $\pm$ 8	  	&	  50 $\pm$ 12	&	 0.39	$\pm$ 0.01	 &		0.51 $\pm$ 0.01	&	0.54 $\pm$ 0.01	&	  7.6 $\pm$ 1.1 \\
$60-70$	 &     26 $\pm$ 7	  	&	  25 $\pm$   9	&	 0.40	$\pm$ 0.01	 &		0.59 $\pm$ 0.01	&	0.62 $\pm$ 0.01	&	  5.6 $\pm$ 1.2 \\
$70-80$	 &     14 $\pm$ 5	  	&	  12 $\pm$   5	&	 0.37	$\pm$ 0.01	 &		0.68 $\pm$ 0.01	&	0.72 $\pm$ 0.01	&	  3.5 $\pm$ 1.2 \\
\hline
\end{tabular}
\label{table:eccent}
\end{table*}

\begin{table*}[!h]
  \centering
  \small
	\caption{$dN/dy$ values for $\pi^{+}$, $\pi^{-}$, $K^{+}$, $K^{-}$, $p$, and $\bar{p}$ from Au+Au collisions at 
	$\sqrt{s_{NN}} = 14.5$ GeV. The uncertainties represent statistical and systematic uncertainties, respectively.}
	\label{tab:dndy}\vspace{0.1in}
	\begin{tabular}{c|c|c|c|c|c|c}
	\hline	
	\rule{0pt}{15pt}
	\textbf{Centra-} &  &  &  &  &  &   \\
	\textbf{lity($\%$)} & $\pi^{+}$ & $\pi^{-}$ & $K^{+}$ & $K^{-}$ & $p$ & $\bar{p}$  \\
	[5pt]
	\hline
        $0-5$ & 141 $\pm$ 0.2 $\pm$ 14 & 145 $\pm$ 0.2 $\pm$ 15   & 26.7 $\pm$ 0.04 $\pm$ 2.8  & 14.9 $\pm$ 0.03 $\pm$ 1.6  &   39.0 $pm$ 0.03 $\pm$ 4.7  & 2.5 $\pm$ 0.02 $\pm$ 0.3 \\  [4pt]         
	$5-10$ & 112  $\pm$ 0.2 $\pm$ 11  & 116  $\pm$ 0.1 $\pm$ 12  &  22.3 $\pm$ 0.04 $\pm$ 2.4  & 12.4 $\pm$ 0.03 $\pm$ 1.2   &     33.1 $\pm$ 0.03 $\pm$ 4.0  & 2.0 $\pm$ 0.01 $\pm$ 0.2  \\  [4pt] 
	$10-20$ & 87.3  $\pm$ 0.1 $\pm$ 9.1  & 90.0  $\pm$ 0.1 $\pm$ 9.2   &16.4 $\pm$ 0.03 $\pm$ 1.7  & 9.5 $\pm$ 0.02 $\pm$ 1.0   &  24.5 $\pm$ 0.03 $\pm$ 2.9  & 1.6 $\pm$ 0.01 $\pm$ 0.2  \\ [4pt] 
	$20-30$ & 59.1  $\pm$ 0.1 $\pm$ 6.1  & 61.3 $\pm$ 0.1 $\pm$ 6.4   & 10.8 $\pm$ 0.02 $\pm$ 1.1  & 6.3 $\pm$ 0.02 $\pm$ 0.6   & 16.2 $\pm$ 0.02 $\pm$ 2.0  & 1.2 $\pm$ 0.01 $\pm$ 0.1  \\  [4pt] 
	$30-40$ & 38.6 $\pm$ 0.1 $\pm$ 4.0  & 40.0 $\pm$ 0.1 $\pm$ 4.1   &  6.8 $\pm$ 0.02 $\pm$ 0.7  & 4.1 $\pm$ 0.01 $\pm$ 0.4   &  10.5 $\pm$ 0.02 $\pm$ 1.3  & 0.9 $\pm$ 0.01 $\pm$ 0.1  \\ [4pt] 
	$40-50$ & 24.6 $\pm$ 0.1 $\pm$ 2.6  & 25.5 $\pm$ 0.1 $\pm$ 2.6   & 4.1 $\pm$  0.01  $\pm$  0.4 & 2.4 $\pm$ 0.01 $\pm$ 0.3 & 6.3 $\pm$ 0.01 $\pm$ 0.8 & 0.60 $\pm$  0.004  $\pm$  0.07  \\ [4pt] 
	$50-60$ & 14.7 $\pm$ 0.04 $\pm$ 1.5 & 15.4 $\pm$ 0.04  $\pm$ 1.5 & 2.2 $\pm$ 0.01 $\pm$  0.2 & 1.3 $\pm$  0.01 $\pm$ 0.1 & 3.5 $\pm$  0.01  $\pm$  0.4 & 0.33 $\pm$ 0.003 $\pm$ 0.04  \\[4pt] 
	$60-70$  & 8.5  $\pm$ 0.03 $\pm$ 0.9 & 8.8 $\pm$ 0.03 $\pm$ 0.9 & 1.2 $\pm$ 0.01 $\pm$ 0.1 & 0.72 $\pm$ 0.01 $\pm$ 0.07 & 1.9 $\pm$ 0.01 $\pm$ 0.2 & 0.16 $\pm$ 0.003 $\pm$ 0.02  \\[4pt] 
	$70-80$  & 4.5  $\pm$ 0.03 $\pm$ 0.5 & 4.8 $\pm$ 0.03 $\pm$ 0.6 & 0.55 $\pm$ 0.01  $\pm$ 0.06 & 0.36 $\pm$ 0.01 $\pm$ 0.04 & 0.6 $\pm$ 0.01 $\pm$ 0.1 & 0.11 $\pm $ 0.002 $\pm$ 0.01  \\
	\hline
	\end{tabular}
\end{table*}


\begin{table*}[!h]
  \centering
  \small
	\caption{The $\left< p_{\mathrm T} \right>$ (MeV/$c$) values for $\pi^{+}$, $\pi^{-}$, $K^{+}$, $K^{-}$, $p$, and $\bar{p}$ from Au+Au collisions 
	at $\sqrt{s_{NN}} = 14.5$ GeV. The uncertainties represent statistical and systematic uncertainties, respectively.}
	\label{tab:mean-pT}\vspace{0.1in}
	\begin{tabular}{c|c|c|c|c|c|c}
	\hline	
	\rule{0pt}{15pt}
	\textbf{Centra-} &  &  &  &  &  &   \\ 
         \textbf{lity($\%$)} & $\pi^{+}$ & $\pi^{-}$ & $K^{+}$ & $K^{-}$ & $p$ & $\bar{p}$  \\ [5pt]
	\hline
	$0-5$ & 395 $\pm$  0.1 $\pm$ 22 & 392 $\pm$  0.1 $\pm$ 21   & 586 $\pm$  0.1 $\pm$ 32 & 560 $\pm$  0.1  $\pm$ 30 & 811 $\pm$ 0.1$\pm$ 53 & 807 $\pm$ 0.2$\pm$ 69 \\[3pt]
	$5-10$ & 393 $\pm$  0.1 $\pm$ 22 & 390 $\pm$  0.1 $\pm$ 21 & 575 $\pm$  0.1  $\pm$ 32 & 559 $\pm$  0.1  $\pm$ 30   & 798 $\pm$  0.1$\pm$ 52 & 800 $\pm$  0.2$\pm$ 68 \\[3pt] 
	$10-20$ & 393 $\pm$  0.2 $\pm$ 22 & 388 $\pm$  0.2 $\pm$ 21&  574 $\pm$  0.1  $\pm$ 32 & 552 $\pm$  0.1  $\pm$ 30   & 781 $\pm$  0.1$\pm$ 51 & 776 $\pm$  0.2$\pm$ 66 \\[3pt]  
	$20-30$ & 389 $\pm$  0.2 $\pm$ 21 & 384 $\pm$  0.2 $\pm$ 20&  560 $\pm$  0.2  $\pm$ 31 & 540 $\pm$  0.1  $\pm$ 29    & 765 $\pm$  0.1 $\pm$ 50 & 751 $\pm$  0.3$\pm$ 64 \\[3pt] 
	$30-40$ & 382 $\pm$  0.2 $\pm$ 21 & 380 $\pm$  0.2 $\pm$ 20&  555 $\pm$  0.2  $\pm$ 31 & 532 $\pm$  0.1  $\pm$ 29    & 730  $\pm$  0.1$\pm$ 48 & 726 $\pm$  0.3$\pm$ 62 \\[3pt] 
	$40-50$ & 374 $\pm$  0.3 $\pm$ 21  & 374 $\pm$  0.3 $\pm$ 20&   530 $\pm$  0.2$\pm$ 29 & 513 $\pm$  0.1$\pm$ 28   & 686 $\pm$  0.1$\pm$ 45 & 675 $\pm$  0.4$\pm$ 57\\[3pt]
	$50-60$ & 365 $\pm$  0.3 $\pm$ 20 & 366 $\pm$  0.3 $\pm$ 19        &      518 $\pm$  0.3 $\pm$ 29 & 495 $\pm$  0.2 $\pm$ 27  & 646 $\pm$  0.1 $\pm$ 42 & 640 $\pm$  0.4$\pm$ 54 \\[3pt] 
	$60-70$ & 356 $\pm$  0.4 $\pm$ 20 & 355 $\pm$  0.4 $\pm$ 19        &      496 $\pm$  0.4 $\pm$ 27 & 475 $\pm$  0.2 $\pm$ 26   & 627 $\pm$  0.1 $\pm$ 40 & 606 $\pm$  0.1 $\pm$ 52  \\[3pt]
	$70-80$ & 349 $\pm$  1 $\pm$ 19 & 347 $\pm$  1 $\pm$ 18        &  484 $\pm$  1 $\pm$ 27 & 460 $\pm$  1 $\pm$ 25   & 587 $\pm$  2 $\pm$ 38 & 573 $\pm$ 1 $\pm$ 49  \\
	\hline
	\end{tabular}
\end{table*}


\begin{table*}[!h] 
  \centering
  \small
	\caption{Kinetic freeze-out parameters $T_{k}$, $\left< \beta \right>$, $n$ and $\chi^{2}/$ndf values from Blast-Wave fits in Au+Au collisions at $\sqrt{s_{NN}} = 14.5$ GeV. The quoted errors are total statistical and systematic uncertainties added in quadrature.}
	\label{tab:bw-para}\vspace{0.1in}
	\setlength{\tabcolsep}{10pt}
	\begin{tabular}{c|c|c|c|c}
	\hline	
	\rule{0pt}{15pt}
	\textbf{Centrality($\%$)} & $T_{k}$ (MeV) & $\left<\beta \right>$ & $n$ & $\chi^{2}/$ndf  \\
	[5pt]
	\hline
	$0-5$ & 114 $\pm$ 7 & 0.485 $\pm$ 0.036 & 0.97 $\pm$ 0.28 & 0.119 \\
	$5-10$ & 116 $\pm$ 7 & 0.442 $\pm$ 0.035 & 0.98 $\pm$ 0.29 & 0.097  \\
	$10-20$ & 118 $\pm$ 7 & 0.429 $\pm$ 0.034 & 0.99 $\pm$ 0.33 & 0.119  \\
	$20-30$ & 122 $\pm$ 7 & 0.401 $\pm$ 0.034 & 1.00 $\pm$ 0.39 & 0.056  \\
	$30-40$ & 124 $\pm$ 8 & 0.371 $\pm$ 0.042 & 1.34 $\pm$ 0.42 & 0.123  \\
	$40-50$ & 130 $\pm$ 6 & 0.312 $\pm$ 0.036 & 1.73 $\pm$ 0.61 & 0.232  \\
	$50-60$ & 134 $\pm$ 6 & 0.238 $\pm$ 0.031 & 2.26 $\pm$ 0.78 & 0.398  \\
	$60-70$ & 136 $\pm$ 6 & 0.194 $\pm$ 0.030 & 2.76 $\pm$ 0.87 & 0.484  \\
	$70-80$ & 139 $\pm$ 7 & 0.168 $\pm$ 0.030 & 2.83 $\pm$ 1.20 & 0.354  \\
	\hline
	\end{tabular}
\end{table*}

\begin{table*}[!h]
  \centering
  \small
  \caption{Inclusive charged particle $v_{1}$ as function of
    $p_{\mathrm T}$
    in Au+Au collisions at $\sqrt{s_{NN}}$ = 14.5~GeV. The uncertainties represent statistical and systematic uncertainties, respectively.}
  \vspace{0.2cm}
  \begin{tabular}{c|c|c|c}
    \hline
      $p_{\mathrm T}$ (GeV/$c$)   &  $v_{1}$( 0--10\%) & $v_{1}$ (10--40\%) & $v_{1}$ (40--80\%)	\\ 	
    \hline
  0.30 & -0.0027 $\pm$ 0.0003 $\pm$ 0.0001 &  -0.0071 $\pm$ 0.0001 $\pm$ 0.0004 & -0.0130 $\pm$ 0.0002 $\pm$ 0.0007 \\ 
 0.50 & -0.0034 $\pm$ 0.0003 $\pm$ 0.0002 &  -0.0093 $\pm$ 0.0001 $\pm$ 0.0005 & -0.0176 $\pm$ 0.0002 $\pm$ 0.0008 \\ 
 0.70 & -0.0032 $\pm$ 0.0004 $\pm$ 0.0002 &  -0.0096 $\pm$ 0.0001 $\pm$ 0.0004 & -0.0201 $\pm$ 0.0003 $\pm$ 0.0009 \\ 
 0.90 & -0.0024 $\pm$ 0.0006 $\pm$ 0.0003 &  -0.0087 $\pm$ 0.0002 $\pm$ 0.0003 & -0.0216 $\pm$ 0.0004 $\pm$ 0.0010 \\ 
 1.10 & 0.0014 $\pm$ 0.0008 $\pm$ 0.0005 &  -0.0070 $\pm$ 0.0003 $\pm$ 0.0004 & -0.0219 $\pm$ 0.0006 $\pm$ 0.0007 \\ 
 1.30 & 0.0019 $\pm$ 0.0010 $\pm$ 0.0005 &  -0.0045 $\pm$ 0.0003 $\pm$ 0.0001 & -0.0226 $\pm$ 0.0008 $\pm$ 0.0005 \\ 
 1.50 & 0.0047 $\pm$ 0.0014 $\pm$ 0.0007 &  -0.0025 $\pm$ 0.0005 $\pm$ 0.0003 & -0.0209 $\pm$ 0.0011 $\pm$ 0.0009 \\ 
 1.70 & 0.0040 $\pm$ 0.0019 $\pm$ 0.0007 &  -0.0002 $\pm$ 0.0006 $\pm$ 0.0003 & -0.0185 $\pm$ 0.0016 $\pm$ 0.0008 \\ 
 1.90 & 0.0079 $\pm$ 0.0026 $\pm$ 0.0007 &  0.0024 $\pm$ 0.0009 $\pm$ 0.0003 & -0.0179 $\pm$ 0.0023 $\pm$ 0.0008 \\ 
 2.10 & 0.0095 $\pm$ 0.0036 $\pm$ 0.0014 &  0.0059 $\pm$ 0.0012 $\pm$ 0.0004 & -0.0173 $\pm$ 0.0032 $\pm$ 0.0013 \\ 
 2.30 & 0.0155 $\pm$ 0.0050 $\pm$ 0.0021 &  0.0097 $\pm$ 0.0017 $\pm$ 0.0007 & -0.0125 $\pm$ 0.0046 $\pm$ 0.0015 \\ 
 2.50 & 0.0089 $\pm$ 0.0069 $\pm$ 0.0024 &  0.0081 $\pm$ 0.0024 $\pm$ 0.0007 & -0.0067 $\pm$ 0.0065 $\pm$ 0.0024 \\ 
\hline
  \end{tabular}
  \label{table:v1pt_data}
\end{table*}

\begin{table*}[!h]
  \centering
  \small
  \caption{Inclusive charged particle $v_{1}$ as function of $\eta$
    in Au+Au collisions at $\sqrt{s_{NN}}$ = 14.5~GeV. The uncertainties represent statistical and systematic uncertainties, respectively.}
  \vspace{0.2cm}
  \begin{tabular}{c|c|c|c}
    \hline
  \textbf{$\eta$}    &  $v_{1}$( 0--10\%) & $v_{1}$ (10--40\%) & $v_{1}$ (40--80\%)        \\
     \hline
-0.95 & 0.0059 $\pm$ 0.0008 $\pm$ 0.0007 &  0.0157 $\pm$ 0.0003 $\pm$ 0.0005 & 0.0334 $\pm$ 0.0006 $\pm$ 0.0000 \\ 
 -0.85 & 0.0048 $\pm$ 0.0008 $\pm$ 0.0006 &  0.0137 $\pm$ 0.0003 $\pm$ 0.0003 & 0.0295 $\pm$ 0.0005 $\pm$ 0.0000 \\ 
 -0.75 & 0.0034 $\pm$ 0.0008 $\pm$ 0.0003 &  0.0119 $\pm$ 0.0002 $\pm$ 0.0001 & 0.0258 $\pm$ 0.0005 $\pm$ 0.0000 \\ 
 -0.65 & 0.0030 $\pm$ 0.0007 $\pm$ 0.0003 &  0.0108 $\pm$ 0.0002 $\pm$ 0.0001 & 0.0211 $\pm$ 0.0005 $\pm$ 0.0000 \\ 
 -0.55 & 0.0019 $\pm$ 0.0007 $\pm$ 0.0003 &  0.0089 $\pm$ 0.0002 $\pm$ 0.0001 & 0.0186 $\pm$ 0.0005 $\pm$ 0.0000 \\ 
 -0.45 & 0.0015 $\pm$ 0.0007 $\pm$ 0.0003 &  0.0073 $\pm$ 0.0002 $\pm$ 0.0001 & 0.0137 $\pm$ 0.0005 $\pm$ 0.0000 \\ 
 -0.35 & 0.0032 $\pm$ 0.0007 $\pm$ 0.0003 &  0.0054 $\pm$ 0.0002 $\pm$ 0.0000 & 0.0111 $\pm$ 0.0005 $\pm$ 0.0002 \\ 
 -0.25 & 0.0015 $\pm$ 0.0007 $\pm$ 0.0003 &  0.0043 $\pm$ 0.0002 $\pm$ 0.0001 & 0.0083 $\pm$ 0.0005 $\pm$ 0.0001 \\ 
 -0.15 & 0.0015 $\pm$ 0.0007 $\pm$ 0.0006 &  0.0026 $\pm$ 0.0002 $\pm$ 0.0001 & 0.0049 $\pm$ 0.0005 $\pm$ 0.0001 \\ 
 -0.05 & 0.0007 $\pm$ 0.0007 $\pm$ 0.0004 &  0.0005 $\pm$ 0.0002 $\pm$ 0.0001 & 0.0013 $\pm$ 0.0005 $\pm$ 0.0001 \\ 
 0.05 & -0.0004 $\pm$ 0.0007 $\pm$ 0.0003 &  -0.0007 $\pm$ 0.0002 $\pm$ 0.0001 & -0.0025 $\pm$ 0.0005 $\pm$ 0.0002 \\ 
 0.15 & -0.0008 $\pm$ 0.0007 $\pm$ 0.0003 &  -0.0028 $\pm$ 0.0002 $\pm$ 0.0001 & -0.0051 $\pm$ 0.0005 $\pm$ 0.0001 \\ 
 0.25 & -0.0013 $\pm$ 0.0007 $\pm$ 0.0004 &  -0.0045 $\pm$ 0.0002 $\pm$ 0.0001 & -0.0072 $\pm$ 0.0005 $\pm$ 0.0001 \\ 
 0.35 & -0.0011 $\pm$ 0.0007 $\pm$ 0.0002 &  -0.0058 $\pm$ 0.0002 $\pm$ 0.0001 & -0.0123 $\pm$ 0.0005 $\pm$ 0.0001 \\ 
 0.45 & -0.0009 $\pm$ 0.0007 $\pm$ 0.0003 &  -0.0078 $\pm$ 0.0002 $\pm$ 0.0002 & -0.0151 $\pm$ 0.0005 $\pm$ 0.0002 \\ 
 0.55 & -0.0035 $\pm$ 0.0007 $\pm$ 0.0003 &  -0.0088 $\pm$ 0.0002 $\pm$ 0.0002 & -0.0182 $\pm$ 0.0005 $\pm$ 0.0003 \\ 
 0.65 & -0.0029 $\pm$ 0.0007 $\pm$ 0.0003 &  -0.0107 $\pm$ 0.0002 $\pm$ 0.0001 & -0.0230 $\pm$ 0.0005 $\pm$ 0.0001 \\ 
 0.75 & -0.0034 $\pm$ 0.0008 $\pm$ 0.0003 &  -0.0123 $\pm$ 0.0002 $\pm$ 0.0000 & -0.0255 $\pm$ 0.0005 $\pm$ 0.0002 \\ 
 0.85 & -0.0041 $\pm$ 0.0008 $\pm$ 0.0008 &  -0.0134 $\pm$ 0.0002 $\pm$ 0.0001 & -0.0285 $\pm$ 0.0005 $\pm$ 0.0004 \\ 
 0.95 & -0.0027 $\pm$ 0.0008 $\pm$ 0.0007 &  -0.0154 $\pm$ 0.0003 $\pm$ 0.0002 & -0.0338 $\pm$ 0.0006 $\pm$ 0.0005 \\ 
 \hline
  \end{tabular}
  \label{table:v1eta_data}
\end{table*}

\end{document}

%% file: author.tex
\affiliation{Abilene Christian University, Abilene, Texas   79699}
\affiliation{AGH University of Science and Technology, FPACS, Cracow 30-059, Poland}
\affiliation{Alikhanov Institute for Theoretical and Experimental Physics, Moscow 117218, Russia}
\affiliation{Argonne National Laboratory, Argonne, Illinois 60439}
\affiliation{American Univerisity of Cairo, Cairo, Egypt}
\affiliation{Brookhaven National Laboratory, Upton, New York 11973}
\affiliation{University of California, Berkeley, California 94720}
\affiliation{University of California, Davis, California 95616}
\affiliation{University of California, Los Angeles, California 90095}
\affiliation{University of California, Riverside, California 92521}
\affiliation{Central China Normal University, Wuhan, Hubei 430079 }
\affiliation{University of Illinois at Chicago, Chicago, Illinois 60607}
\affiliation{Creighton University, Omaha, Nebraska 68178}
\affiliation{Czech Technical University in Prague, FNSPE, Prague 115 19, Czech Republic}
\affiliation{Technische Universit\"at Darmstadt, Darmstadt 64289, Germany}
\affiliation{E\"otv\"os Lor\'and University, Budapest, Hungary H-1117}
\affiliation{Frankfurt Institute for Advanced Studies FIAS, Frankfurt 60438, Germany}
\affiliation{Fudan University, Shanghai, 200433 }
\affiliation{University of Heidelberg, Heidelberg 69120, Germany }
\affiliation{University of Houston, Houston, Texas 77204}
\affiliation{Huzhou University, Huzhou, Zhejiang  313000}
\affiliation{Indian Institute of Science Education and Research (IISER), Berhampur 760010 , India}
\affiliation{Indian Institute of Science Education and Research, Tirupati 517507, India}
\affiliation{Indian Institute Technology, Patna, Bihar, India}
\affiliation{Indiana University, Bloomington, Indiana 47408}
\affiliation{Institute of Physics, Bhubaneswar 751005, India}
\affiliation{University of Jammu, Jammu 180001, India}
\affiliation{Joint Institute for Nuclear Research, Dubna 141 980, Russia}
\affiliation{Kent State University, Kent, Ohio 44242}
\affiliation{University of Kentucky, Lexington, Kentucky 40506-0055}
\affiliation{Lawrence Berkeley National Laboratory, Berkeley, California 94720}
\affiliation{Lehigh University, Bethlehem, Pennsylvania 18015}
\affiliation{Max-Planck-Institut f\"ur Physik, Munich 80805, Germany}
\affiliation{Michigan State University, East Lansing, Michigan 48824}
\affiliation{National Research Nuclear University MEPhI, Moscow 115409, Russia}
\affiliation{National Institute of Science Education and Research, HBNI, Jatni 752050, India}
\affiliation{National Cheng Kung University, Tainan 70101 }
\affiliation{Nuclear Physics Institute of the CAS, Rez 250 68, Czech Republic}
\affiliation{Ohio State University, Columbus, Ohio 43210}
\affiliation{Institute of Nuclear Physics PAN, Cracow 31-342, Poland}
\affiliation{Panjab University, Chandigarh 160014, India}
\affiliation{Pennsylvania State University, University Park, Pennsylvania 16802}
\affiliation{NRC "Kurchatov Institute", Institute of High Energy Physics, Protvino 142281, Russia}
\affiliation{Purdue University, West Lafayette, Indiana 47907}
\affiliation{Pusan National University, Pusan 46241, Korea}
\affiliation{Rice University, Houston, Texas 77251}
\affiliation{Rutgers University, Piscataway, New Jersey 08854}
\affiliation{Universidade de S\~ao Paulo, S\~ao Paulo, Brazil 05314-970}
\affiliation{University of Science and Technology of China, Hefei, Anhui 230026}
\affiliation{Shandong University, Qingdao, Shandong 266237}
\affiliation{Shanghai Institute of Applied Physics, Chinese Academy of Sciences, Shanghai 201800}
\affiliation{Southern Connecticut State University, New Haven, Connecticut 06515}
\affiliation{State University of New York, Stony Brook, New York 11794}
\affiliation{Temple University, Philadelphia, Pennsylvania 19122}
\affiliation{Texas A\&M University, College Station, Texas 77843}
\affiliation{University of Texas, Austin, Texas 78712}
\affiliation{Tsinghua University, Beijing 100084}
\affiliation{University of Tsukuba, Tsukuba, Ibaraki 305-8571, Japan}
\affiliation{United States Naval Academy, Annapolis, Maryland 21402}
\affiliation{Valparaiso University, Valparaiso, Indiana 46383}
\affiliation{Variable Energy Cyclotron Centre, Kolkata 700064, India}
\affiliation{Warsaw University of Technology, Warsaw 00-661, Poland}
\affiliation{Wayne State University, Detroit, Michigan 48201}
\affiliation{Yale University, New Haven, Connecticut 06520}

\author{J.~Adam}\affiliation{Brookhaven National Laboratory, Upton, New York 11973}
\author{L.~Adamczyk}\affiliation{AGH University of Science and Technology, FPACS, Cracow 30-059, Poland}
\author{J.~R.~Adams}\affiliation{Ohio State University, Columbus, Ohio 43210}
\author{J.~K.~Adkins}\affiliation{University of Kentucky, Lexington, Kentucky 40506-0055}
\author{G.~Agakishiev}\affiliation{Joint Institute for Nuclear Research, Dubna 141 980, Russia}
\author{M.~M.~Aggarwal}\affiliation{Panjab University, Chandigarh 160014, India}
\author{Z.~Ahammed}\affiliation{Variable Energy Cyclotron Centre, Kolkata 700064, India}
\author{I.~Alekseev}\affiliation{Alikhanov Institute for Theoretical and Experimental Physics, Moscow 117218, Russia}\affiliation{National Research Nuclear University MEPhI, Moscow 115409, Russia}
\author{D.~M.~Anderson}\affiliation{Texas A\&M University, College Station, Texas 77843}
\author{R.~Aoyama}\affiliation{University of Tsukuba, Tsukuba, Ibaraki 305-8571, Japan}
% ...

\collaboration{STAR Collaboration}\noaffiliation